\documentclass[useAMS,usenatbib]{mn2e}
\usepackage{amssymb}
\usepackage{multirow}
\usepackage{graphicx}

\def\lya{Ly$\alpha$}
\def\lyc{LyC}
\def\filteru{\textit{U}}
\def\filterb{\textit{B}}
\def\filterv{\textit{V}}
\def\filterr{\textit{R}}
\def\filteri{\textit{i}$'$}
\def\filterz{\textit{z}$'$}
\def\filterlyc{\textit{NB359}}
\def\filterlya{\textit{NB497}}
\def\filterbv{\textit{BV}}
\def\HI{H\,{\sevensize I}}
\def\popiii{{\sevensize Pop}\,{\sevensize III}}
\def\starburst{{\sevensize SB}\,{\sevensize 99}}
\def\ew{$EW($\lya$)$}
\def\ssa{{\sevensize SSA22}}
\def\flycfuv{$\left<f_{\lyc}/f_{UV}\right>_{obs}$}
\def\viobs{$\left<\textit{V}-\textit{i}'\right>_{obs}$}

\title[\lyc~sources at $z\sim 3$]
  {Searching for candidates of Lyman continuum sources - revisiting the SSA22 field\thanks{Based on data collected at Subaru Telescope, which is operated by the National Astronomical Observatory of Japan}}
\author[Micheva et al.]
  {Genoveva Micheva,$^1$
  Ikuru Iwata,$^{1,2}$
  Akio K. Inoue,$^3$
  Yuichi Matsuda,$^{2,4}$\\
  \newauthor 
  Toru Yamada,$^5$
  and Tomoki Hayashino$^6$\\
  $^1$Subaru Telescope, National Astronomical Observatory of Japan,
  650 North A'ohoku Place, Hilo, HI 96720, USA\\
  $^2$Graduate University for Advanced Studies (SOKENDAI), Osawa 2-21-1, Mitaka, Tokyo 181-8588, Japan \\
  $^3$College of General Education, Osaka Sangyo University,
  3-1-1, Nakagaito, Daito, 574-8530 Osaka, Japan.\\
  $^4$National Astronomical Observatory of Japan, Osawa 2-21-1, Mitaka, Tokyo 181-8588, Japan\\
  $^5$Astronomical Institute, Tohoku University, Aramaki, Aoba-ku, Sendai, Miyagi 980-8578, Japan\\
  $^6$Research Center for Neutrino Science, Graduate School of Science, Tohoku University, Aramaki, Aoba-ku,\\ Sendai, Miyagi 980-8578, Japan }
\date{Released 2015 Xxxxx XX}

\pagerange{\pageref{firstpage}--\pageref{lastpage}} \pubyear{2015}

\def\LaTeX{L\kern-.36em\raise.3ex\hbox{a}\kern-.15em
    T\kern-.1667em\lower.7ex\hbox{E}\kern-.125emX}

\begin{document}

\label{firstpage}

\maketitle

\begin{abstract}
We present the largest to date sample of hydrogen Lyman continuum (\lyc) emitting galaxy candidates at any redshift, with $18$ \lya~Emitters (LAEs) and $7$ Lyman Break Galaxies (LBGs), obtained from the SSA22 field with Subaru/Suprime-Cam. The sample is based on the $159$ LAEs and $136$ LBGs observed in the field, all with spectroscopically confirmed redshifts, and these \lyc~candidates are selected as galaxies with counterpart in a narrow-band filter image which traces \lyc~ at $z\geq 3.06$. Many \lyc~candidates show a spatial offset between the rest-frame non-ionizing ultraviolet (UV) detection and the \lyc-emitting substructure or between the \lya~emission and \lyc. The possibility of foreground contamination complicates the analysis of the nature of \lyc~emitters, although statistically it is highly unlikely that all candidates in our sample are contaminated by foreground sources. Many viable \lyc~LAE candidates have flux density ratios inconsistent with standard models, while also having too blue UV slopes to be foreground contaminants. Stacking reveals no significant \lyc~detection, suggesting that there is a dearth of objects with marginal \lyc~signal strength, perhaps due to a bimodality in the \lyc~emission. The foreground contamination-corrected $3\sigma$ upper limits of the observed average flux density ratios are $f_{\lyc}/f_{UV}<0.08$ from stacking LAEs and $f_{\lyc}/f_{UV}<0.02$ from stacking LBGs. There is a sign of a positive correlation between \lyc~and \lya, suggesting that both types of photons escape via a similar mechanism. The \lyc~detection rate among proto-cluster LBGs is seemingly lower compared to the field.
\end{abstract}


\section{Introduction}
The escape fraction $f_{esc}$ of Lyman continuum (\lyc) is the fraction of hydrogen ionizing radiation that escapes into the intergalactic medium (IGM) and contributes either to reionizing the Universe at redshifts $z\gtrsim6$ or to keeping it ionized at lower redshifts. Despite copious amount of research using ground-based and space telescopes~\citep[][]{1995ApJ...454L..19L,2001ApJ...546..665S,2006ApJ...651..688S, 2007ApJ...668...62S,2010ApJ...723..241S,2015ApJ...804...17S,2009ApJ...692.1476C, 2009ApJ...692.1287I,2010ApJ...725.1011V,2015A&A...576A.116V,2011ApJ...736...18N,2013ApJ...765...47N,2011A&A...532A.107L,2013A&A...553A.106L,2013ApJ...779...65M,2015ApJ...810..107M,2016arXiv160201555S,2016A&A...587A.133G} $f_{esc}$ is still poorly constrained due to its sensitivity to neutral hydrogen (\HI). With increasing redshifts the Universe becomes more opaque to \lyc~radiation and directly measuring $f_{esc}$ from individual galaxies becomes almost impossible beyond $z\sim5$~\citep{2008MNRAS.387.1681I,2014MNRAS.442.1805I}, although indirect measurements should be possible well into the epoch of reionization with the upcoming James Webb Space Telescope~\citep{2013ApJ...777...39Z}. At low redshifts the IGM is much more transparent. However, efforts to measure $f_{esc}$ from $z\lesssim2$ galaxies have so far resulted only in upper limits~\citep[e.g.,][]{1995ApJ...454L..19L,2009ApJ...692.1476C,2007ApJ...668...62S,2010ApJ...723..241S}, with only few detections of \lyc~leakage from $z\sim0$ galaxies~\citep[][]{2011A&A...532A.107L,2013A&A...553A.106L,2016arXiv160306779L,2016Natur.529..178I}. On a side note, space-based instruments, whose availability is relatively limited, are required for studies at $z\lesssim2$, while at $z\gtrsim3$ the more numerous ground-based telescopes are well-suited for this type of research. The redshift that has proven most fruitful in the search for \lyc~leaking candidates is $z\sim3$, with tens of candidates found in the \ssa~proto-cluster at $z\sim3.1$~\citep{2009ApJ...692.1287I,2011ApJ...736...18N,2013ApJ...765...47N}, or in the {\sevensize HS1549+1933} field at $z\sim2.85$~\citep{2013ApJ...779...65M}, although the latter detections have been recently revised~\citep{2015ApJ...810..107M}. Evidence of the evolution of $f_{esc}$ can be found in the literature~\citep[e.g.,][]{2006MNRAS.371L...1I,2009ApJ...692.1287I,2010ApJ...723..241S,2013A&A...554A..38B,2014MNRAS.440.3309D,2016arXiv160201555S}, with low-$z$ galaxies showing a practically zero escape fraction in spite of the usually much fainter limiting magnitude in surveys of local galaxies, and the higher redshift galaxies showing large ionizing emissivities. 
\subsection{The \ssa~proto-cluster}
First discovered by~\citet{1998ApJ...492..428S} the \ssa~proto-cluster at $z\sim3.1$ is one of the most dense concentrations of galaxies known to date. The abundance of galaxies populating this field lends itself well to the search and analysis of various types of galaxies and their properties, e.g. the formation of massive galaxies~\citep{2012ApJ...750..116U,2013ApJ...778..170K}, the so-called Lyman $\alpha$ Blobs~\citep{2004AJ....128..569M}, or large scale structure and clustering~\citep{2004AJ....128.2073H,2012AJ....143...79Y}. With improved observing strategies, specially designed filters with precise transmission ranges, and with the combining of years of existing observations, the number of significantly detected Lyman $\alpha$ Emitters (LAEs) and Lyman Break Galaxies (LBGs) in the \ssa~field has been revised and increased. Such a rich, well-defined, $z\sim3.1$ sample can in turn be used to search for galaxies showing any evidence of \lyc~leakage.\\ 

\noindent One such work is that by~\citet{2009ApJ...692.1287I} who compile a sample of \lyc~candidates using direct detections in a narrowband filter. We re-analyze the data, including newer deeper data in \filterb\filterv\filterr\filteri\filterz of the same field and present a larger catalog of additional \lyc~detections, with updated photometry and improved astrometry in the \lyc~filter. We now have a larger base sample of both LAEs and LBGs, and a larger \lyc~candidate sample. Some of the \lyc~candidates in~\citet{2009ApJ...692.1287I} already show observed flux density ratios $f_{\lyc}/f_{UV}\gtrsim1$. A possible solution to this was already presented in~\citet{2010MNRAS.401.1325I} with the Lyman Bump model, in which nebular \lyc~escapes through matter-bound nebulae along the same paths as stellar~\lyc~and boosts the ionizing to non-ionizing flux density ratio. Additionally,~\citet{2011MNRAS.411.2336I} assume a two component model in which a primordial stellar population (PopIII) and a normal stellar population with subsolar metallicity and Salpeter initial mass function (IMF) coexist. This model simultaneously reproduces the observed non-ionizing UV slope and the large flux density ratios of the \lyc~candidates. \\

\noindent Detailed analysis of the observed flux density ratios is beyond the scope of this paper, however, and here we only present the new \lyc~candidate sample and the extended base sample of the $295$ LAEs and LBGs at spectroscopically confirmed redshifts of $z\geq3.06$, with observed flux density ratios from average stacking of the LAEs and LBGs. This work is accompanied by an online photometric catalog of all galaxies in our sample in the \filterb\filterv\filterr\filteri\filterz~broadband and \filterlya,~\filterlyc~narrowband filters. \\
\subsection{Structure of this paper}
\noindent In Section~\ref{sec:obs} we present the observations and image quality of the data. Section~\ref{sec:sampleselection} shows the data selection and contamination analysis. The photometry of the full sample is in Section~\ref{sec:photometry}. In Section~\ref{sec:results} we examine the properties of the \lyc~sample. Sections~\ref{sec:discuss} and~\ref{sec:conclusions} show the discussion and conclusions. Appendix~\ref{appendix:notes} shows available HST images and complementary unpublished spectra for the \lyc~sample, as well as notes on individual objects. Appendix~\ref{appendix:brightvar} shows light curves from the brightness variability test. Throughout this paper we adopt a flat $\Lambda$CDM cosmology with $H_{0}=70$ km s${}^{-1}$ Mpc${}^{-1}$, $\Omega_{M}=0.3$, and $\Omega_{\Lambda}=0.7$. All magnitudes are in the AB magnitude system.

\setcounter{table}{0}
\begin{table*}
\begin{minipage}{206mm}
\caption{\lyc~candidates in the \ssa~field. Objects with AGN spectral features are excluded.}
\protect\label{tab:coordinates}
\scriptsize
\begin{tabular}{l c c c l l}
ID  &  $\alpha$   & $\delta$    &   $z$  &   $z$ Reference${}^{\dagger}$ & NED name \\
\hline
\multicolumn{6}{c}{LAE}\\
LAE01${}^{\star}$  &  22~16~46.9  &  00~26~25.5  &  3.099  & FOCAS 2003/2010  &- \\
LAE02${}^{\star}$  &  22~16~51.4  &  00~25~02.4  &  3.127  & M04,FOCAS 2010  & LAB 06\\
LAE03${}^{\star}$  &  22~16~52.7  &  00~27~05.0  &  3.090  & FOCAS 2010  & - \\
LAE04${}^{\star}$  &  22~16~53.9  &  00~21~37.1  &  3.085  & Y12 & LAE J221653.9+002137\\
LAE05${}^{\star}$  &  22~16~59.3  &  00~25~00.6  &  3.088  & Y12 &  LAE J221659.3+002501 \\
LAE06              &  22~17~08.0  &  00~19~32.0  &  3.075  & I11 &  [IKI2011] f \\
LAE07${}^{\star}$  &  22~17~16.7  &  00~23~08.3  &  3.065  & I11,Y12 & LAE J221716.7+002309 \\
LAE08${}^{\star}$  &  22~17~29.4  &  00~06~28.2  &  3.080  & I11 & [IKI2011] h \\
LAE09${}^{\star}$  &  22~17~34.8  &  00~15~41.1  &  3.099  & N13 [LAE 038] &- \\
LAE10${}^{\star}$  &  22~17~39.0  &  00~17~25.7  &  3.090  & Y12 & LAE J221739.0+001726 \\
LAE11${}^{\star}$  &  22~17~39.2  &  00~22~41.7  &  3.098  &  Y12 & LAE J221739.2+002242 \\
LAE12${}^{\star}$  &  22~17~40.3  &  00~11~28.8  &  3.065  & Y12  & LAE J221740.3+001129  \\
LAE13${}^{\star}$  &  22~17~43.3  &  00~21~48.5  &  3.097  &  Y12 & LAE J221743.3+002149 \\
LAE14              &  22~17~45.9  &  00~23~18.7  &  3.095  & I11,Y12 & LAE J221745.9+002319 \\
LAE15              &  22~17~53.2  &  00~12~37.4  &  3.094  & I11,Y12 & LAE J221753.2+001238  \\
LAE16              &  22~18~ 8.0  &  00~11~50.8  &  3.096  & Y12 & LAE J221808.0+001151  \\
LAE17              &  22~18~13.9  &  00~22~21.8  &  3.089  &  Y12 & LAE J221813.9+002222 \\
LAE18${}^{\star}$  &  22~18~20.2  &  00~12~35.6  &  3.087  & Y12 & LAE J221820.8+001241  \\
\hline
\multicolumn{6}{c}{LBG}\\
LBG01             &  22~16~47.1  &  00~18~43.2  &  3.680  &  VIMOS 2006 & -  \\
LBG02             &  22~17~01.4  &  00~27~07.8  &  3.113  &  VIMOS 2008/FOCAS 2010  &- \\
LBG03             &  22~17~08.0  &  00~09~58.3  &  3.287  &  VIMOS 2006 &- \\
LBG04             &  22~17~23.5  &  00~03~57.3  &  3.311  &  S01,S03  &SSA 22b oct96D08\\
LBG05${}^{\star}$   &  22~17~23.7  &  00~16~01.4  &  3.102  & FOCAS 2008/S03  &SSA 22a MD032 \\ 
LBG06${}^{\star}$   &  22~17~27.3  &  00~18~09.9  &  3.080  & FOCAS 2008/VIMOS 2008/S03/N13  & SSA 22a MD046 \\
LBG07${}^{\star}$   &  22~17~37.9  &  00~13~44.2  &  3.094  & LRIS 2010/ S03 & SSA 22a MD014 \\
\hline
\multicolumn{6}{c}{Confirmed contaminants in the \lyc~image}\\
ID & $\alpha$ & $\delta$ &z &NED name&Ref${}^{\ddagger}$\\
LBG08  &  22~17~19.8  &  00~18~19.0  &  3.151  &SSA 22a C049&S15\\
LAE19  &  22~17~24.8  &  00~17~17.0  &3.100& - &N13, Na13 \\
LAE20  &  22~17~26.2  &  00~13~19.3  &3.100& [IKI2011] d &N13\\
LBG09  &  22~17~30.9  &  00~13~10.7 &3.285& SSA 22a aug96M016&N13\\
LAE21  &  22~17~36.7  &  00~16~28.8  &  3.091&[NSS2011] LAE 025&N13  \\
\hline\end{tabular}
\end{minipage}
\medskip\begin{flushleft}
$\dagger$: Redshift reference: I11 - \citet{2011MNRAS.411.2336I}, N13 - \citet{2013ApJ...765...47N}, S01 - \citet{2001ApJ...562...95S}, S03 - \citet{2003ApJ...592..728S}, M04 - \citet{2004AJ....128..569M}, Y12 - \citet{2012ApJ...751...29Y} \\
$\ddagger$: Contamination confirmation reference: N13 - \citet{2013ApJ...765...47N}, S15 - \citet{2015ApJ...804...17S}, Na13 - \citet{2013ApJ...766..122N}\\
$\star$: possible but unconfirmed contaminant\end{flushleft}
\end{table*}

\section{Observations and data calibration}\protect\label{sec:obs}
The data we use in our analysis have been previously compiled and published~\citep{2011MNRAS.412.2579N,2004AJ....128.2073H,2009ApJ...692.1287I}. While two of those studies concentrate on the search for high redshift \lya~emitters in the direction of the \ssa~field,~\citet{2009ApJ...692.1287I} perform an analysis similar to ours, using the same set of filters. Our re-analysis, however, utilizes the deeper available data from~\citet{2011MNRAS.412.2579N} and improved astrometry in the \filterlyc~filter, resulting in a larger \lyc~candidate sample. The astrometric uncertainty in the \filterlyc~band is now $\sim0.2$\arcsec.

\subsection{Subaru/Suprime-Cam Data}
The primary optical imaging data were collected with the Subaru $8.2$ meter telescope on Maunakea during the years $2001-2008$, and consist of Suprime-Cam~\citep{2002PASJ...54..833M} broadband~\citep{2011MNRAS.412.2579N} and narrowband~\citep[][]{2004AJ....128.2073H,2009ApJ...692.1287I} observations of the \ssa~proto-cluster field ($\alpha=22^{h}17^{m}34^{s},\delta=+00\degr17'00\arcsec$; J2000) in the filters Johnson-Cousins \filterb, \filterv, \filterr ~and SDSS \filteri, \filterz, as well as those with \filterlyc~(FWHM$=15$ nm, central $\lambda=359$ nm) and \filterlya~(FWHM$=7.7$ nm, central $\lambda=497.7$ nm). The broadband filters cover the rest-frame UV regime of redshift $\sim3$ sources, while the narrowband \filterlyc~samples the \lyc~region for galaxies at $z\geq3.06$ and \filterlya~samples the redshifted Lyman $\alpha$ (\lya) line for galaxies at $3.06<z<3.13$. It is noteworthy that the \filterlyc~filter was especially designed by our group for the search of ionizing radiation from redshift $z\geq3.06$ sources and has a transmission of less than $0.01\%$ beyond a wavelength of $400$ nm in the laboratory. The size of the surveyed field is $530$ arcmin${}^2$. \\

\subsubsection{Calibration check}
We compared our photometry to the literature using AGN in our field of view~\citep{2008A&A...492..637G} in the \filterb, \filterv, and \filterr~bands, and data from the Gemini Deep Deep Survey~\citep[GDDS,][]{2004AJ....127.2455A} in the \filterv~band. We found no systematic offsets in our measurements. In the \filteri~and \filterz~bands we compared our stellar photometry to the SDSS. We found no systematic differences, however we note that since the Suprime-Cam detector becomes non-linear above $\sim30000$ counts all bright stars in our frames (\filteri$\lesssim 20.5$ mag) are either saturated or in the non-linear regime and hence could not be used for comparison. At the same time the number of faint stars in our field of view with reliable SDSS photometry decreases rapidly beyond \textit{i} $\gtrsim21.5$ mag, hence the comparison region was limited to a $1.5$ magnitude range and a few hundred stars. As a final check we verified that our stellar photometry is consistent with the Kurucz and Pickles synthetic stellar libraries in all filters~\citep{1993yCat.6039....0K,1998yCat..61100863P}.\\

\setcounter{figure}{0}
\begin{figure}
\centering\tiny
\includegraphics[width=85mm]{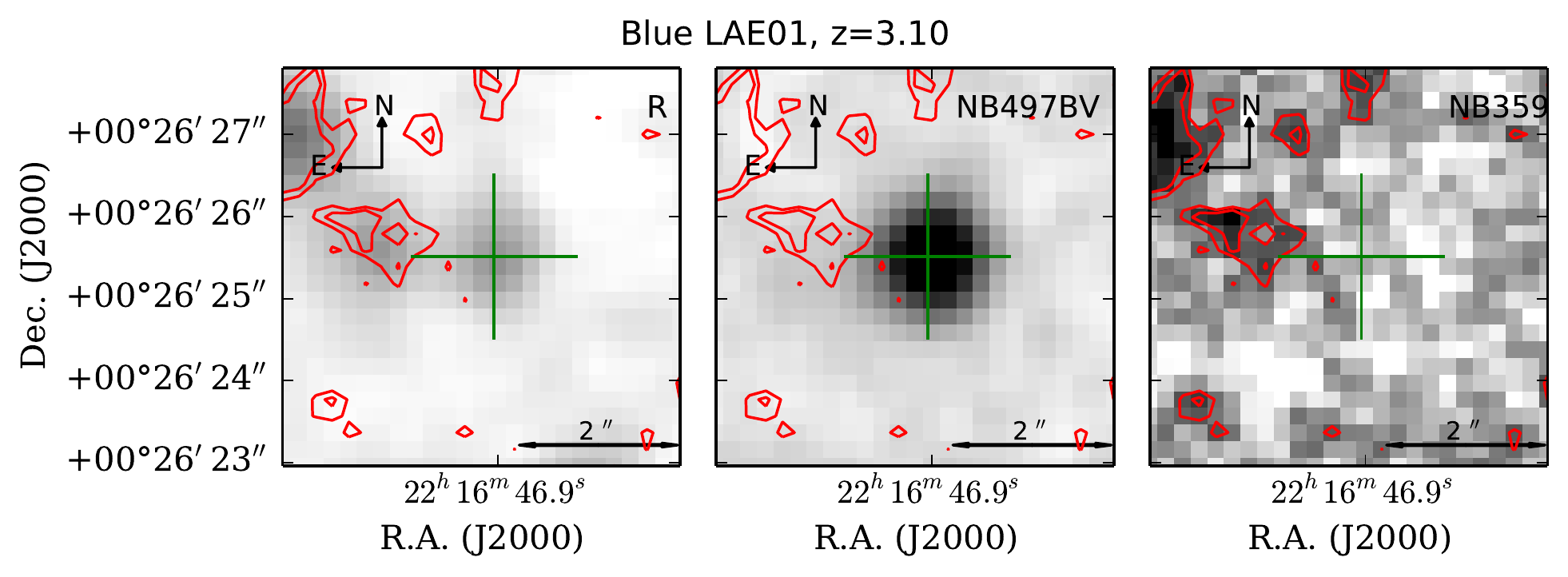}
\includegraphics[width=85mm]{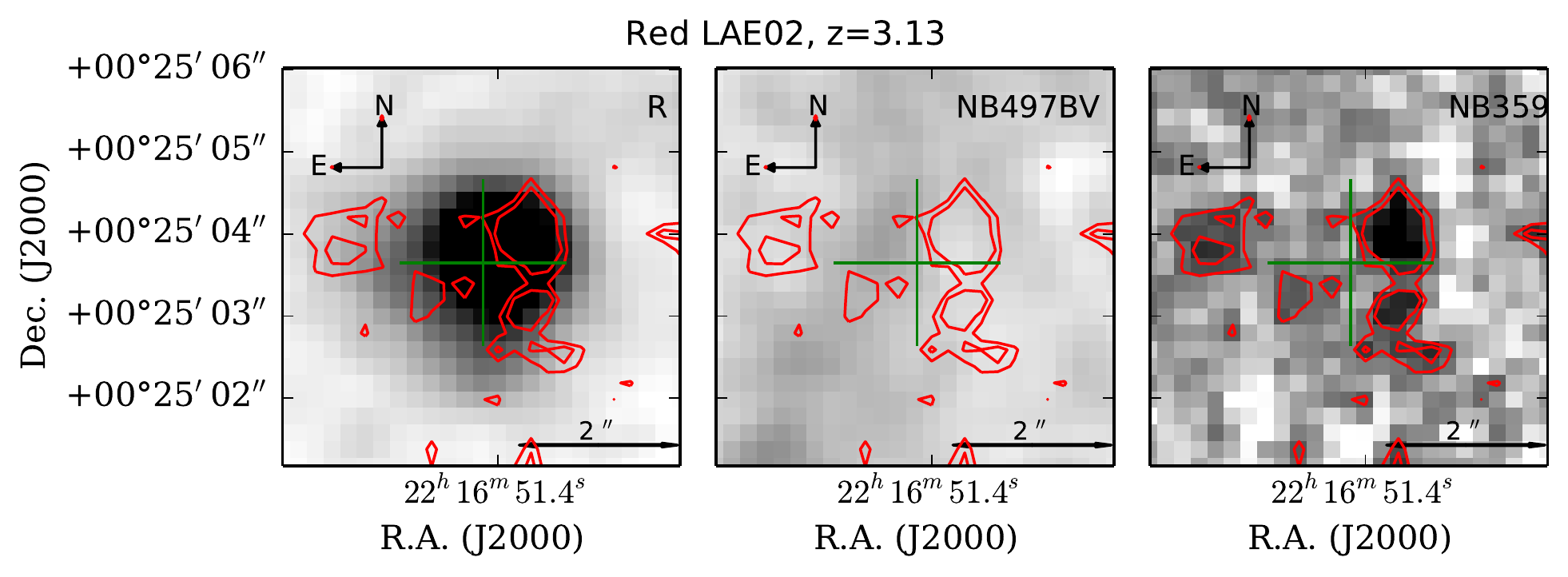}
\includegraphics[width=85mm]{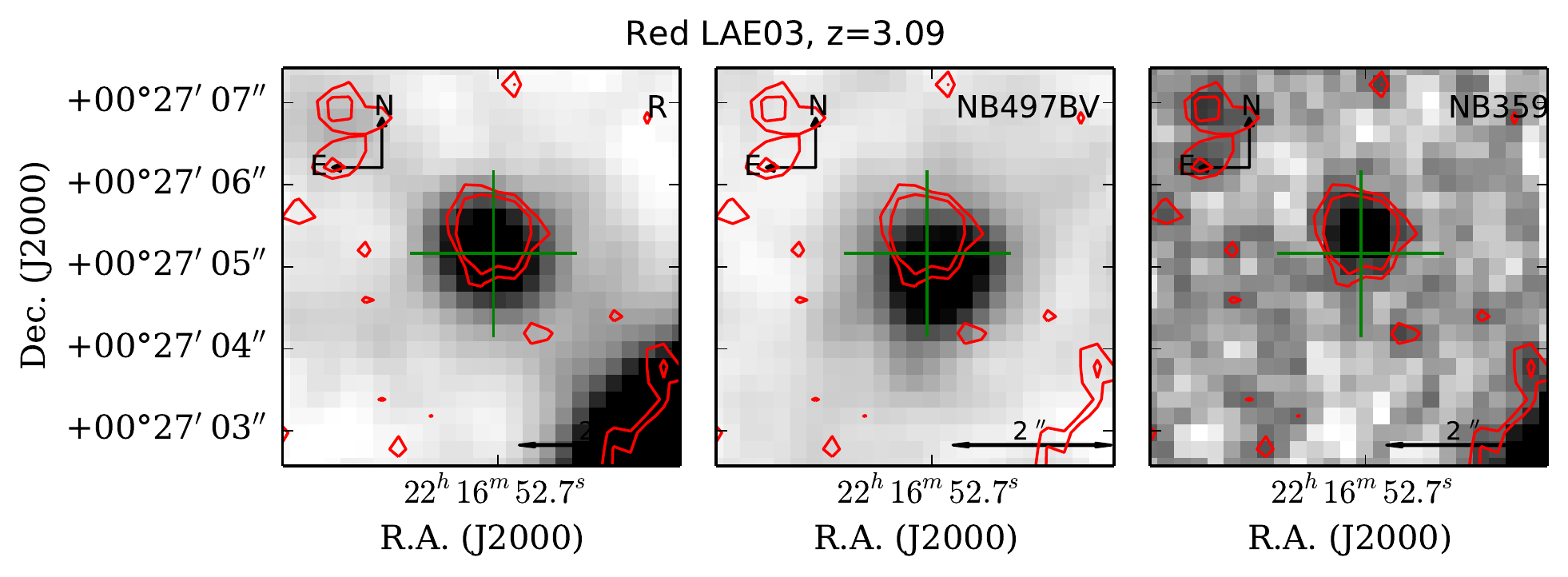}
\includegraphics[width=85mm]{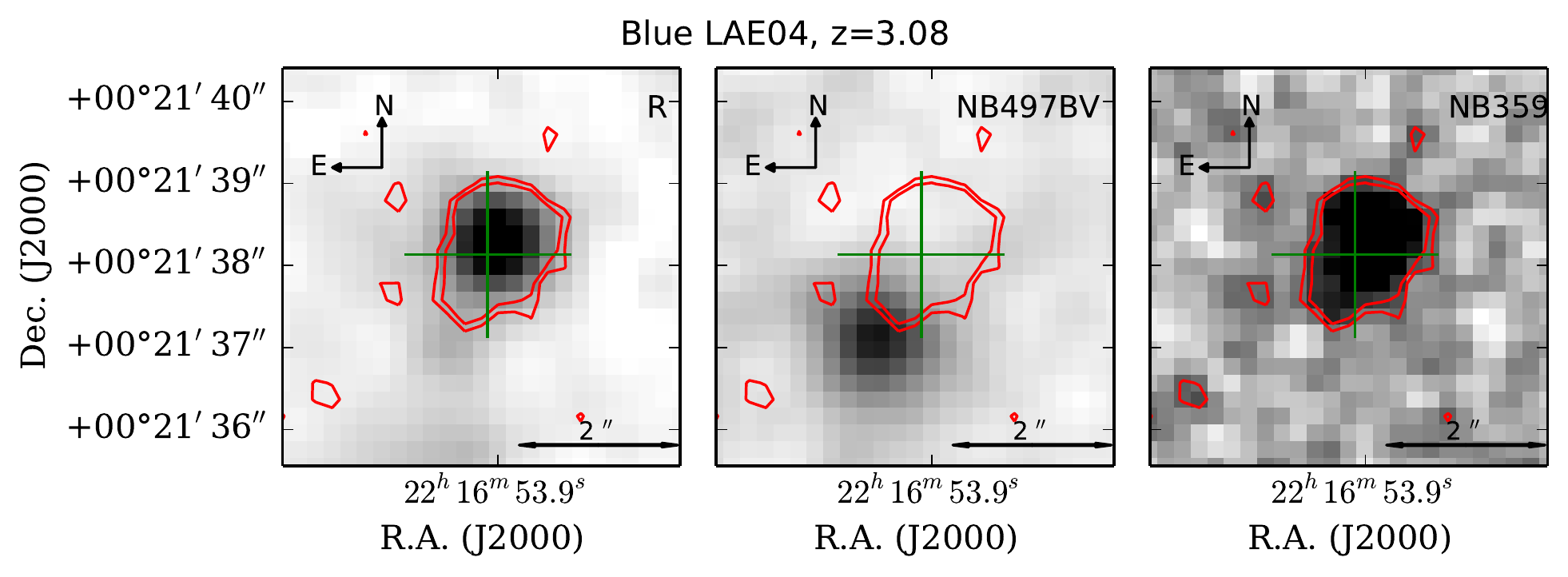}
\caption{LAE \lyc~candidates. The green cross marks the location of the \filterr~band detection. The filter order is always \filterr, \filterlya$-$\filterbv~(continuum-subtracted~\lya), \filterlyc~(\lyc). The $2$ and $3\sigma$ contours trace the \lyc~detection in the \filterlyc~band, and are overplotted in all other bands. For display purposes each filter has global contrast settings, so that direct brightness comparison between LAE objects is possible. For example LAE02 is very faint in \lya~relative to LAE01 or LAE03. The scale bar is 2\arcsec in all images. The color map is inverted.}\protect\label{fig:contours}
\end{figure}

\subsubsection{Image quality and limiting magnitude}\protect\label{image_quality}
The PSF FWHM was measured in all filters by fitting a Moffat function to stars in the field. The number of stars varied from filter to filter but the final FWHM was always the median of $\gtrsim1200$ stars. The \filterb~band PSF had FWHM$=0.78\arcsec\pm0.12\arcsec$. For the \filterv, \filterr, \filteri, \filterz, \filterlyc, and \filterlya~bands the measurements were $0.82\arcsec\pm0.12\arcsec$, $1.06\arcsec\pm0.12\arcsec$, $0.78\arcsec\pm0.22\arcsec$, $0.77\arcsec\pm0.18\arcsec$, $0.80\arcsec\pm0.27\arcsec$, and $1.00\arcsec\pm0.11\arcsec$, respectively. When performing analysis involving all seven filters, we therefore convolve all images to a common PSF of $\sim1.00\arcsec$. However, since the PSF is not only better in the majority of filters but also fairly similar, we also separately analyze the unconvolved images in the \filterb,\filterv, \filteri, \filterz, and \filterlyc~bands. Although there is scatter, there is no discernible systematic pattern in the spatial distribution of stellar FWHMs across the frames in any filter. The $3\sigma$ limiting magnitudes were obtained from $\sim20000$ rectangular apertures with area equivalent to a $\diameter=1.2\arcsec$ circular aperture (difference $\sim1\%$) on empty sky regions and measured to be \filterb~$=28.6^{m}$, \filterv~$=28.3^{m}$, \filterr~$=28.1^{m}$, \filteri~$=27.8^{m}$, \filterz~$=27.2^{m}$, \filterlyc~$=27.4^{m}$, and \filterlya~$=27.7^{m}$ on $1.0\arcsec$ PSF convolved and Galactic extinction corrected frames.\\

\noindent All photometry has been corrected for Galactic extinction following~\citet{2011ApJ...737..103S}, with individual $A_V$ values obtained for each object from the galactic Dust Reddening and Extinction calculator at the NASA/IPAC Infrared Science Archive. The individual extinction $A_V$ corrections in magnitudes are provided in the online catalog.

\subsubsection{Images and Contours}\protect\label{sec:contours}
In Figures~\ref{fig:contours} and~\ref{fig:contours2} we present images of our \lyc~candidates in the reference \filterr~band, and the two other most informative bands, \filterlyc~(\lyc) and \filterlya~(\lya~for sources at $z\sim3.09$). To highlight the \lya~emission we actually show \filterlya$-$\filterbv, which subtracts the continuum making it easier to detect any offsets between \lya~and \lyc. The \filterbv~image was constructed following $(2B+V)/3$~\citep{2004AJ....128.2073H}. Overplotted on all images are the $2$ and $3\sigma$ contours of the \filterlyc~detection, and the position of the \filterr~detection is marked with a green cross.  

\subsection{HST data}
\noindent For some targets Hubble Space Telescope (HST) Advanced Camera for Surveys (ACS) and/or Wide Field Camera 3 (WFC3) data is available from the archive in filters overlapping our observed wavelength range. We use this data only for visual analysis of the visible substructure since ground-based resolution precludes such insight. World coordinate system (WCS) inconsistencies in the HST images compared to our data were corrected beforehand with a full plate solution where possible. The LyC candidates with available HST data are summarized in Table~\ref{tab:HST}, the images shown in Figure~\ref{fig:hst}.\\

\section{Sample selection}\protect\label{sec:sampleselection}
We found a total of $308$ galaxies ($136$ LBGs, $159$ LAEs, and $13$ AGN) in the \ssa~field, all with spectroscopically confirmed redshifts of $z\ge3.06$. To qualify as a \lyc~candidate a source had to be detected at $\geq3\sigma$ significance with $1.2\arcsec$ diameter aperture and it had to be within a radius $r=1.4\arcsec$ ($\sim10$ kpc at $z\sim3.1$) from the \filterr~band detection position. The obtained $30$ \lyc~candidates, listed in Table~\ref{tab:coordinates}, contained five confirmed contaminants which we include in our contamination estimation (Sec.~\ref{sec:contamination}) but exclude from the all other analysis and figures. Excluding the confirmed contaminants the sample of \lyc~candidates consists of $7$ LBGs and $18$ LAEs. Suprime-Cam images are presented in Figures~\ref{fig:contours} (LAEs) and \ref{fig:contours2} (LBGs), and the accompanying spectra in Figure~\ref{fig:spectra} in the appendix.\\

\begin{figure*}
\centering\tiny
\includegraphics[width=85mm]{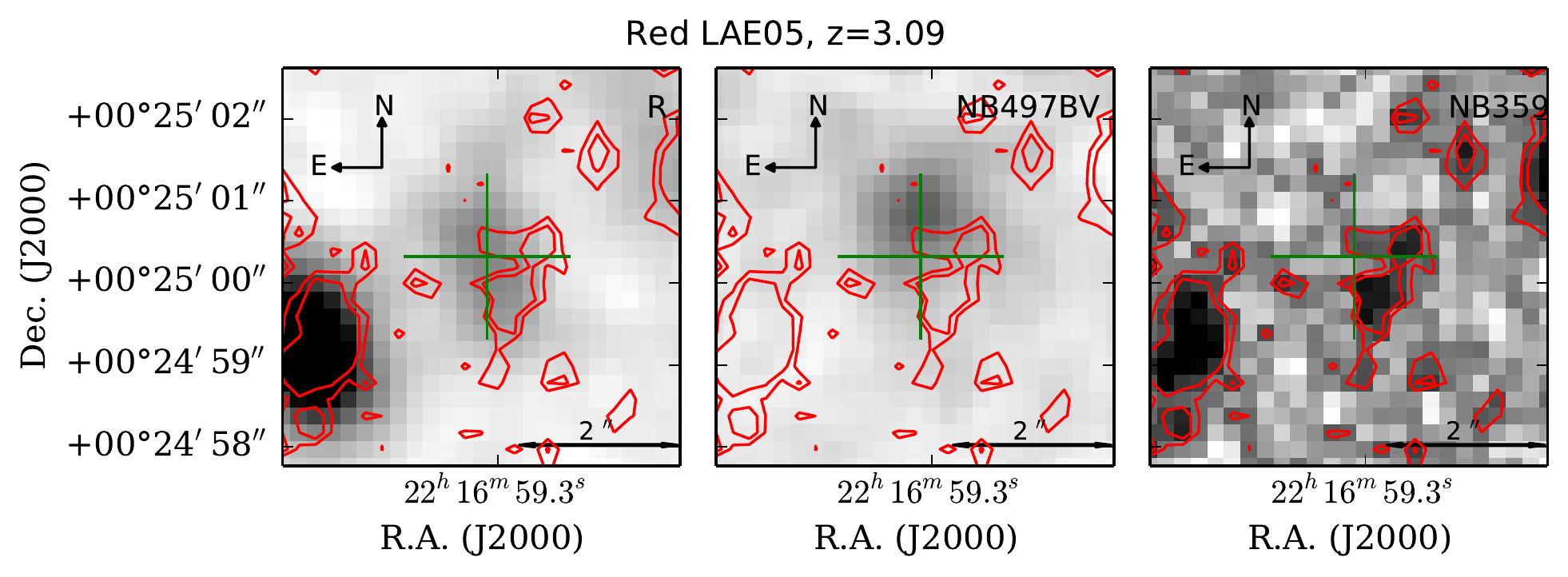}
\includegraphics[width=85mm]{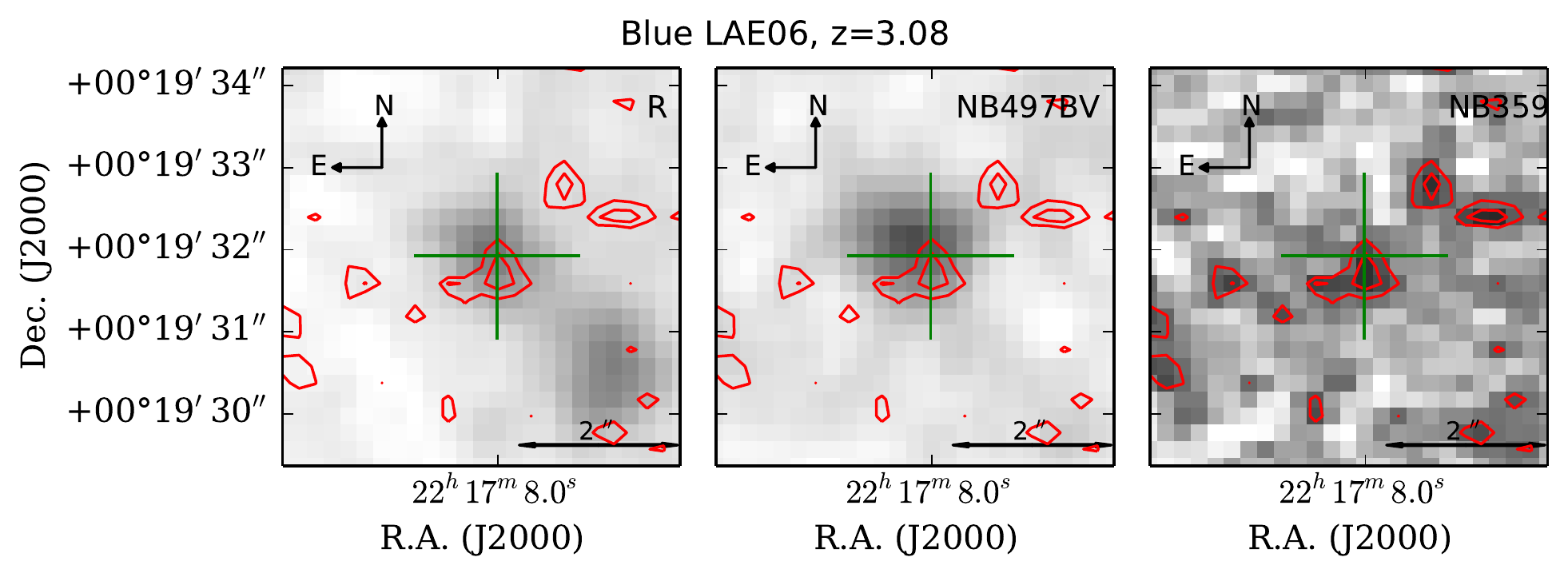}
\includegraphics[width=85mm]{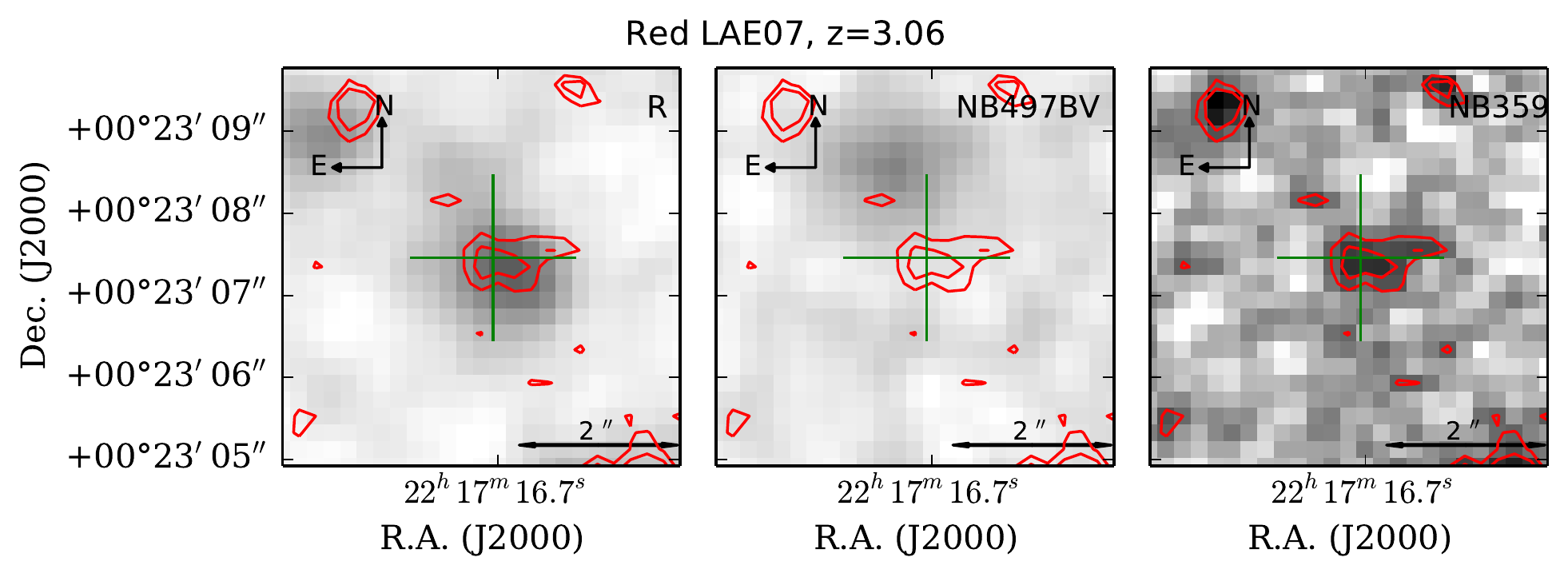}
\includegraphics[width=85mm]{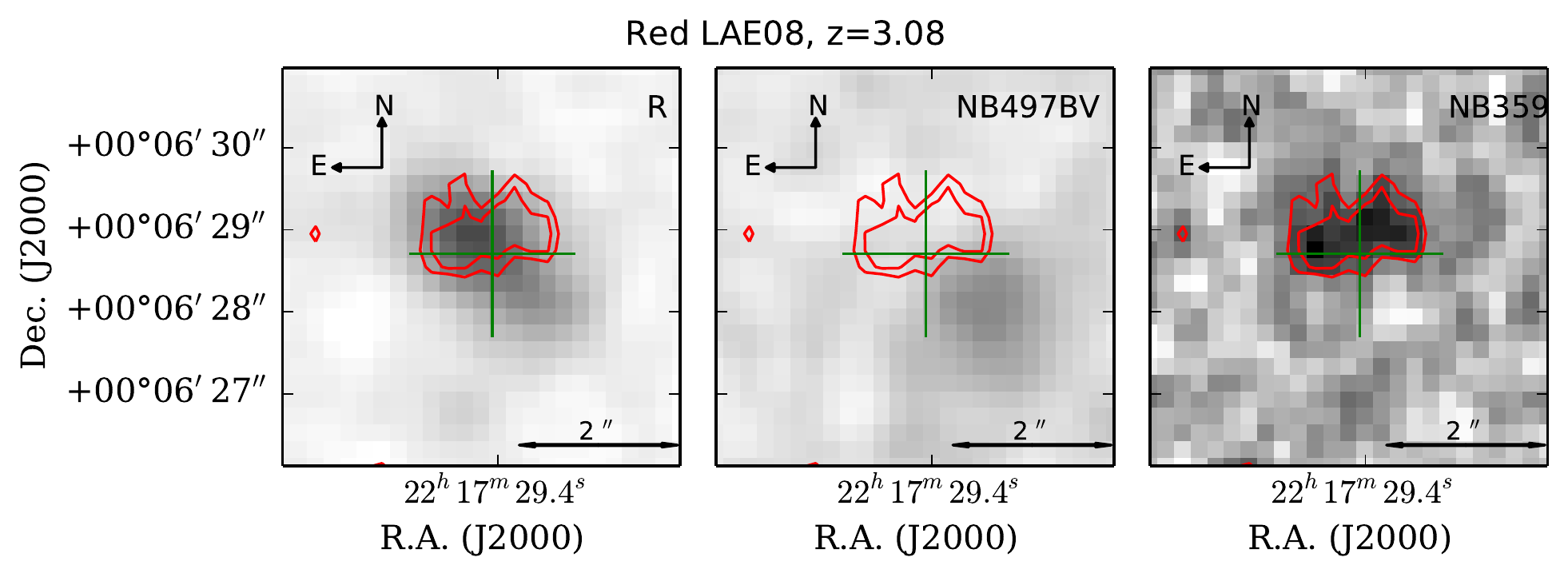}
\includegraphics[width=85mm]{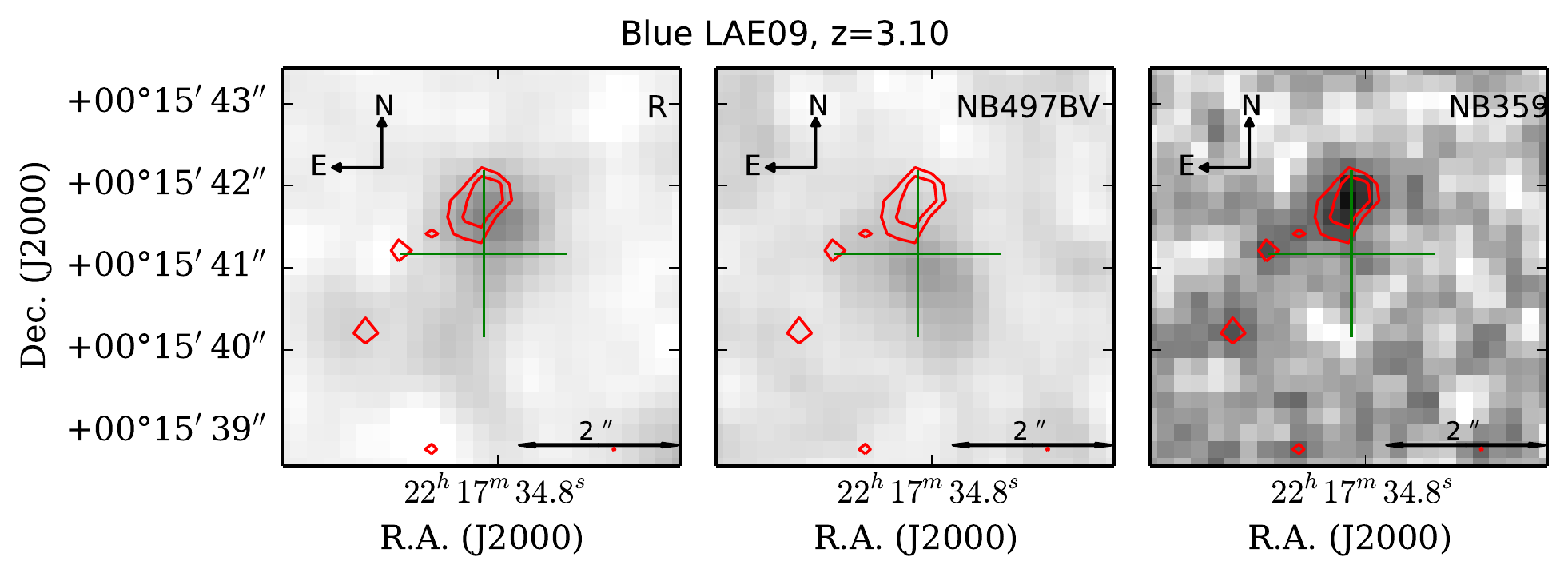}
\includegraphics[width=85mm]{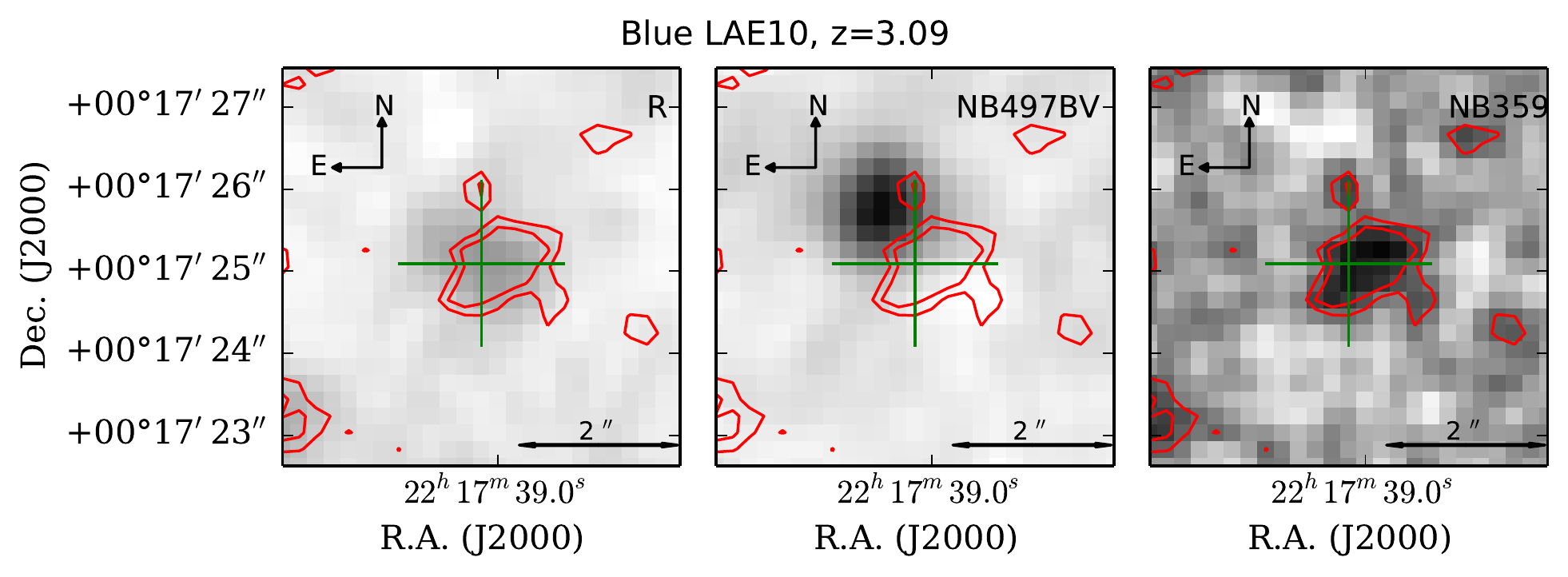}
\includegraphics[width=85mm]{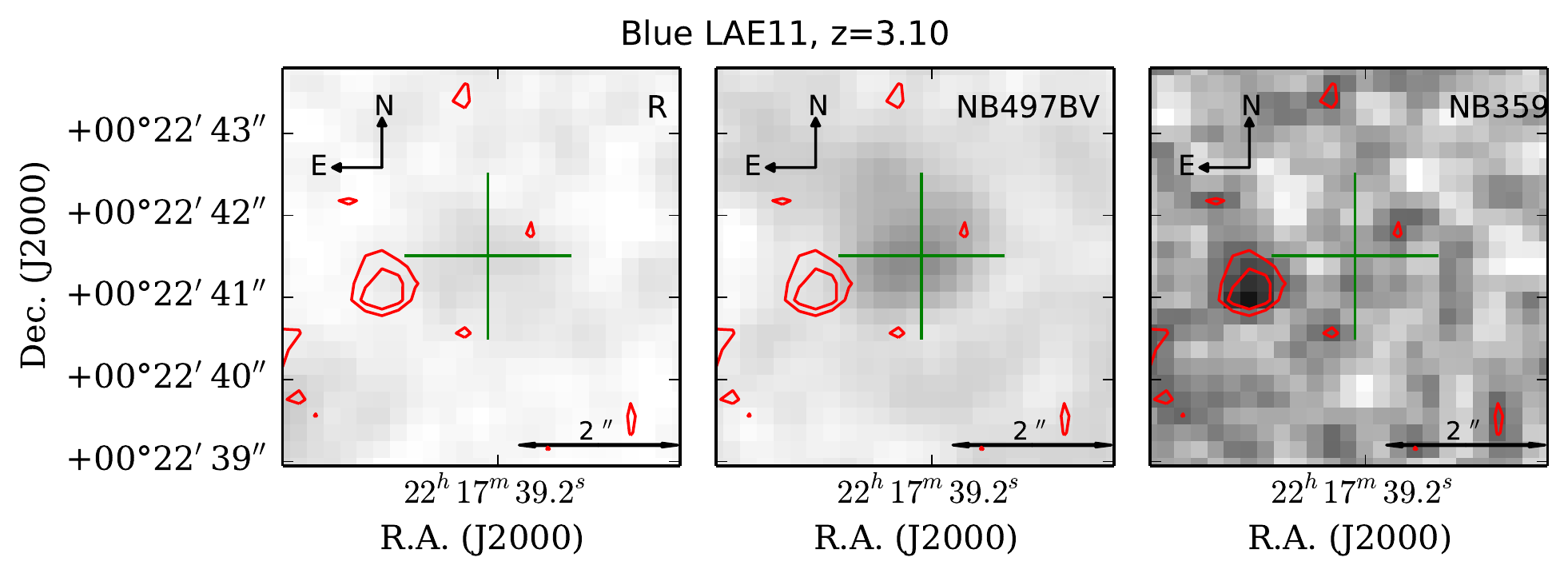}
\includegraphics[width=85mm]{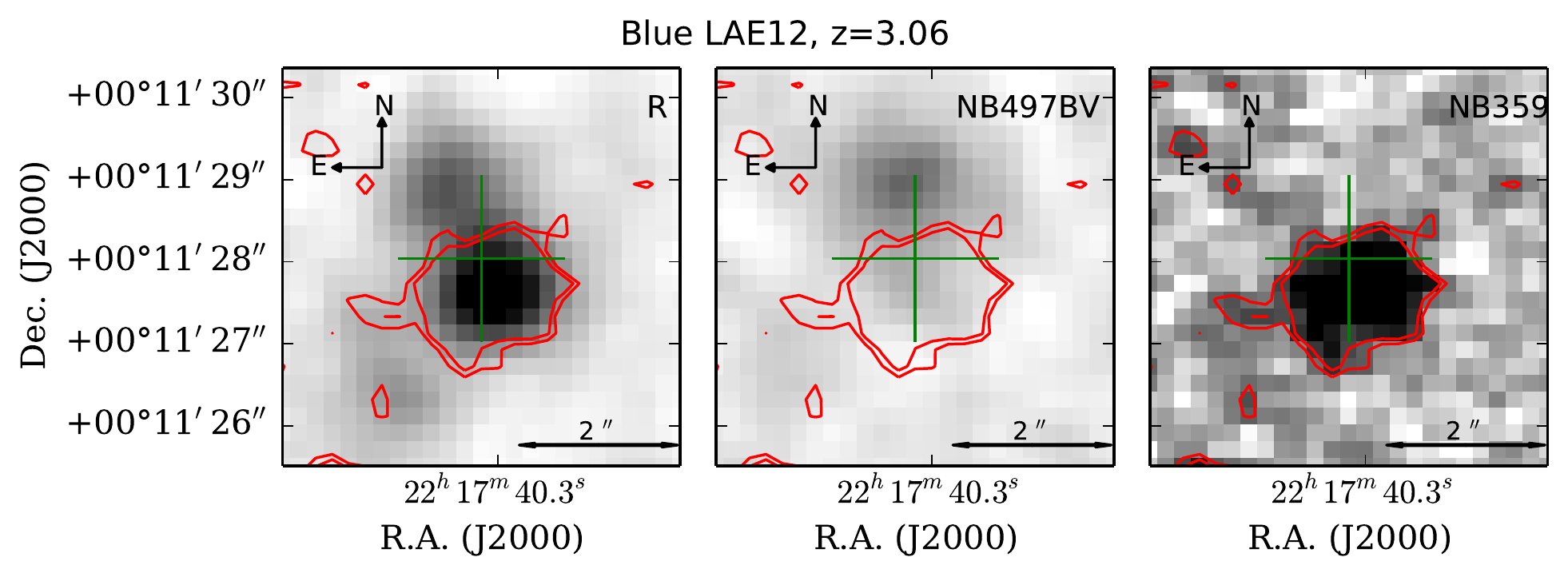}
\includegraphics[width=85mm]{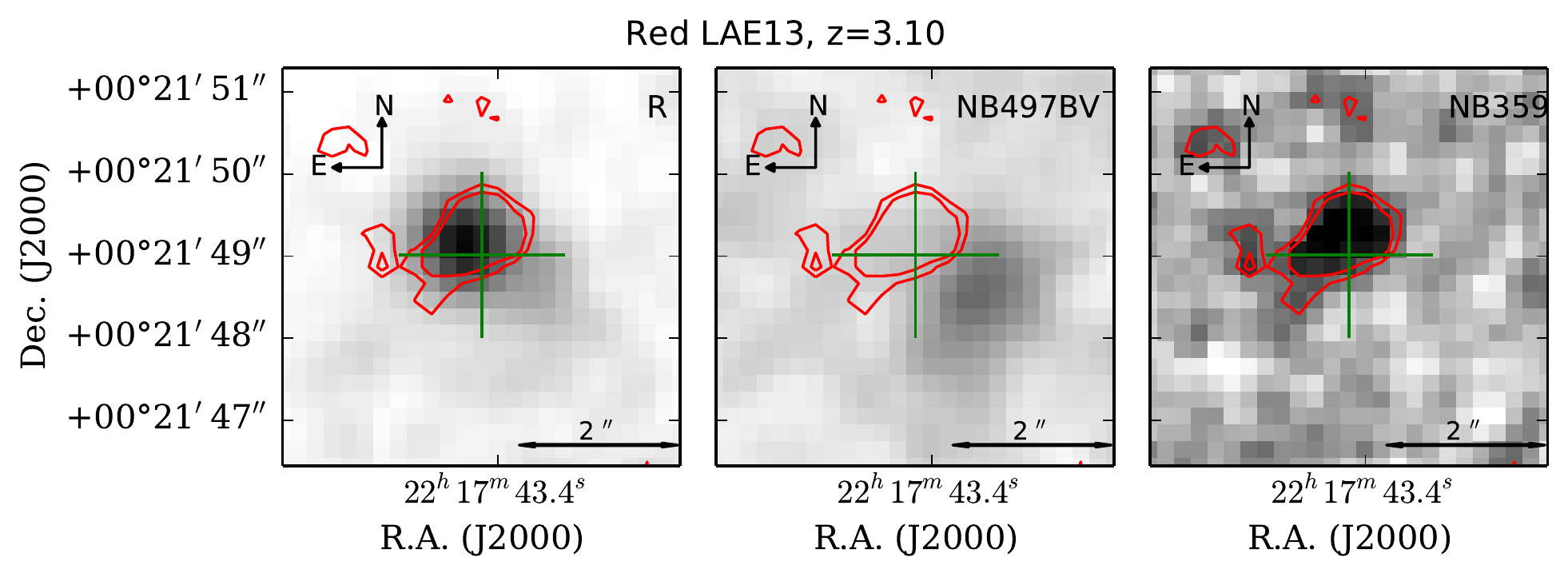}
\includegraphics[width=85mm]{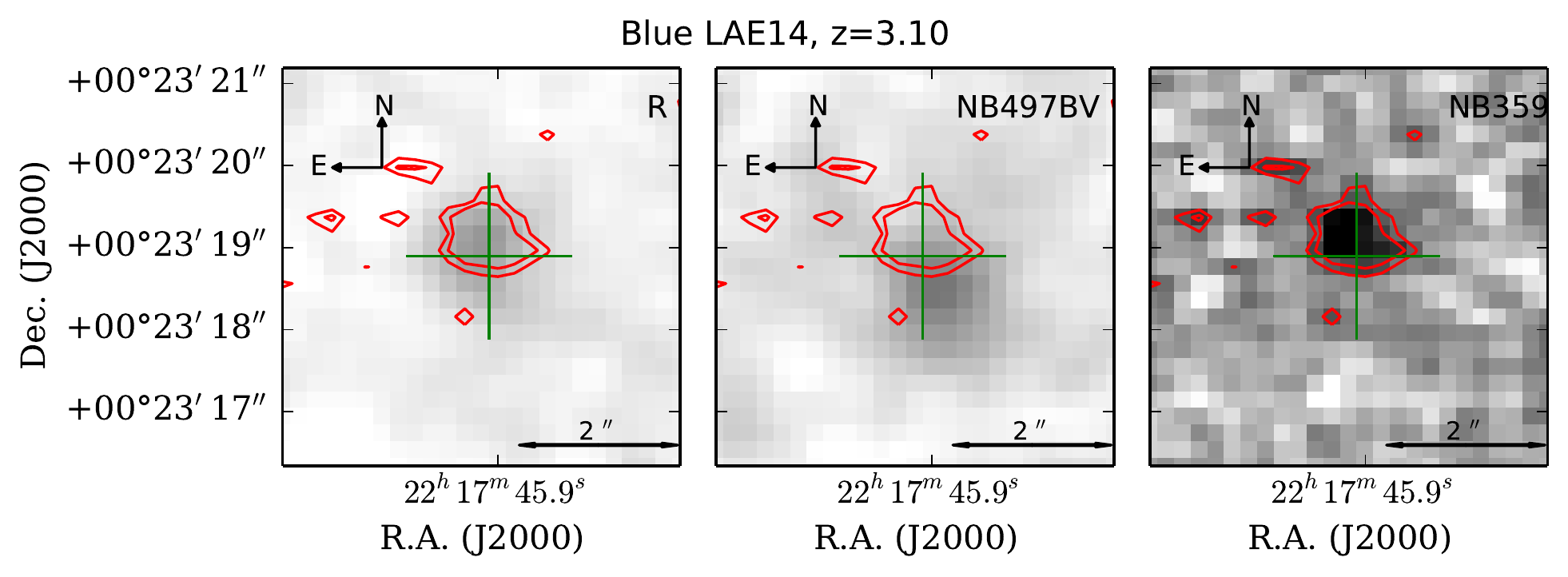}
\includegraphics[width=85mm]{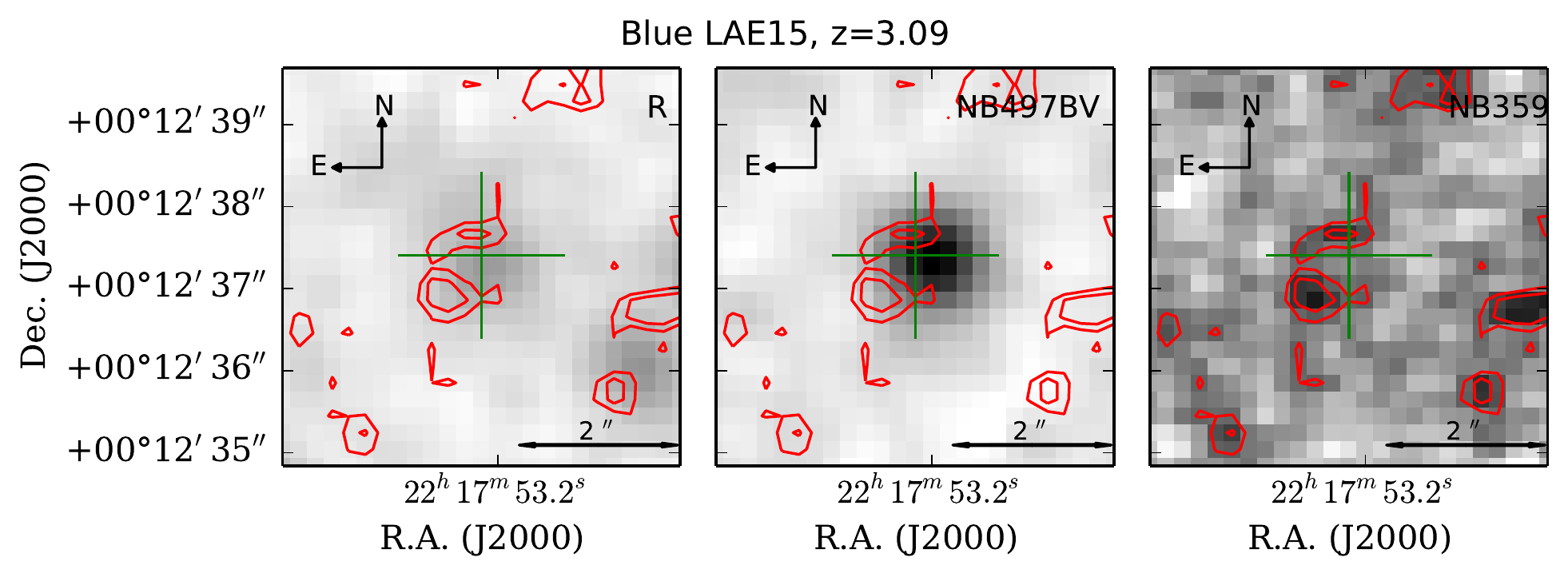}
\includegraphics[width=85mm]{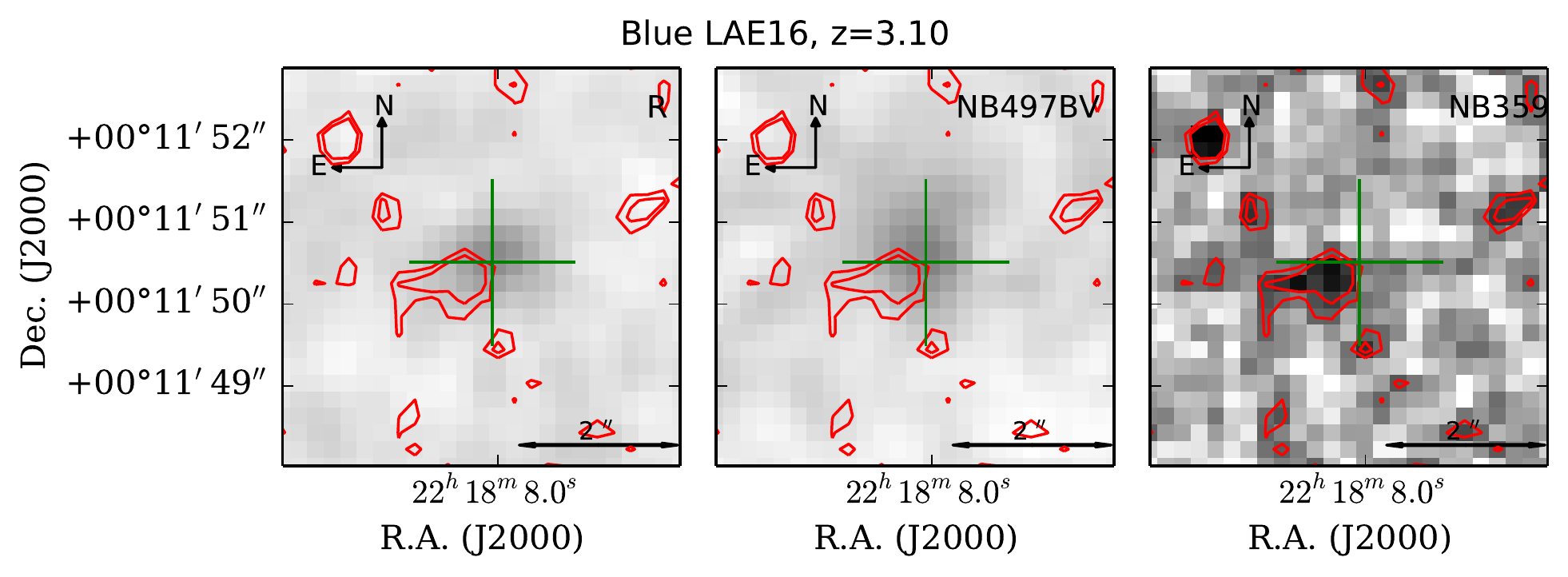}
\includegraphics[width=85mm]{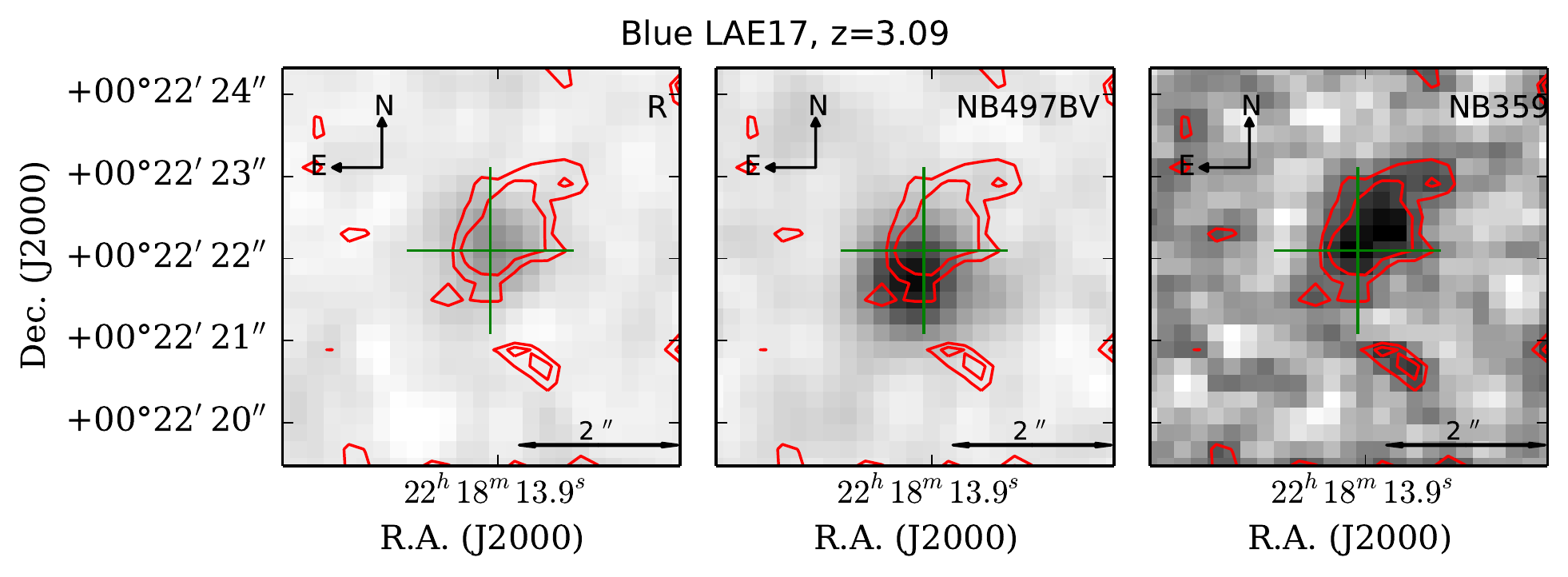}
\includegraphics[width=85mm]{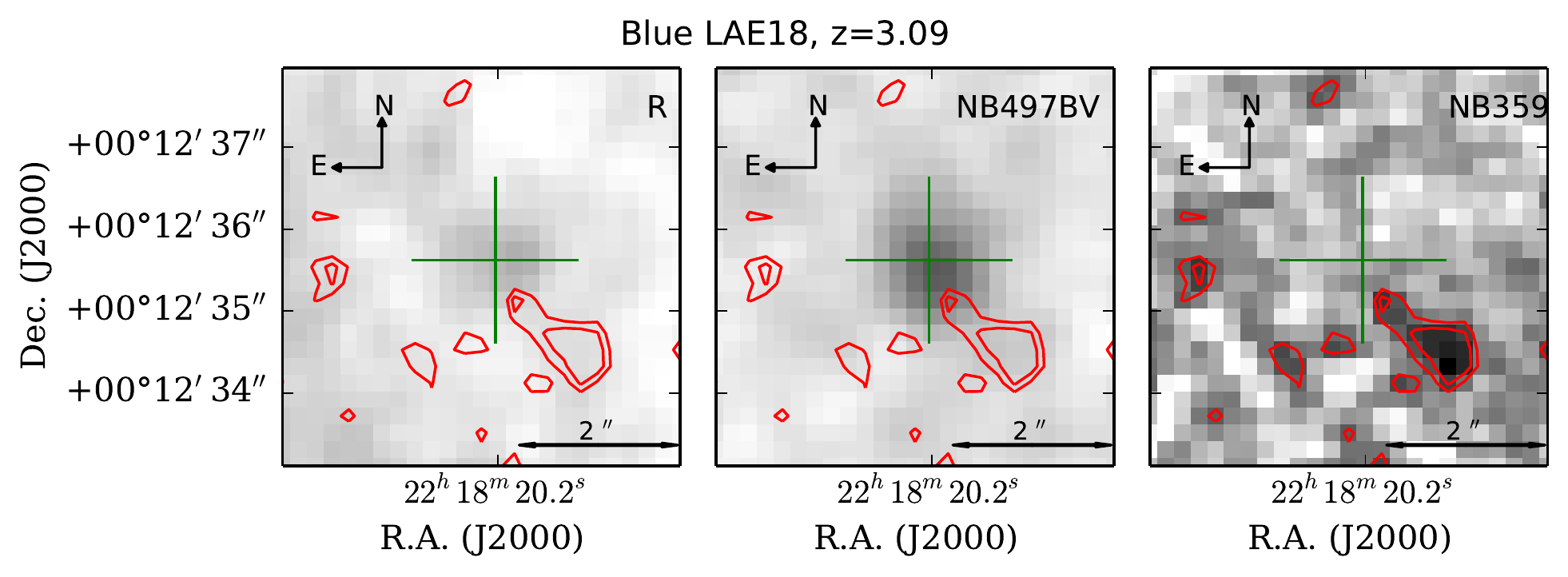}
\contcaption{}
\end{figure*}

\noindent SourceExtractor detections in the \filterr~band were considered to be the reference positions of the sources. For some targets the position of the \lyc~emission was offset from the reference position or from the position of the \lya~emission. Similar offsets have been found in previous studies~\citep[e.g.,][]{2009ApJ...692.1287I,2013ApJ...779...65M,2011ApJ...736...18N,2013ApJ...765...47N}. The reason we considered offset \lyc-emitting objects to possibly be a part of the main galaxy at the reference position is because strongly star-forming galaxies are likely to have disturbed morphology with clumpy substructures spread throughout the galaxy. There are also arguments in the literature that dust and gas geometry plays a significant role in creating escape tunnels for both \lya~and \lyc~\citep[e.g.,][]{2011MNRAS.418.1115R}, and consequently the \lyc~leakage would not be expected to be isotropic and may only be detectable from an individual star-forming region in the galaxy. In such a case, restricting ourselves to non-offset \lyc~candidates, as often done in the literature~\citep[e.g.][]{2010ApJ...725.1011V,2015ApJ...804...17S,2016A&A...587A.133G}, would cause an underestimation of the number density of \lyc~emitting galaxies and the cosmic average $f_{esc}$. However, our approach indeed allows for foreground contamination of the \lyc~candidates, and thus, we correct for contamination statistically in Sec~\ref{sec:contamination_stat}. Note that these two approaches are both meaningful and complementary.\\

\noindent We have flagged $13$ out of $18$ \lyc~LAEs and $3$ out of $7$ \lyc~LBGs as possible but unconfirmed contaminants. Our reasoning and justification is summarized in Section~\ref{appendix:notes} in the appendix, together with available HST images and spectra. The distribution of \lyc~offsets for the entire sample is shown in Figure~\ref{fig:offset_hist}, where \filterlyc~traces \lyc, \filterlya~traces \lya, and \filterlya-\filterb\filterv~is the continuum subtracted \lya~emission. \\

\begin{figure}
\centering\tiny
\includegraphics[width=85mm]{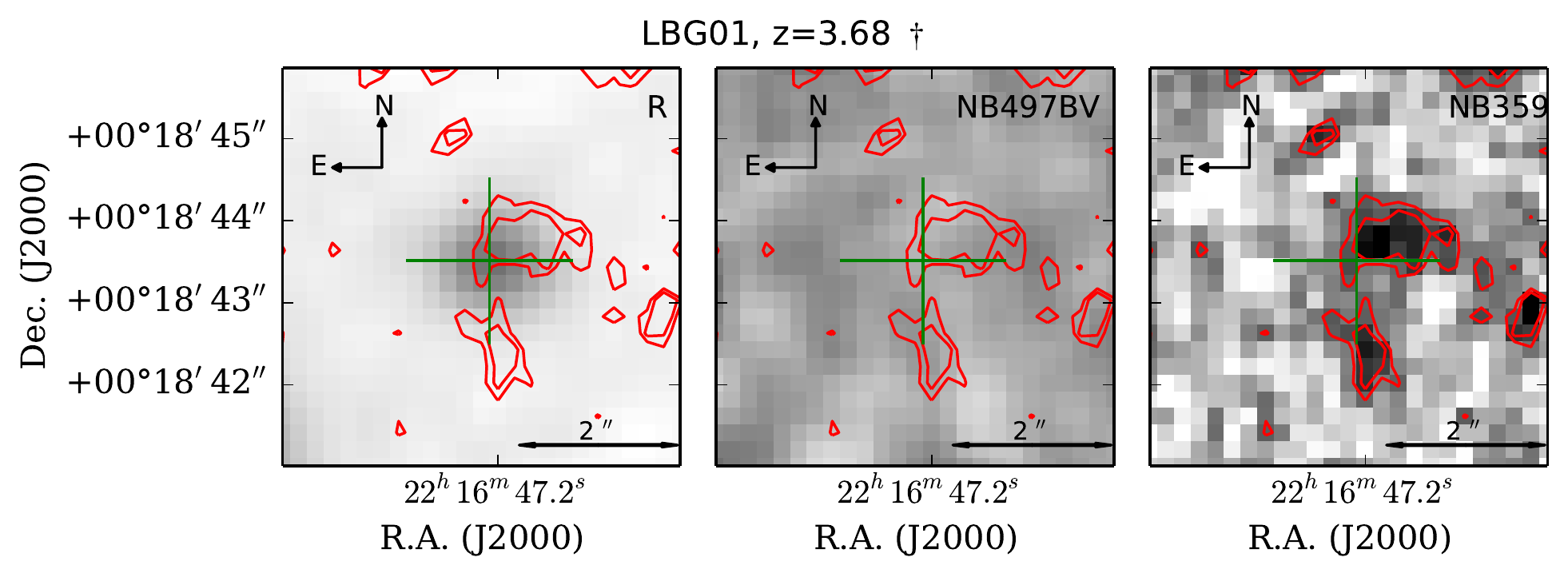}
\includegraphics[width=85mm]{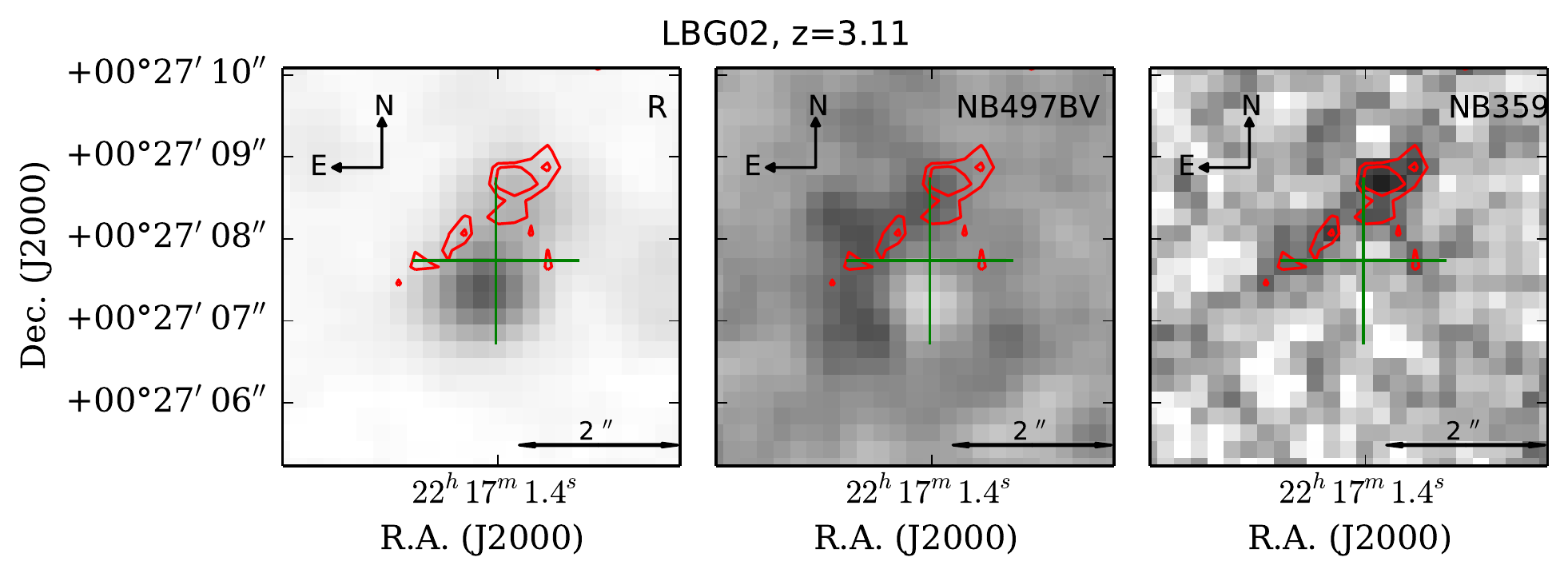}
\includegraphics[width=85mm]{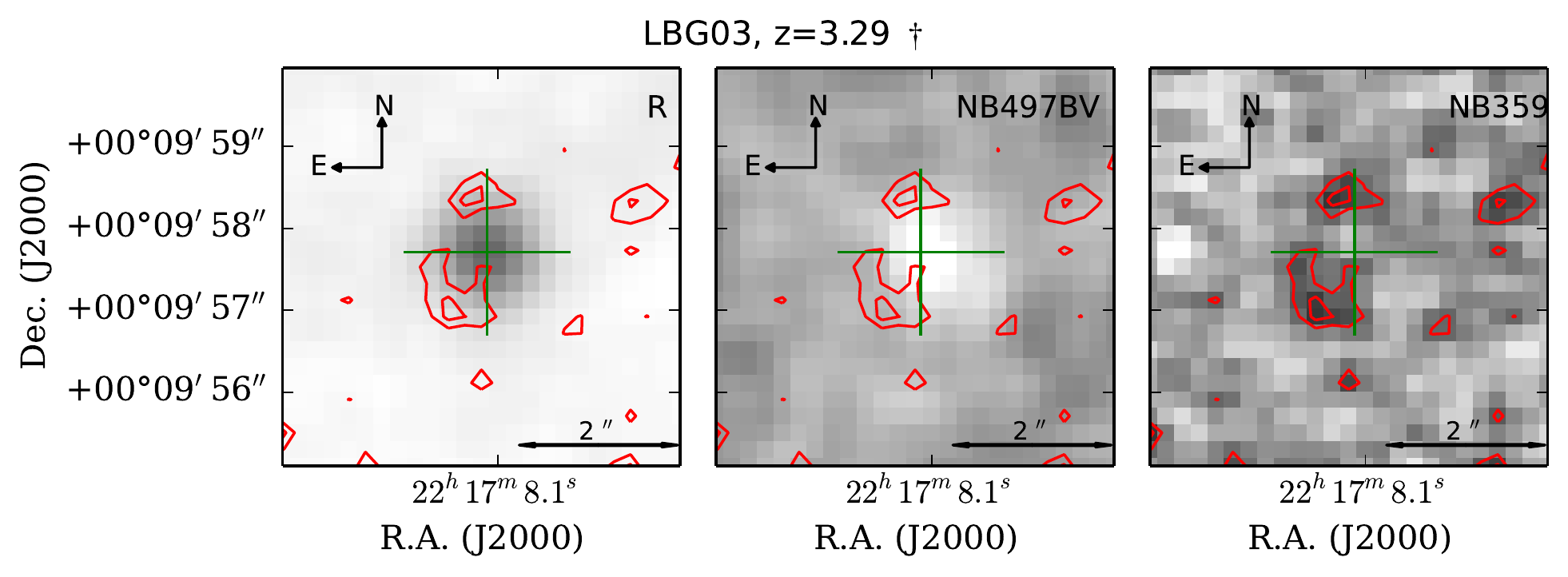}
\includegraphics[width=85mm]{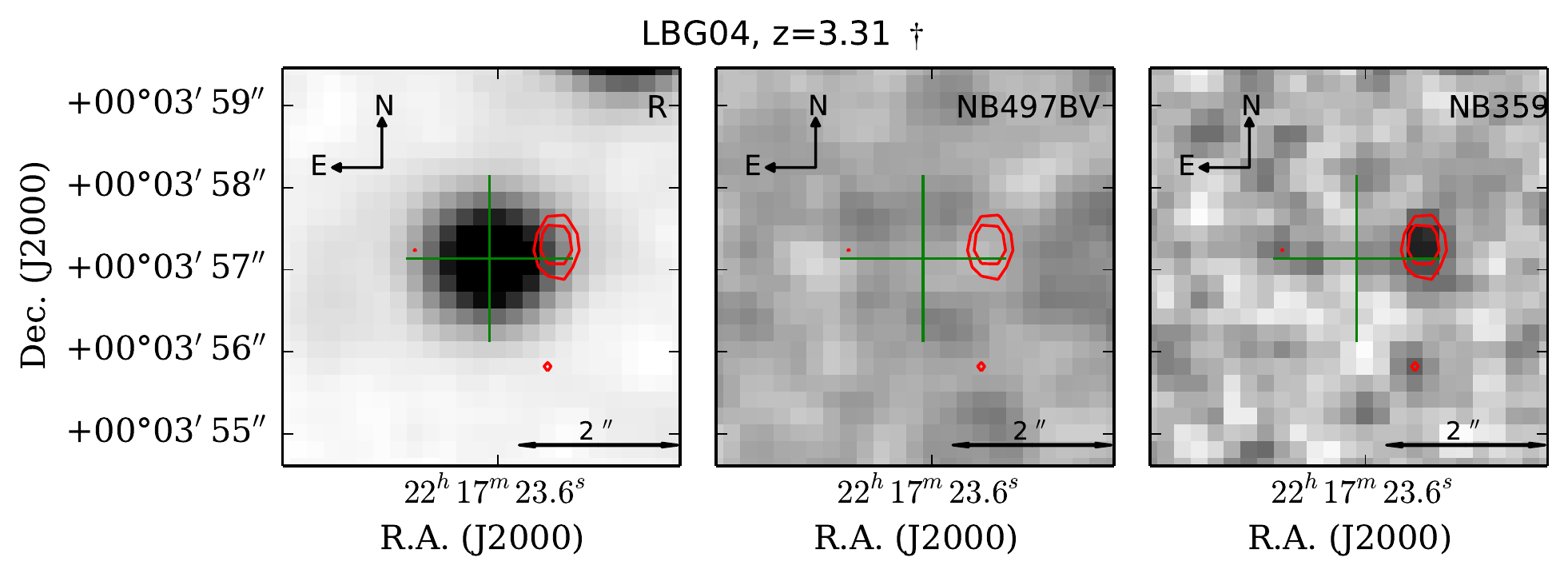}
\includegraphics[width=85mm]{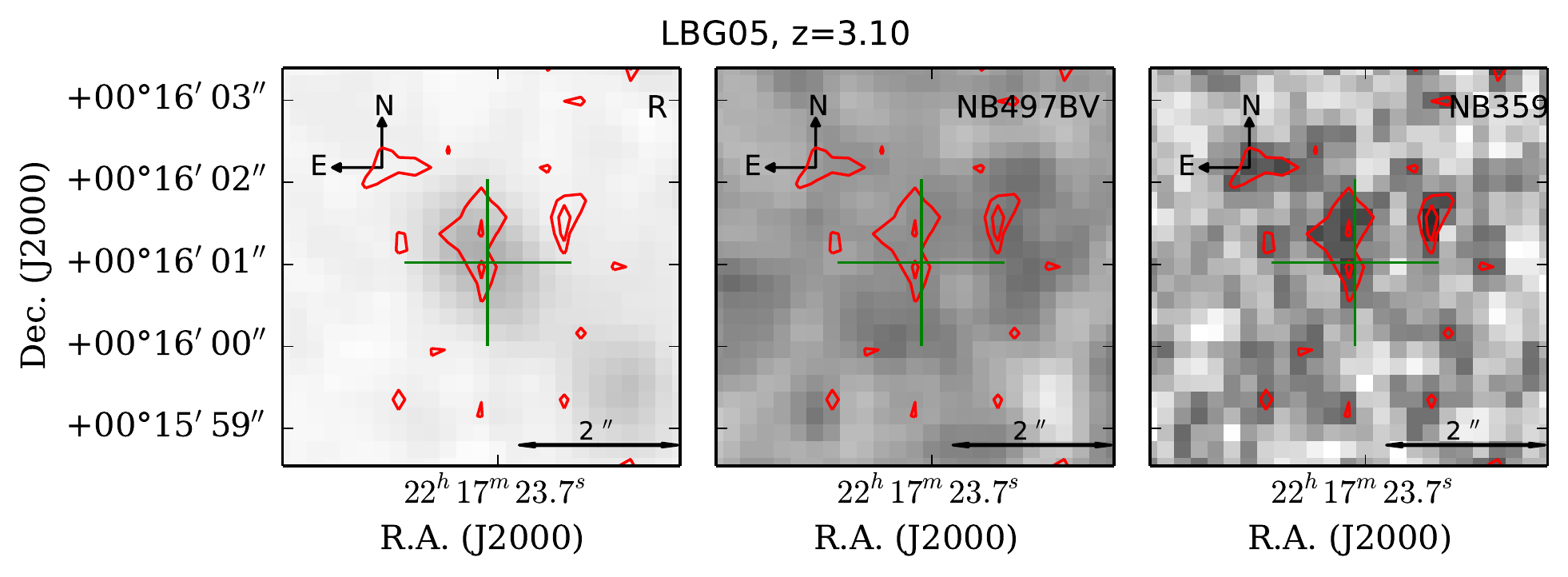}
\includegraphics[width=85mm]{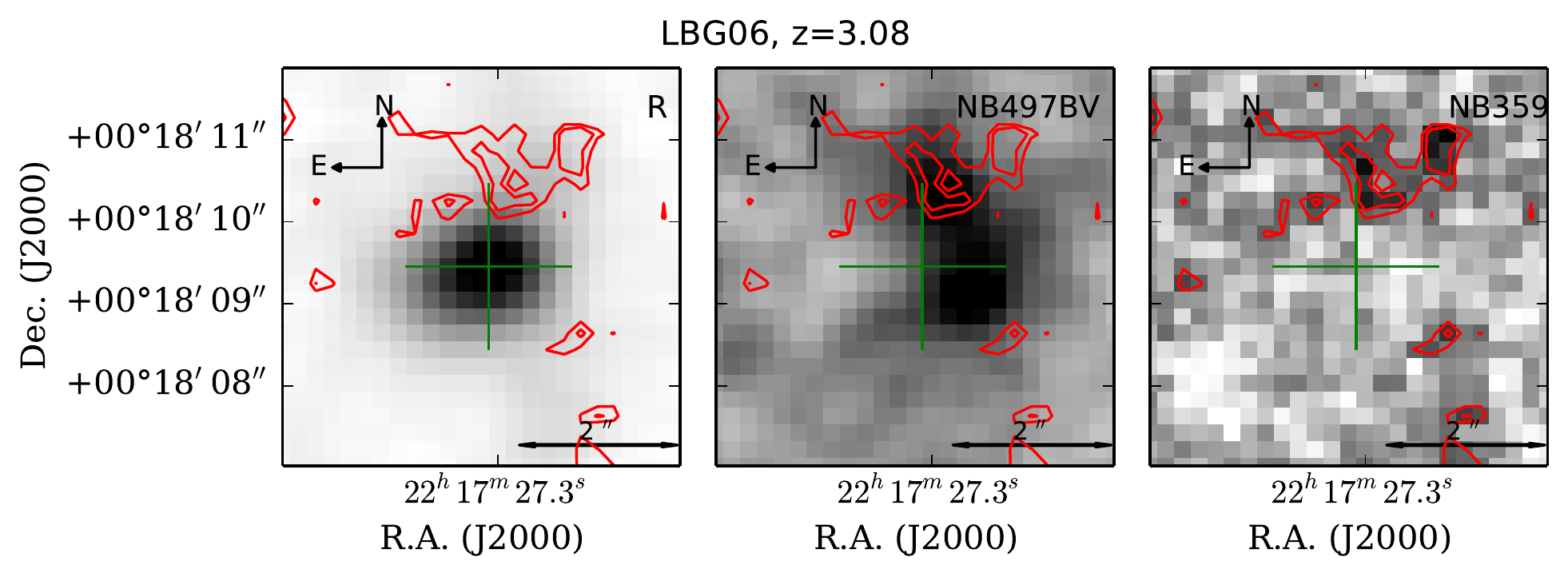}
\includegraphics[width=85mm]{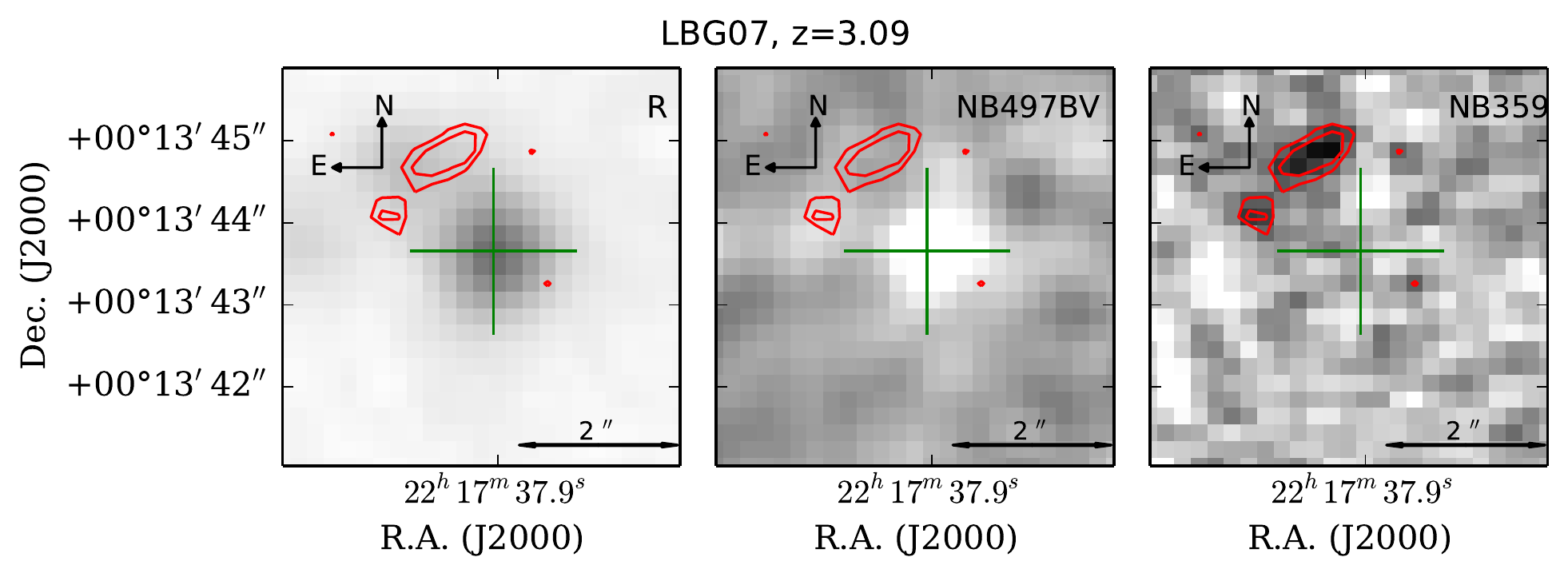}
\caption{Same as in Fig.~\ref{fig:contours} but for the LBG \lyc~candidates. Note that for LBGs with redshifts outside the \filterlya~filter range ($3.06<z<3.13$) the \filterlya$-$\filterbv~image does not represent \lya~emission. These redshifts are marked with $\dagger$.}\protect\label{fig:contours2}
\end{figure}

\subsection{Contamination}\protect\label{sec:contamination}
The contamination rate in general would depend on the depth of the data, the spatial resolution, and the redshift~\citep{2015ApJ...804...17S}. It was not possible to spectroscopically determine if there are any more foreground contaminants in our sample due to the lack of any convincing feature in their spectra. We therefore turn to statistical estimates. 
\subsubsection{False detection rate} \protect\label{sec:contamination_fdr}
We estimated the false detection rate by using a SourceExtractor catalog of detections obtained from a negative of the \filterlyc~frame, crossmatched with the catalog of \filterr~band detections. The search radius for spurious detections was $1.4$\arcsec, as in the original detection procedure for \lyc~candidates. The false detection rate for sources above our detection limit is $1.9\%$, or $2.6$ false detections among $136$ LBGs, and $3.0$ false detections among $159$ LAEs. Additionally, we have checked  the reliability of the detections in the \filterlyc~image. We have confirmed that all \lyc~candidates are detected (at the $2\sigma$ level) separately in the two images comprising the final stacked \filterlyc~image.
\begin{figure*}
\includegraphics[width=184mm]{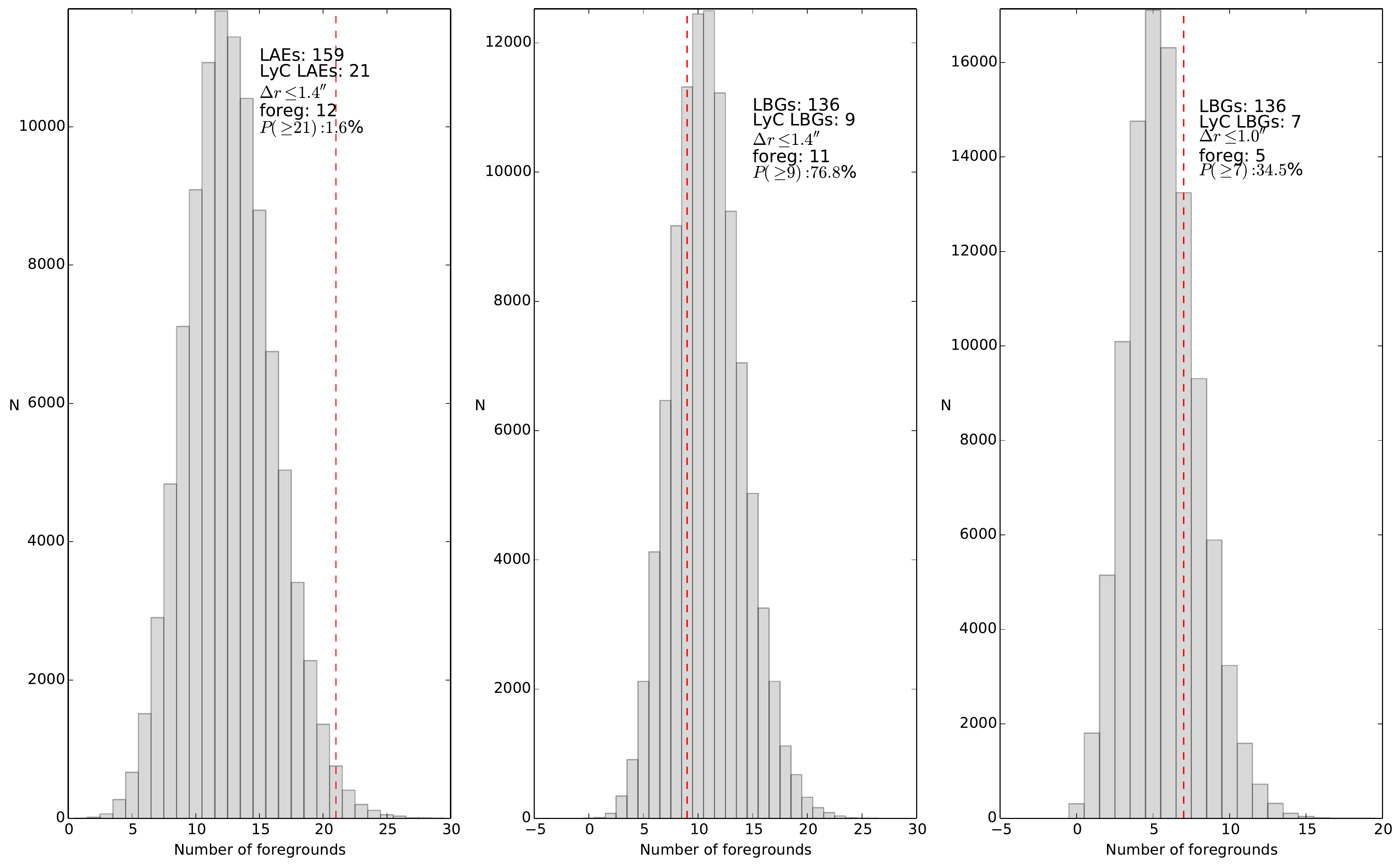}
\caption{Expected number of contaminants from $100,000$ MC realizations for different $\Delta r$ offset measurements. The total number of objects in the current base sample, the total number of \lyc~candidates, the assumed offset $\Delta r$, the median number of contaminants in MC tests 'foreg', and the probability $P$ that all \lyc~candidates are contaminants are indicated in the inset of each panel. The number of input objects in each simulation is marked by the dashed red line. }
\protect\label{fig:foreg}
\end{figure*}

\begin{figure}
\includegraphics[width=86mm]{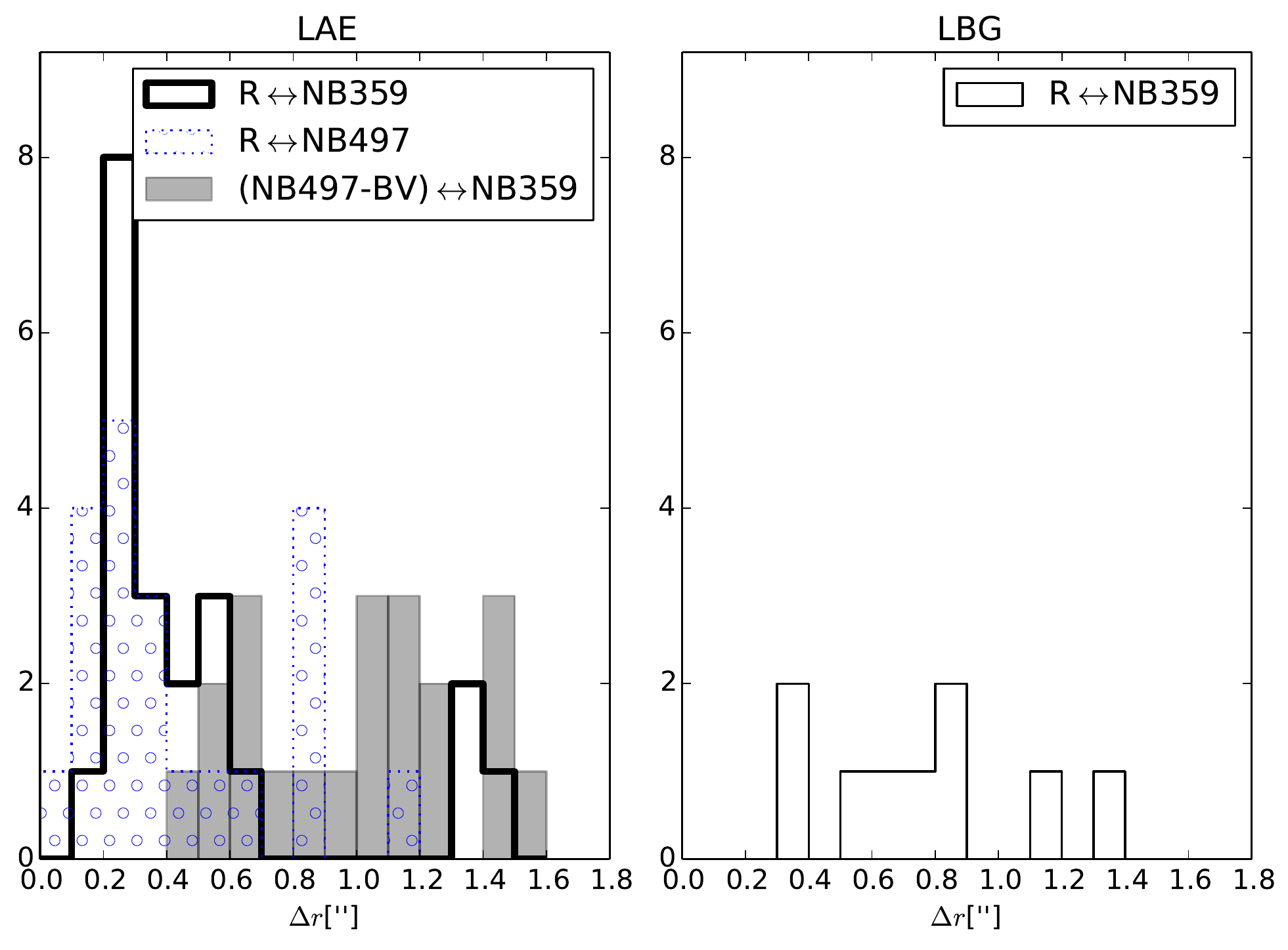}
\caption{Offset distributions for LAEs and LBGs. For LAEs we have made use of the available multiwavelength data to investigate additional offset measurements.}
\protect\label{fig:offset_hist}
\end{figure}

\subsubsection{Statistical contamination rate} \protect\label{sec:contamination_stat}
We estimate the statistical probability of contamination in three different ways. Our primary procedure relies on the actual detections in the \filterlyc~(\lyc) image. The effect of possible spurious detections is therefore taken into account. This procedure assumes that foreground objects and our target LAEs and LBGs are not spatially correlated and is based on a Monte-Carlo (MC) simulation. In each MC realization we randomly place $N$ number of sample galaxies in the \filterlyc~frame, where $N$ is $159$ for LAEs, and $136$ for LBGs. We search a circular area with a radius of $\Delta r$ around each randomly placed position for objects with detection significance $\geq3\sigma$ in the SourceExtractor catalog. For each position if at least one object is detected the total number of contaminants for that realization is increased by one. This procedure is repeated $100,000$ times. The resulting distribution of expected number of contaminants for the $N=159$ LAEs and $\Delta r=1.4\arcsec$ is shown in the left panel of Figure~\ref{fig:foreg}. For LAEs the contaminant distribution shows that it is statistically unlikely for all $21$ \lyc~LAEs to be foreground contaminated. The median number of the contaminants is $12$ out of $21$. For $N=136$ LBGs and $\Delta r=1.4\arcsec$ the distribution is shown in the middle panel of Figure~\ref{fig:foreg}. The probability that there are at least $9$ foreground contaminants is $P(\geq9)\sim77\%$. All except two \lyc~LBG candidates have offsets lower than $1\arcsec$ (see Figure~\ref{fig:offset_hist}), so in the right panel of Figure~\ref{fig:foreg} we show the expected number of contaminants for an assumed offset of $\Delta r\le 1\arcsec$. The probability that all seven \lyc~LBGs are contaminated is $P(\geq7)\sim35\%$, still high, but significantly reduced.\\

\noindent We also investigated the possible effect of large scale structure variations on the results. To remove any large scale fluctuations in the spatial distribution of \filterlyc~detections we performed the same MC simulation but this time assuming a uniform distribution of the \filterlyc~detections, while preserving their total number. The expected number distribution of contaminants from this test did not show any significant differences, so in our case large scale fluctuations are not expected to alter the results.\\

\noindent We perform two additional tests. We adopt the same method as~\citet{2010MNRAS.404.1672V}, who used number counts from the ultradeep VIMOS \filteru~band imaging of the GOODS-South field~\citep{2009ApJS..183..244N} to estimate the chance of foreground contamination of redshift $z\gtrsim3$ galaxies. The magnitude range in our \filterlyc~band is $24.0\leq AB\leq27.0$. The corresponding \filteru~band surface number density of sources in our magnitude range is $\rho=248000 $ deg$^{-2}$ according to~\citet{2009ApJS..183..244N}. With our resolution of $1\arcsec$, the chance that an individual object is contaminated by a foreground galaxy is thus $P_{cont,obj}=6.0\%$. This is the Poisson probability to get one contaminant within a circular aperture with a radius of $1\arcsec$ under the expected number of $0.06$ according to the~\citet{2009ApJS..183..244N} number counts. The probability that at least $10$ out of $295$ objects are contaminated by foregrounds is very high, $P_{cont,10}\sim98.5\%$. This is a binomial distribution probability assuming a success probability of $P_{cont,obj}$. The probability that all $30$ \lyc~candidates are foreground contaminants is negligible, $P_{cont,ALL}\sim0.40\%$. Similarly, the probability for all $21$ \lyc~LAEs to be contaminants is $P_{cont, LAEs}\sim0.06\%$. For $P_{cont, ALL}$ to go up to $\sim50\%$ the \filteru~band number counts in~\citet{2009ApJS..183..244N} would have to be underestimated by a factor of $1.67$, or $67\%$. The number counts in the literature can vary, for example for the same magnitude range the number counts reported by~\citet{2009A&A...507..195R} differ by those of~\citet{2009ApJS..183..244N} by $10\%$. A $67\%$ uncertainty is therefore unlikely. This second contamination estimation is fully consistent with the results from the MC procedure based on our actual \filterlyc~image. \\

\noindent As a final test, we have compared our primary MC procedure to the method presented in~\citet{2013ApJ...765...47N}, which takes the individual offsets for each object into account but assumes a constant \filteru~band number density across the field of view. A test run of $1000$ MC realizations of the Nestor method with a search radius of $\Delta r=1.4$\arcsec~results in five (three) expected contaminants out of nine ($21$) \lyc~LBGs (\lyc~LAEs) using the offset $\Delta r=$\filterr$ \leftrightarrow $ \lyc. Using the $\Delta r=$\lya$\leftrightarrow$ \lyc~offset for LAEs results in $15$ expected contaminants out of $21$ \lyc~LAEs. These results are qualitatively consistent with our method if we also assume constant number counts. As demonstrated they are highly dependent on which offset measurement one uses (e.g., $\Delta r=$\filterr$ \leftrightarrow $ \lyc, $\Delta r=$\filterr$\leftrightarrow$ \lya, or $ \Delta r=$\lya$ \leftrightarrow$ \lyc, where the double arrow indicates the offset is measured between the corresponding two filters).\\

\noindent In conclusion, it is highly unlikely at a $98\%$ confidence level that all $21$ \lyc~LAEs are contaminants. The situation for the \lyc~LBGs is statistically much less convincing, however this group contains some of our most viable candidates, with no spatial offsets in \lyc~emission, and clean spectra with no mysterious emission lines. \\

\begin{figure}
\centering\tiny
\includegraphics[width=55mm]{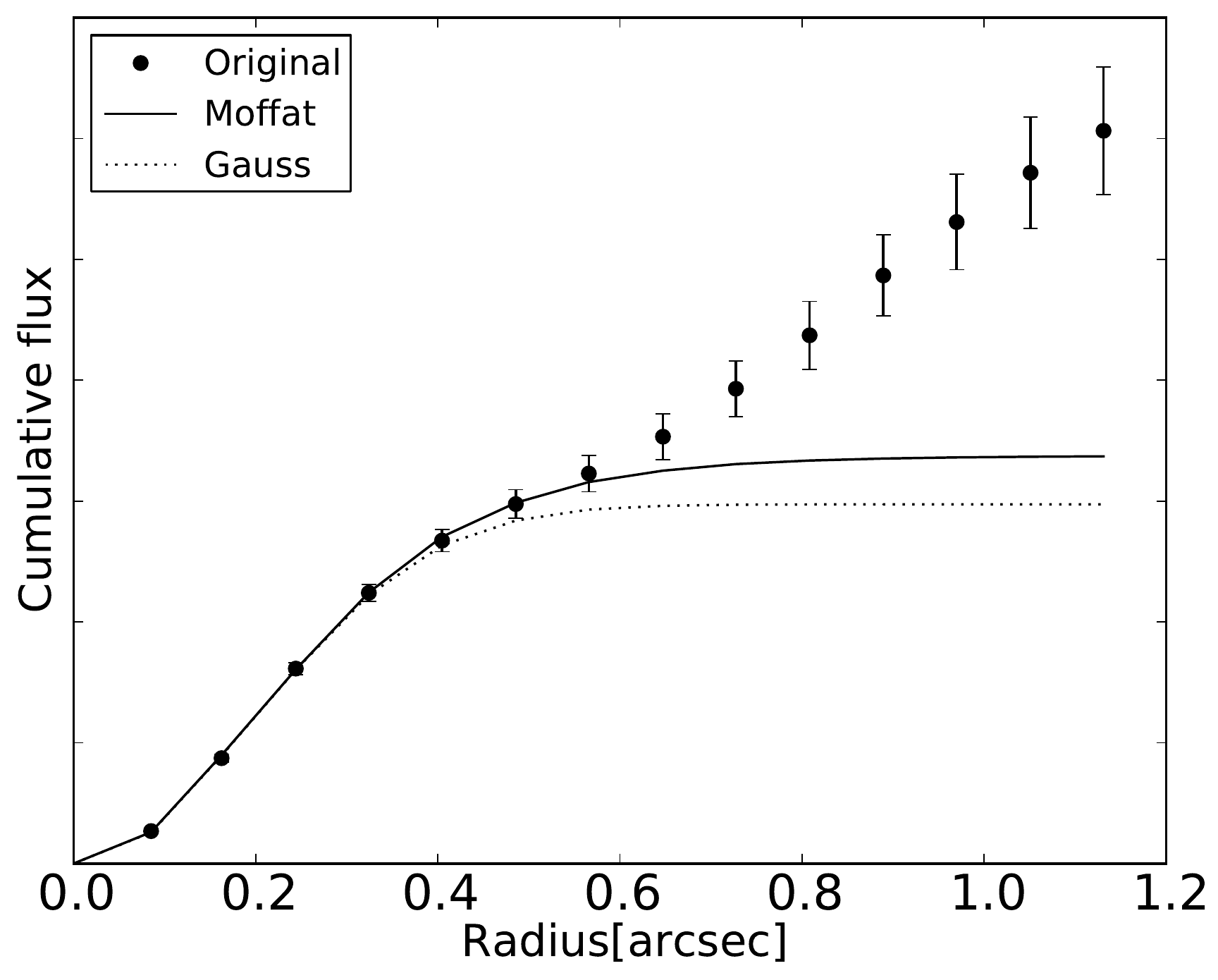}
\caption{Example growth curve modeling, here for LAE06 in the \filteri~band.}\protect\label{fig:growthcurve}
\end{figure}

\begin{table*}
\begin{minipage}{180mm}
\caption{Photometry of \lyc~candidates at the position of the \filterr~band detection. FWHM($\arcsec$) is measured from the reference \filterr~band. The omitted uncertainties for the \filterr($1.2\arcsec$) column are $<0.1$ mag. Galactic extinction by~\citet{2011ApJ...737..103S} has been applied to all values. The $(f_{1500}/f_{900})$ column at the position of the \filterr~band is the observed total flux density ratio. Possible (but unconfirmed) contaminants are indicated with a '${}^{\star}$'.}
\protect\label{tab:phot}
\begin{tabular}{l c c c c c c r}
\hline
\multicolumn{8}{c}{Red LAEs}\\
 ID &z& FWHM(\arcsec) &R & \filterr($1.2\arcsec$) & \filterlyc & \filterlyc($1.2\arcsec$) & $(f_{1500}/f_{900})_{obs}$  \\
\hline
LAE01${}^{\star}$ & $3.099$ & $1.3$  &  $25.81\pm 0.04$ & $26.79$  &                        &                     &                    \\
LAE02${}^{\star}$ & $3.127$ & $1.4$  &  $23.53\pm 0.02$ & $24.79$  &  $24.58\pm 0.06$ & $26.56\pm 0.10$ & $  2.64\pm  0.16$\\
LAE03${}^{\star}$ & $3.090$ & $1.1$  &  $24.41\pm 0.03$ & $25.29$  &  $25.44\pm 0.08$ & $26.32\pm 0.09$ & $  2.59\pm  0.20$\\
LAE05${}^{\star}$ & $3.088$ & $1.4$  &  $25.26\pm 0.07$ & $26.43$  &  $25.19\pm 0.11$ & $26.85\pm 0.11$ & $  0.94\pm  0.11$\\
LAE07${}^{\star}$ & $3.065$ & $1.3$  &  $25.19\pm 0.06$ & $26.28$  &  $25.91\pm 0.10$ & $26.83\pm 0.11$ & $  1.94\pm  0.20$\\
LAE08${}^{\star}$ & $3.080$ & $1.4$  &  $24.84\pm 0.04$ & $26.04$  &  $25.09\pm 0.09$ & $26.49\pm 0.10$ & $  1.25\pm  0.12$\\
LAE13${}^{\star}$ & $3.097$ & $1.1$  &  $24.93\pm 0.02$ & $25.85$  &  $25.31\pm 0.09$ & $26.27\pm 0.09$ & $  1.42\pm  0.11$\\
\multicolumn{8}{c}{Blue LAEs}\\
 ID &z& FWHM(\arcsec) &R & \filterr($1.2\arcsec$) & \filterlyc & \filterlyc($1.2\arcsec$) & $(f_{1500}/f_{900})_{obs}$  \\
\hline
LAE04${}^{\star}$ & $3.085$ & $1.0$  &  $24.74\pm 0.02$ & $25.61$  &  $24.37\pm 0.08$ & $25.52\pm 0.06$ & $  0.71\pm  0.06$\\
LAE06 & $3.075$ & $1.1$  &  $25.64\pm 0.05$ & $26.53$  &  $25.45\pm 0.12$ & $26.76\pm 0.11$ & $  0.85\pm  0.10$\\
LAE09${}^{\star}$ & $3.099$ & $1.5$  &  $25.37\pm 0.05$ & $26.73$  &  $25.63\pm 0.12$ & $27.24\pm 0.13$ & $  1.27\pm  0.15$\\
LAE10${}^{\star}$ & $3.090$ & $1.2$  &  $25.69\pm 0.07$ & $26.72$  &  $25.34\pm 0.10$ & $26.47\pm 0.09$ & $  0.72\pm  0.08$\\
LAE11${}^{\star}$ & $3.098$ & $1.6$  &  $26.53\pm 0.08$ & $27.78$  &  $25.89\pm 0.10$ & $27.93\pm 0.18$ & $  0.56\pm  0.07$\\
LAE12${}^{\star}$ & $3.065$ & $1.7$  &  $23.96\pm 0.05$ & $25.53$  &  $24.40\pm 0.09$ & $25.73\pm 0.07$ & $  1.49\pm  0.14$\\
LAE14 & $3.095$ & $1.1$  &  $25.78\pm 0.07$ & $26.73$  &  $25.26\pm 0.10$ & $26.39\pm 0.09$ & $  0.62\pm  0.07$\\
LAE15 & $3.094$ & $1.3$  &  $25.67\pm 0.08$ & $26.81$  &  $25.35\pm 0.12$ & $26.82\pm 0.11$ & $  0.75\pm  0.10$\\
LAE16 & $3.096$ & $0.8$  &  $26.15\pm 0.04$ & $26.93$  &  $25.74\pm 0.14$ & $27.09\pm 0.13$ & $  0.69\pm  0.09$\\
LAE17 & $3.089$ & $0.8$  &  $26.19\pm 0.04$ & $26.78$  &  $25.27\pm 0.10$ & $26.32\pm 0.09$ & $  0.43\pm  0.04$\\
LAE18${}^{\star}$ & $3.087$ & $1.1$  &  $26.45\pm 0.06$ & $27.22$  &  $25.51\pm 0.17$ & $28.22\pm 0.21$ & $  0.42\pm  0.07$\\
\multicolumn{8}{c}{LBGs}\\
 ID &z& FWHM(\arcsec) &R & \filterr($1.2\arcsec$) & \filterlyc & \filterlyc($1.2\arcsec$) & $(f_{1500}/f_{900})_{obs}$  \\
\hline
LBG01 & $3.680$ & $0.9$  &  $24.66\pm 0.02$ & $25.53$  &  $25.49\pm 0.12$ & $26.76\pm 0.11$ & $  2.15\pm  0.25$\\
LBG02 & $3.113$ & $1.2$  &  $24.31\pm 0.02$ & $25.46$  &  $26.05\pm 0.22$ & $27.18\pm 0.13$ & $  4.98\pm  1.00$\\
LBG03 & $3.287$ & $0.8$  &  $24.69\pm 0.02$ & $25.44$  &  $25.18\pm 0.11$ & $27.22\pm 0.13$ & $  1.58\pm  0.17$\\
LBG04 & $3.311$ & $1.0$  &  $23.44\pm 0.01$ & $24.30$  &  $26.64\pm 0.27$ & $27.83\pm 0.18$ & $ 19.01\pm  4.72$\\
LBG05${}^{\star}$ & $3.102$ & $1.1$  &  $25.05\pm 0.05$ & $26.11$  &  $25.87\pm 0.14$ & $27.17\pm 0.13$ & $  2.13\pm  0.28$\\
LBG06${}^{\star}$ & $3.080$ & $1.4$  &  $23.44\pm 0.01$ & $24.61$  &  $24.54\pm 0.30$ & $28.83\pm 0.27$ & $  2.76\pm  0.76$\\
LBG07${}^{\star}$ & $3.094$ & $1.4$  &  $24.25\pm 0.02$ & $25.43$  &  $25.63\pm 0.34$ & $28.07\pm 0.20$ & $  3.57\pm  1.12$\\
\hline
\end{tabular}
\end{minipage}
\end{table*}
\section{Suprime-Cam Photometry}\protect\label{sec:photometry}

Two types of \filterb\filterv\filterr\filteri\filterz, \filterlya, and \filterlyc~measurements are used in the analysis and made available in the online catalog, along with FWHM, Galactic extinction correction, and \lyc~offset. One magnitude measurement represents the total magnitude of the source with individual aperture size, and one with a fixed $\diameter=1.2\arcsec$ aperture size for all sources. UV and \lyc~photometry are summarized in Table~\ref{tab:phot}. Additionally, in Table~\ref{tab:clrclr} we list the \filterv$-$\filteri~and \filterlyc$-$\filterr~colors for each object, along with probability of contamination $P$ based on measured offsets between the \filterlyc~detection and either \filterlya$-$\filterbv~(for LAEs) and \filterr~band (for LBGs). For the calculation of $P$ we obtain the surface number density $\rho=177327$ deg${}^{-2}$ from the \filterlyc~image.\\

\noindent The existence of spatial offsets implies that for some \lyc~candidates the $\diameter=1.2\arcsec$ aperture photometry at the \filterr~band position will not represent the total strength of the \lyc~emission. Therefore the online catalog also provides the $\diameter=1.2\arcsec$ aperture photometry at the position of the \filterlyc~detection. This fixed aperture is not large enough to measure the total magnitude in every case but it could highlight any intrinsic differences between the sources on equal spatial scales (for objects at the same redshift). For the position of the \filterr~band we perform both types of photometry in all filters, while for the position of the \filterlyc~band we only obtain the $\diameter=1.2\arcsec$ aperture photometry. \\

\noindent To measure the total magnitude of a given source we first construct its growth curve with the IRAF PHOT task and a range of apertures. Then we model a synthetic object with the same axis ratio as the original object, same total magnitude and Gaussian radial profile, convolve it with the seeing of the frame, and measure its growth curve. We do this for a range of total magnitudes and FWHMs, centered at the preliminary results obtained through SExtractor photometry, which also provides the axis ratio of the synthesized object. For each galaxy in the sample all model growth curves are examined visually and an individual decision on the cutoff radius is made, resulting in one cutoff radius per galaxy. The growth curve which gives the smallest $\chi^2$ compared to the original growth curve is chosen as the best model for that object and is used to provide its asymptotic total magnitude. In an attempt to accommodate differences in object shapes this procedure is performed for two initial models, one with a synthetic object following a Gaussian profile, convolved with a Gaussian PSF, and the other with a synthetic object following a Gaussian profile, convolved with a Moffat PSF. One model is therefore a pure Gaussian even after convolution, while the other is a Gaussian/Moffat hybrid. For brevity we refer to the latter simply as the Moffat model. For each object we therefore have two best model growth curves. When these two show significant differences in their asymptotic magnitudes we visually examine which growth curve best follows the original growth curve and choose the appropriate asymptotic total magnitude for that object. The concept is illustrated in Figure~\ref{fig:growthcurve}. The same procedure gives the best FWHMs for each object. Here we note that the resolution is such that the measured FWHM is unreliable since it is often comparable to those of the point sources. We therefore cannot use the best-fit FWHM as a measure of the object size in the absence of atmospheric distortion.
\subsection{Uncertainties}
The photometric uncertainties for the fixed $\diameter=1.2\arcsec$ aperture are obtained from the output of IRAF PHOT, with some modifications. Initial tries showed unsatisfactory estimation of the sky background value with IRAF PHOT for a number of targets due to the far wings of bright saturated stars in the sky annuli. The calculated mode of the values inside the sky annulus did not take into account the cases where the sky histogram had several nearby peaks with similar amplitudes. To estimate the background more reliably we instead measured for each target the weighted average of all data points within $\pm5\%$ of the mode in the sky annulus histogram. This constant sky value was then used in IRAF PHOT to obtain new error measurements. To correct for the effect that a constant sky value has on the IRAF PHOT errors, we additionally corrected these with the uncertainty in the sky level (Section~\ref{image_quality}), which was assumed constant throughout the frame. These are the errors listed in our catalog for fixed $\diameter=1.2\arcsec$ aperture photometry.\\

\noindent The uncertainties in the asymptotic total magnitudes (``model mag'') are obtained by an MC simulation with 1000 runs per object per model per filter, where each realization perturbs each data point of the original growth curve by a random value drawn from a $1\sigma$ normal distribution, where $\sigma$ is the error of the current data point. This gives a range of model growth curves which best fit each perturbed realization. The asymptotic magnitude uncertainty for a given target, model, and filter is then the standard deviation of the $1000$ asymptotic magnitudes from the MC simulation. The MC uncertainties for Gaussian models are on average larger than those for the Moffat models by $\lesssim10\%$. For derived quantities, such as the flux density ratios and the equivalent width, the uncertainties were obtained through error propagation.

\begin{figure}
\includegraphics[width=85mm]{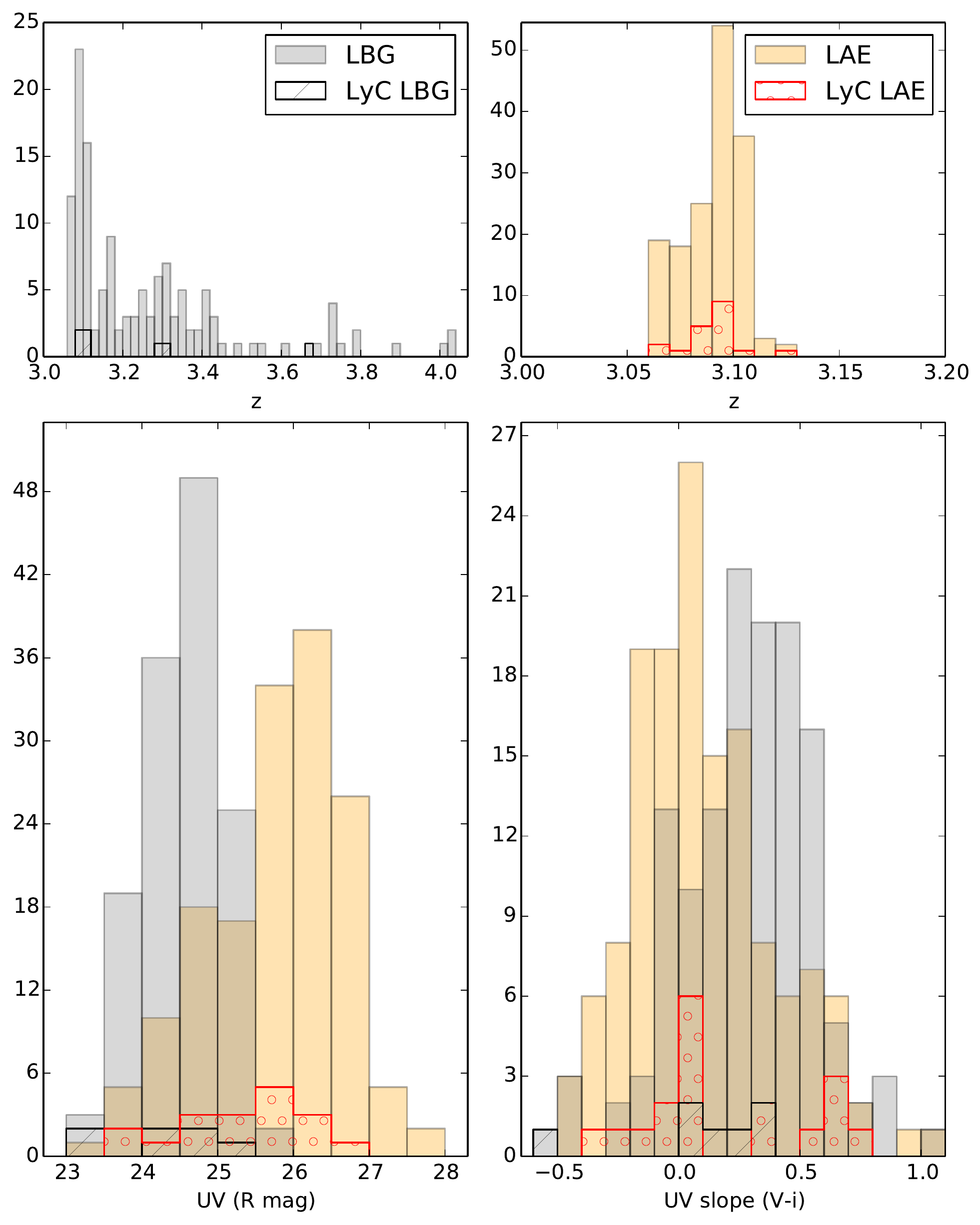}
\caption{Rest-frame UV magnitude (\filterr~band, left) and the distribution of the measured UV slopes (\filterv$-$\filteri, right) for all LAEs (orange bars) and all LBGs (gray bars) in the sample. \lyc~LAE (dotted bars) and \lyc~LBG (hatched bars) candidates are also shown. AGN and confirmed foreground contaminants are not included. All measurements are of the total magnitudes. The redshift distributions of the full LAE and LBG samples, as well as the \lyc~subsamples, are shown in the top panels.}
\label{fig:statistics}
\end{figure}
\begin{figure}
\includegraphics[width=84mm]{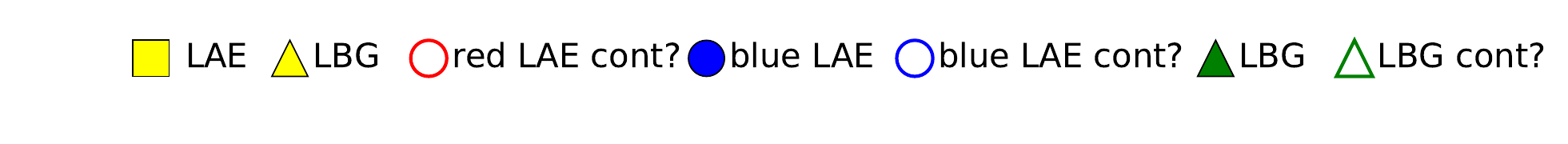}
\includegraphics[width=84mm]{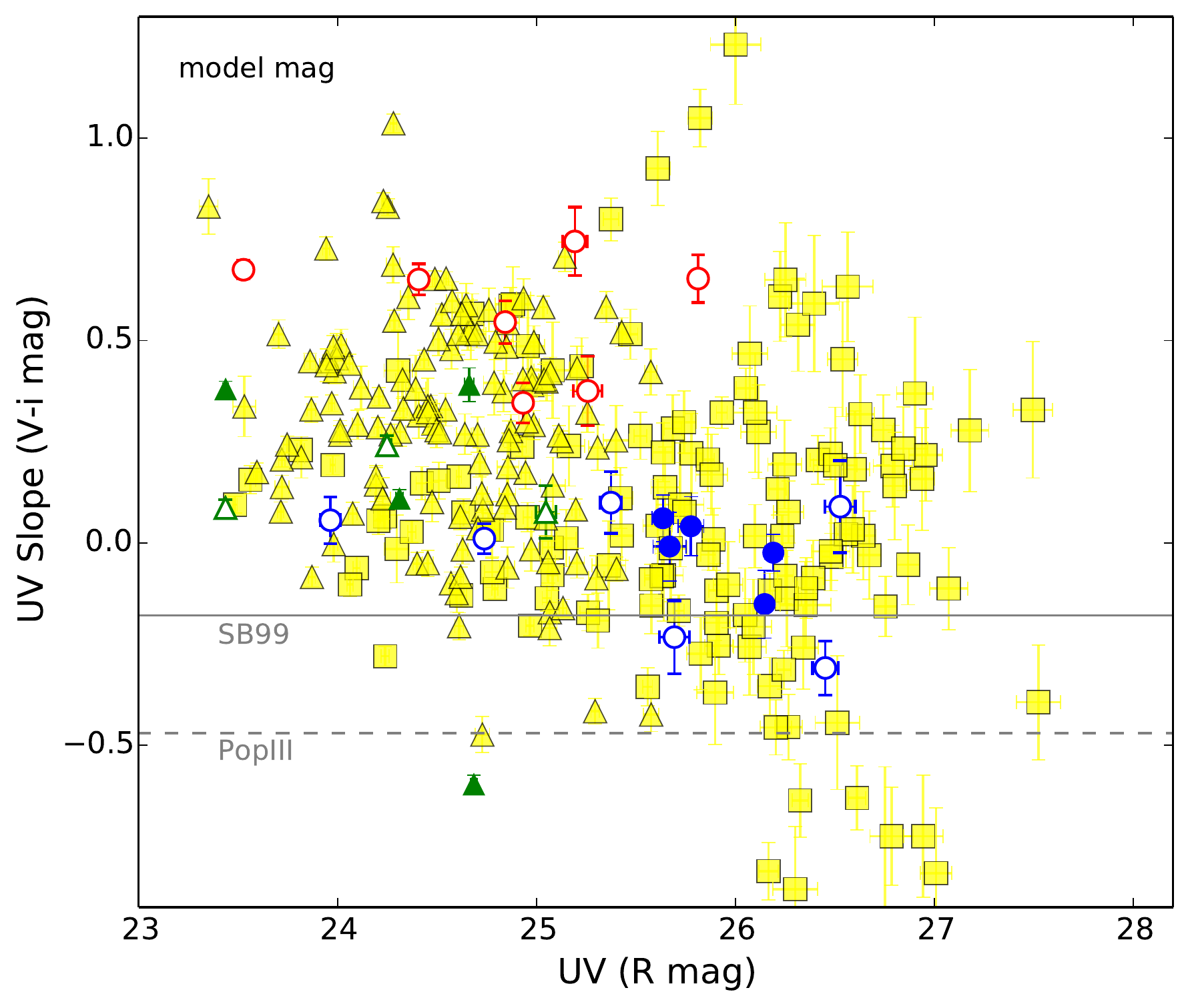}
\caption{Rest frame UV slope versus non-ionizing total UV magnitude at the position of the \filterr~band for the full sample. The control sample of non-\lyc~are shown in yellow markers (triangle - LBG, square - LAE). Red and blue \lyc~LAEs are colored accordingly, \lyc~LBGs with green triangles. Possible but unconfirmed \lyc~contaminants are indicated with open markers of the corresponding shape and color. Bluest \filterv$-$\filteri~model predictions from a \starburst~and a \popiii~model are plotted with solid, respectively dashed lines.}
\label{fig:Vi vs R}
\end{figure}
\begin{figure}
\includegraphics[width=84mm]{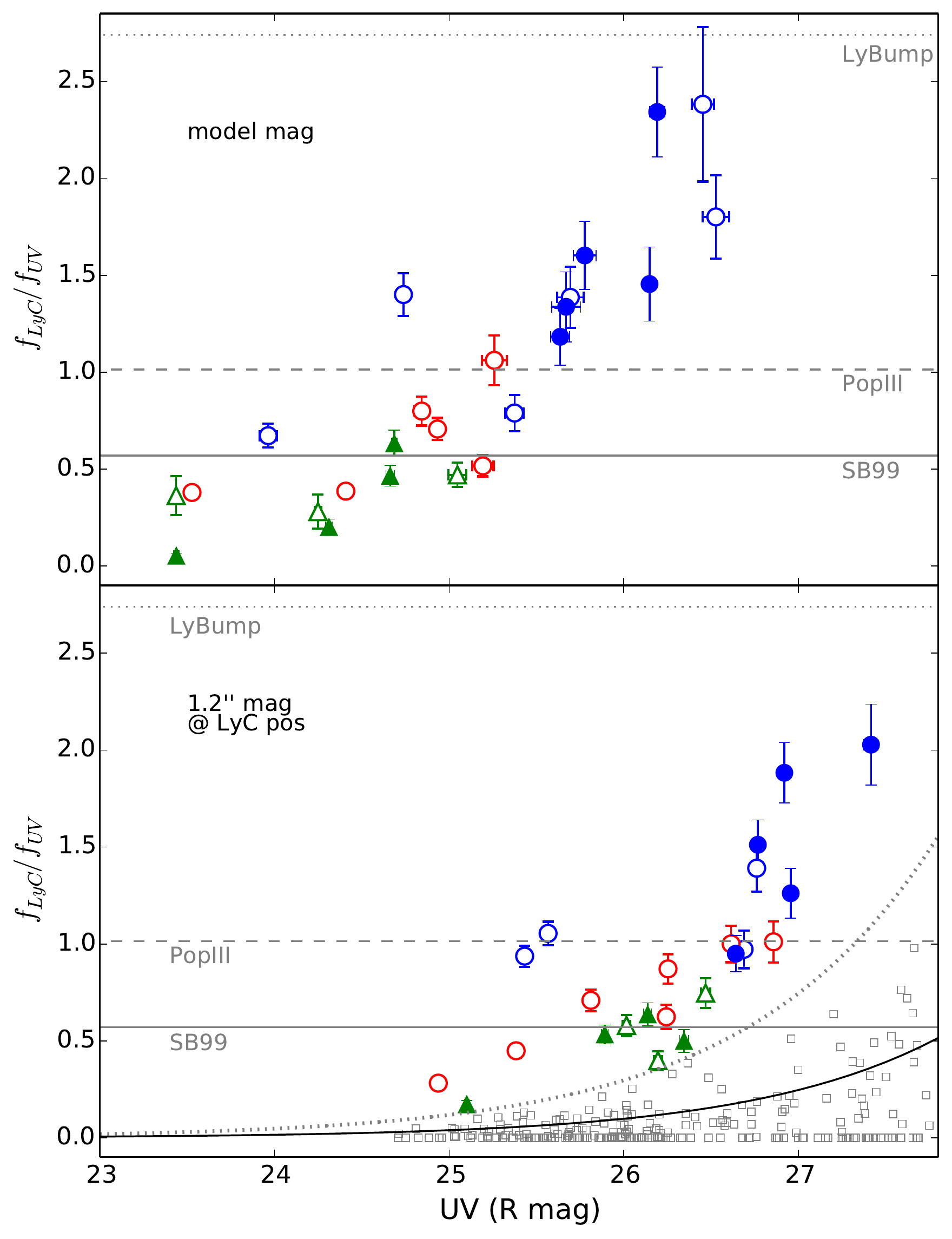}
\caption{Observed flux density ratios for total brightness at the \filterr~position (upper panel) and $1.2\arcsec$ fixed aperture measurements at the \filterlyc~position (lower panel) for the \lyc~candidates. The horizontal lines and the marker legend are the same as in Fig.~\ref{fig:Vi vs R}, with an added line for the LyBump \popiii~model (see Sec.~\ref{sec:flux density ratios} for details). For convenience the $1\sigma$ (black curve) and $3\sigma$ (gray dotted curve) detection limits from $1.2\arcsec$ apertures in the \filterlyc~(LyC) band are also shown. LAE11 and LAE18 are omitted from the lower panel because they are too faint in $1.2\arcsec$ \filterr~band photometry at the position of the \filterlyc~band. The open gray squares are the \lyc~non-detections from both the LAE and LBG base samples.}
\label{fig:flux density ratios}
\end{figure}
\subsection{Variability}
\citet{2015A&A...576A.116V} suggest that low luminosity AGNs hidden in star forming galaxies could be responsible for the detected \lyc~emission from these galaxies. We performed a brightness variability test on all objects in our samples to detect potential AGNs. None of the \lyc~candidates show any variability in brightness over the span of 6 years, however it is still possible that low luminosity AGNs could be present and either be non-variable, vary on amplitudes below our detection level, or vary with a shorter time-scale than our sampling ($\sim1$ year). The individual light curves are shown in Figure~\ref{fig:lightcurves} in the appendix.\\

\section{Properties of \lyc~sources}\protect\label{sec:results}

\noindent With respect to the UV continuum on average the LBGs in the sample are brighter than the LAEs, which is expected since the strong \lya~emission line deepens the detection limit in spectroscopically confirmed samples of galaxies. The distribution of total rest-frame UV brightness for the sample is shown in Figure~\ref{fig:statistics}, where we also show the distribution of \filterv$-$\filteri~colors and redshifts from Table~\ref{tab:clrclr}. We use \filterv$-$\filteri~color (approximately flux density ratio between 133nm and 188nm for a galaxy at $z=3.1$) to investigate rest-frame UV slope. Despite the difference in brightness ranges, the color distribution has an overlapping range for LAEs and LBGs, with the LBGs being on average redder than the LAEs, which is consistent with the literature~\citep[e.g.][]{2007ApJ...667...79G} and reflects the different nature of these types of galaxies. Note that the fraction of \lyc~LBG candidates seems to be relatively small in the proto-cluster. A comparatively similar amount of \lyc~LBG candidates are found at higher redshifts, unassociated with the proto-custer. The \lyc~LAEs show a broad range of colors. To highlight any effects such color differences may have on the properties of these galaxies, we divide them into red LAE (\filterv$-$\filteri$>0.2$) and blue LAE (\filterv$-$\filteri$\lesssim0.2$). Note that this separation is quite arbitrary, we simply aim to distinguish between the very red and very blue LAEs. We feel such a separation may be meaningful because the number of \lyc~candidates is overrepresented among the extremely red LAEs. The two photometry tables (Tab.~\ref{tab:phot} and~\ref{tab:clrclr}) show the blue/red labels for the LAE sample. \\

\begin{table}
\begin{minipage}{84mm}
\caption{\lyc~candidates \filterv$-$\filteri~and \filterlyc$-$\filterr~colors at the position of the \filterr~band detection. Redshift $z$ and \lyc~offset $\Delta r$ (\arcsec) are indicated for convenience. The individual probability for contamination $P$, indicated in per cent, is calculated as $P=\rho\pi\Delta r^2$, where $\rho=1.77327e5$ deg${}^{-2}$ is the surface number density we obtain from the \filterlyc~image. }\protect\label{tab:clrclr}
\begin{tabular}{ l c r r r r}
ID&z&\filterv$-$\filteri&\filterlyc$-$\filterr&$\Delta r^{\dagger}$ & $P[\%]$\\
\hline
\multicolumn{6}{|c|}{Red LAE viable candidates}\\
\multicolumn{6}{|c|}{-}\\
\multicolumn{6}{|c|}{Red LAE possible contaminants}\\
LAE01&$3.099$&$  0.65\pm  0.06$&                 &$1.4$ &  $8.4$\\
LAE02&$3.127$&$  0.67\pm  0.02$&$  1.05\pm  0.07$&$0.9$ &  $3.5$\\
LAE03&$3.090$&$  0.65\pm  0.04$&$  1.03\pm  0.08$&$0.5$ &  $1.1$\\
LAE05&$3.088$&$  0.38\pm  0.09$&$ -0.06\pm  0.13$&$1.1$ &  $5.2$\\
LAE07&$3.065$&$  0.74\pm  0.08$&$  0.72\pm  0.11$&$1.3$ &  $7.3$\\
LAE08&$3.080$&$  0.55\pm  0.05$&$  0.24\pm  0.10$&$1.5$ &  $9.7$\\
LAE13&$3.097$&$  0.35\pm  0.05$&$  0.38\pm  0.09$&$1.2$ &  $6.2$\\
\multicolumn{6}{|c|}{Blue LAE viable candidates}\\
LAE06&$3.075$&$  0.06\pm  0.06$&$ -0.18\pm  0.13$&$0.4$ & $0.7$ \\
LAE14&$3.095$&$  0.04\pm  0.07$&$ -0.51\pm  0.12$&$0.8$ & $2.8$ \\
LAE15&$3.094$&$ -0.01\pm  0.08$&$ -0.32\pm  0.15$&$0.5$ & $1.1$ \\
LAE16&$3.096$&$ -0.15\pm  0.08$&$ -0.41\pm  0.14$&$0.7$ & $2.1$ \\
LAE17&$3.089$&$ -0.02\pm  0.05$&$ -0.92\pm  0.11$&$0.7$ & $2.1$ \\
\multicolumn{6}{|c|}{Blue LAE possible contaminants}\\
LAE04&$3.085$&$  0.01\pm  0.04$&$ -0.37\pm  0.09$&$1.4$ & $8.4$ \\
LAE09&$3.099$&$  0.10\pm  0.08$&$  0.26\pm  0.13$&$0.8$ & $2.8$ \\
LAE10&$3.090$&$ -0.23\pm  0.09$&$ -0.35\pm  0.12$&$1.0$ & $4.3$ \\
LAE11&$3.098$&$  0.09\pm  0.11$&$ -0.64\pm  0.13$&$1.2$ & $6.2$ \\
LAE12&$3.065$&$  0.06\pm  0.06$&$  0.43\pm  0.10$&$1.1$ & $5.2$ \\
LAE18&$3.087$&$ -0.31\pm  0.07$&$ -0.94\pm  0.18$&$1.2$ & $6.2$ \\
\multicolumn{6}{|c|}{LBG viable candidates}\\
LBG01&$3.680$&$  0.39\pm  0.04$&$  0.83\pm  0.12$&$0.5$ & $1.1$ \\
LBG02&$3.113$&$  0.11\pm  0.02$&$  1.74\pm  0.22$&$0.9$ & $3.5$ \\
LBG03&$3.287$&$ -0.60\pm  0.02$&$  0.50\pm  0.12$&$0.8$ & $2.8$ \\
LBG04&$3.311$&$  0.38\pm  0.02$&$  3.20\pm  0.27$&$0.8$ & $2.8$ \\
\multicolumn{6}{|c|}{LBG possible contaminants}\\
LBG05&$3.102$&$  0.08\pm  0.07$&$  0.82\pm  0.15$&$0.3$ & $0.4$ \\
LBG06&$3.080$&$  0.09\pm  0.02$&$  1.10\pm  0.30$&$1.1$ & $5.2$ \\
LBG07&$3.094$&$  0.24\pm  0.02$&$  1.38\pm  0.34$&$1.3$ & $7.3$ \\
\hline
\end{tabular}
\end{minipage}
\medskip
$\dagger$ For LAEs, $\Delta r=($\filterlya$-$\filterbv$)\leftrightarrow$\filterlyc. For LBGs $\Delta r=$\filterr$\leftrightarrow$\filterlyc. The double arrow indicates the offset is measured between the corresponding filters.
\end{table}

\subsection{The UV slope}\protect\label{sec:uv slope}
In Figure~\ref{fig:Vi vs R} we show the color-magnitude distribution of the \lyc~candidates and the control sample using modeled asymptotic growth curve magnitude measurements from Tables~\ref{tab:phot} and~\ref{tab:clrclr}. The red LyC LAEs are on average brighter than the blue \lyc~LAEs. Note that our morphology/spectrum analysis in Sec.~\ref{sec:sampleselection} flagged all red \lyc~LAEs as possible but unconfirmed contaminants independently of their extreme red colors. The \lyc~candidates in general occupy the same parameter space as the non-\lyc~sample and in both groups there are some galaxies that show extremely blue UV slopes independent of magnitude. These violate the model predictions from Starburst 99~\citep[][\starburst]{1999ApJS..123....3L} shown as a gray horizontal line, and also those from a Population III model~\citep[\popiii]{2003A&A...397..527S}, shown as a gray dashed line. Both models are for zero age, with mass range $1$-$100M_{\sun}$, a Salpeter IMF ($\alpha=2.35$), no nebular emission, and either zero or low ($Z=0.0004$) metallicity. These are the bluest stellar evolutionary models for this IMF since increased age and metalicity, as well as accounting for nebular emission, will make the colors redder. Although the larger uncertainty at fainter magnitudes affects the colors, fainter galaxies appear to have bluer UV slopes. These galaxies are dominated by LAEs because of the faint limit detection bias favoring LAEs over LBGs. \\

\subsection{Ionizing to non-ionizing flux density ratios}\protect\label{sec:flux density ratios}
Figure~\ref{fig:flux density ratios} shows the flux density ratios of the ionizing to non-ionizing radiation as it varies with UV brightness using data from Table~\ref{tab:phot} and our online catalog. For both axes we use the \filterr~band to sample the non-ionizing radiation. It is immediately obvious that some \lyc~candidates, mostly the blue LAEs, have extremely high flux density ratios. While these galaxies are among the faintest, their \lyc~detections are well above the $3\sigma$ limit. The extreme ratios are preserved when measuring the flux at the offset position of the \filterlyc~(\lyc) band (bottom panel). The possible contaminants LAE11 and LAE18 are omitted from the bottom panel because they are very faint already at the \filterr~band position and are well below the detection limit in \filterr~at the position of the \filterlyc~band. \\

\noindent As Figure~\ref{fig:flux density ratios} demonstrates, many of the observed LAE flux density ratios imply at least a flat spectrum with no Lyman Break, i.e. a ratio of $f_{LyC}/f_{UV}=1$ (\filterlyc$-$\filterr~$=0$) which is inconsistent with current ``standard'' population synthesis models, as indicated by the {\sevensize SB99} line in the figure. The presence of IGM is expected to further decrease the observed flux density ratio. \\

\noindent There are, however, physically consistent models which predict higher flux density ratios than the standard models, such as the so-called Lyman limit 'bump' model~\citep[][]{2010MNRAS.401.1325I}. The model proposes that, instead of being completely absorbed, nebular \lyc~can escape through matter-bound nebulae along the same paths as stellar \lyc. This nebular \lyc~contribution would compensate for the Lyman break by significantly contributing to the flux bluewards of the Lyman limit, causing the so-called 'bump' and making the \filterlyc$-$\filterr~much bluer than the ``standard'' models. The bluest \filterlyc$-$\filterr~color prediction from this model is shown with a dotted line, for {\sevensize PopIII} with a high mass range (50-500 $M_{\sun}$). \\

\noindent The existence of {\sevensize PopIII} stars at $z\sim2\mbox{-}3$ has been suggested by several theoretical works in the literature~\citep[e.g.][]{2007MNRAS.382..945T,2010MNRAS.404.1425J}, and pristine gas clouds in the IGM have already been found at $z\sim3$~\citep[e.g.][]{2011Sci...334.1245F} and~\citet{2011MNRAS.411.2336I} present a two component model which offers an explanation for the co-existence of {\sevensize PopIII} stars and dust. In addition, there are other scenarios resulting in an effectively top-heavy IMF, which would contribute to a boost of the observed $f_{\lyc}/f_{UV}$. Namely, runaway massive stars~\citep{2012ApJ...755..123C}, and stochasticity in the IMF sampling~\citep{2013MNRAS.428.2163F}. The most massive stars from these scenarios, as well as the massive binary model, BPASS~\citep{2009MNRAS.400.1019E}, can give intrinsic $f_{\lyc}/f_{UV}\sim1$. This, in combination with the Lyman Bump mechanism, may help explain our observations. \\

\noindent Many of the \lyc~LAEs with flux density ratios $f_{LyC}/f_{UV}\gtrsim1$ have small to insignificant spatial offsets, and were therefore not flagged as possible contaminants. We have shown in Sec.~\ref{sec:contamination_stat} that it is statistically highly unlikely for the full \lyc~LAE sample to be foreground contaminants: the cumulative probability to have $21$ or more contaminants is $1.6\%$, with a probability of $1\%$ to have exactly $21$ contaminants. Our observations are one single trial. We find no compelling reason to assume that in one single trial we have managed to observe an event which has $1\%$ probability of realization.\\

\noindent Consider further that even if all \lyc~LAE candidates with $f_{LyC}/f_{UV}\gtrsim1$ are assumed to be foreground contaminants, this would not automatically explain their observed flux density ratios. Most of these objects have $1.5\lesssim f_{LyC}/f_{UV}\lesssim2$. As a reminder, \lyc~is measured in the \filterlyc~band with effective wavelength $\lambda_{eff}=359$ nm, and UV in the \filterr~band with $\lambda_{eff}=650$ nm. These objects therefore have $f_\nu \propto \nu^{0.7}$ to $f_\nu \propto \nu^{1.2}$, respectively. The corresponding UV slope range is $-2.7\lesssim \beta_{UV}\lesssim -3.2$. The average $\beta_{UV}$ for galaxies at $1\le z<3$ is $\beta_{UV}=-1.5$~\citep[e.g][]{2009ApJ...705..936B,2014ApJ...793L...5K}. Even if such extreme foreground objects exist, they would be very rare, further decreasing the probability that all of them are foreground contaminants.\\

\begin{figure}
\includegraphics[width=84mm]{figures/legend_ALL_horizontal.pdf}
\includegraphics[width=84mm]{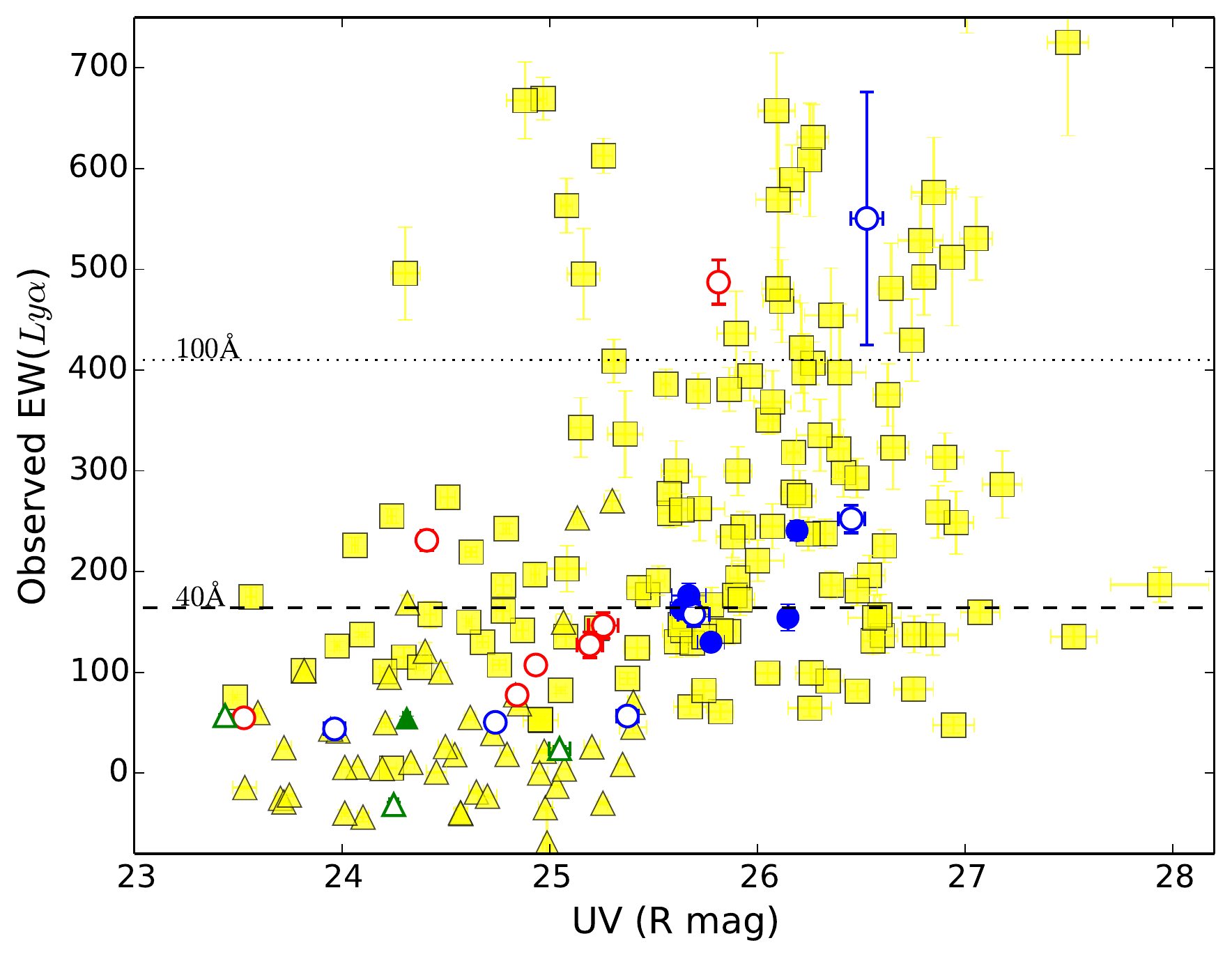}
\caption{\ew~vs. total UV magnitude. Rest-frame \ew~for $z=3.1$ indicated with dashed/dotted lines for convenience. Only objects with redshift $3.06 \le z \le 3.13$ are shown.}
\label{fig:lya vs R}
\end{figure}

\subsection{Observed Ly$\alpha$~Equivalent width}
The \filterlya~band measures the \lya~line for galaxies in the redshift range $3.06<z<3.13$.~\citet{2009MNRAS.400L..66M} derive a conversion formula for the observed equivalent width using the same \filterlya~band and a so-called \textit{BV} image (\textit{BV}$=(2$\filterb$+$\filterv$)/3$) with the same effective wavelength as the \filterlya~band:
\[
EW(Ly_{\alpha})=\frac{(887+936)\times77\times(10^{0.4(BV-NB497)}-1)}{(887+936)-0.6\times77\times10^{0.4(BV-NB497)}}
\]
where $887$ \AA, $936$ \AA, and $77$ \AA~are the FWHMs of the \filterb, \filterv, and \filterlya~respectively. Using their formula in Figure~\ref{fig:lya vs R} we show the observed \ew~for objects associated with the proto-cluster. The redshift constraint imposed by the \lya~filter limits the non-LyC LBG sample to $46$ galaxies, and the non-\lyc~LAEs to $132$. Similar to previous results in the literature~\citep[e.g.,][]{2003ApJ...588...65S, 2006ApJ...645L...9A}, we find a statistically significant correlation at the $9\sigma$ level between decreasing galaxy brightness and increasing \lya~emission strength using both Spearman's $\rho$ and Kendall's $\tau$, which are $0.59$ and $0.41$, respectively.~\citet{2011ApJ...736...18N} find that LAE galaxies with \lyc~detection tend to have lower \ew~than non-\lyc~LAE. In our sample of non-\lyc~LAEs the average is $\left<EW(Ly\alpha)\right>\sim279\pm16$\AA, where the uncertainty is the standard error. For \lyc~LAEs it is $\left<EW(Ly\alpha)\right>\sim178\pm33$\AA, which is significantly lower than the general population at a $>99\%$ confidence level, although the sample scatter is large. For the LBGs it is difficult to draw conclusions because there are only four \lyc~LBG candidates within the redshift range of the \lya~filter (Fig.~\ref{fig:lya vs R}). However small the sample, we can note that contrary to the literature, the average ($\sim26$\AA), weighted average ($\sim34$\AA, the individual uncertainty of each data point was used as weight), and median ($\sim40$\AA) \ew~for these four \lyc~LBGs are consistent with or larger than those of the full LBG sample (\ew$~\sim34$\AA, $~\sim2$\AA, $~\sim19$\AA). We cannot confirm or refute the claim that galaxies with \lyc~detection tend to have lower \lya~based on this investigation. However, even if true, this does not preclude the possibility of a positive \lyc/\lya~correlation among the subgroup of galaxies with \lyc~emission.
\subsection{Ly$\alpha$ vs LyC}
\begin{figure}
\includegraphics[width=84mm]{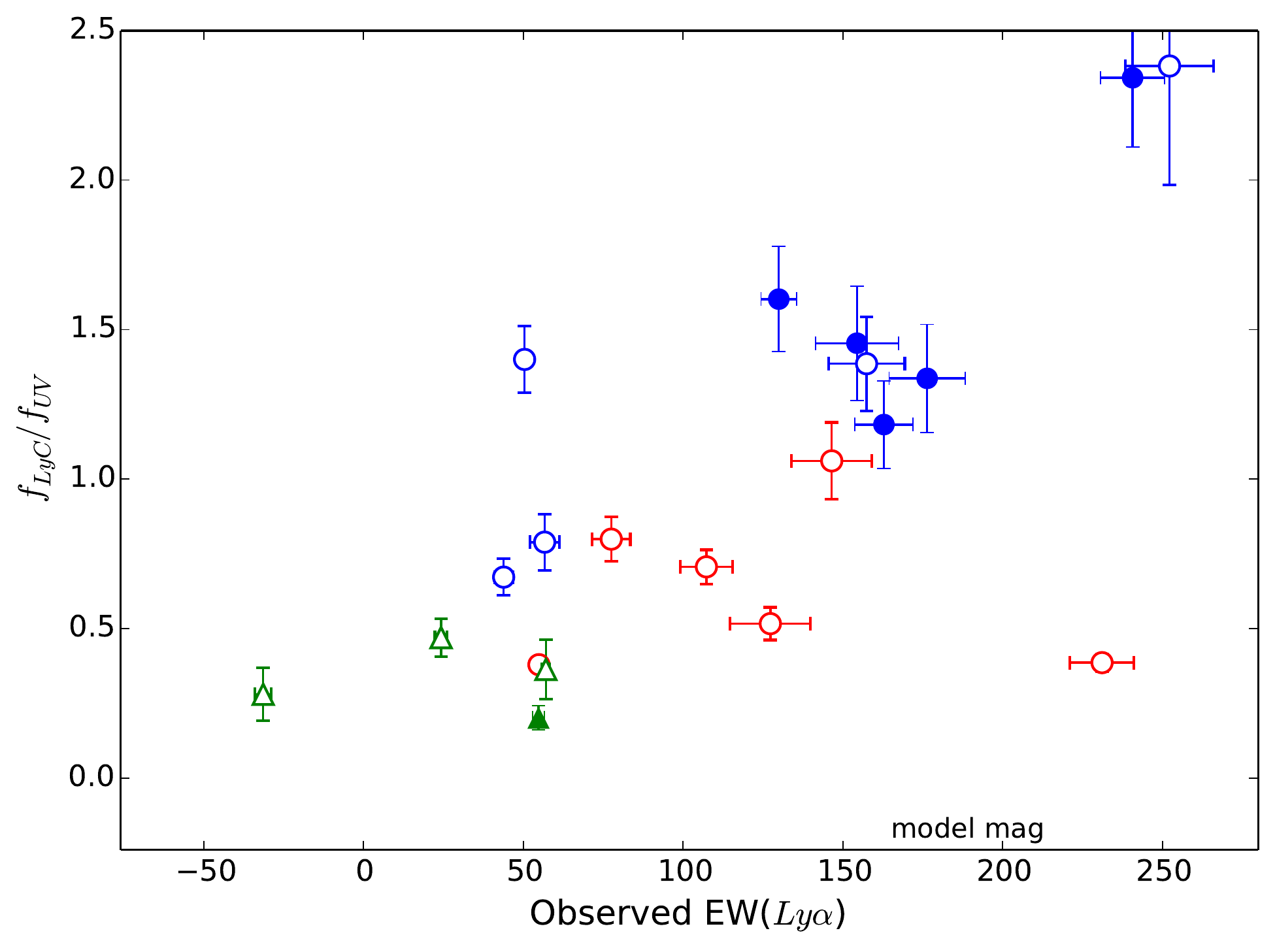}
\caption{\lyc~versus \lya~equivalent width. LAE11 is omitted because it is too faint in \filterb,~\filterv~for reliable conversion to \ew. Spearman's correlation coefficient $\rho$ is $0.67$, and Kendall's $\tau=0.5$, however the sample size is too small to infer a statistically significant correlation. If the correlation is real it suggests that with decreasing \ew~the flux density ratio also decreases. Marker legend is the same as in Fig.~\ref{fig:Vi vs R}.}
\label{fig:lyc vs lya}
\end{figure}

 The mechanisms that allows \lya~to escape from the confines of a galaxy are likely to also fascilitate the escape of \lyc, especially if the main reason behind a successful escape is the geometry of the dust and gas distribution in the galaxy, and the presence of possible tunnels of low dust and gas density through which both \lya~and \lyc~can escape~\citep{2011MNRAS.418.1115R}. On the other hand, a simplified view of the post-production fate of ionizing radiation states that if most of the ionizing photons produce \lya~then the amount of \lyc~will be significantly decreased, if not zero. The intrinsic \lya~or \lyc~cannot be measured but it is nonetheless interesting to examine the observed \lya~and observed \lyc~for any possible correlations. 
  There is a lot of scatter in Figure~\ref{fig:lyc vs lya}, in part surely due to the \ew~conversion formula, and a statistically significant correlation cannot be inferred from e.g. Spearman's $\rho=0.67$ or Kendall's $\tau=0.5$ because the number of available data points is too small. Taken at face value, the figure would suggest that with decreasing \ew~the flux density ratio also decreases and the most extreme $f_{LyC}/f_{UV}$ objects have significant \ew. On the one hand this seeming correlation may be secondary, a product from the correlations between flux density ratio and luminosity in Figure~\ref{fig:flux density ratios}, and \ew~and luminosity in Figure~\ref{fig:lya vs R}. On the other hand, such a correlation would be consistent with the idea that both a \lya~and a \lyc~photon benefit from the same environmental conditions to escape. This is further supported by the fact that on average LAEs, which have a stronger \lya~emission, show a \lyc~emission greater than that of LBGs, as Figure~\ref{fig:lyc vs lya} clearly indicates. Another possible explanation for this correlation may be a selection bias of the Lyman Break method which preferentially rejects galaxies with small amount of IGM attenuation and strong \lyc~emission~\citep{2014MNRAS.441..837C}. Either way, we cannot reject the null hypothesis that $f_{LyC}/f_{UV}$ and \ew~in Figure~\ref{fig:lyc vs lya} are statistically independent with any confidence level worth mentioning. \\

\begin{figure}
\includegraphics[width=84mm]{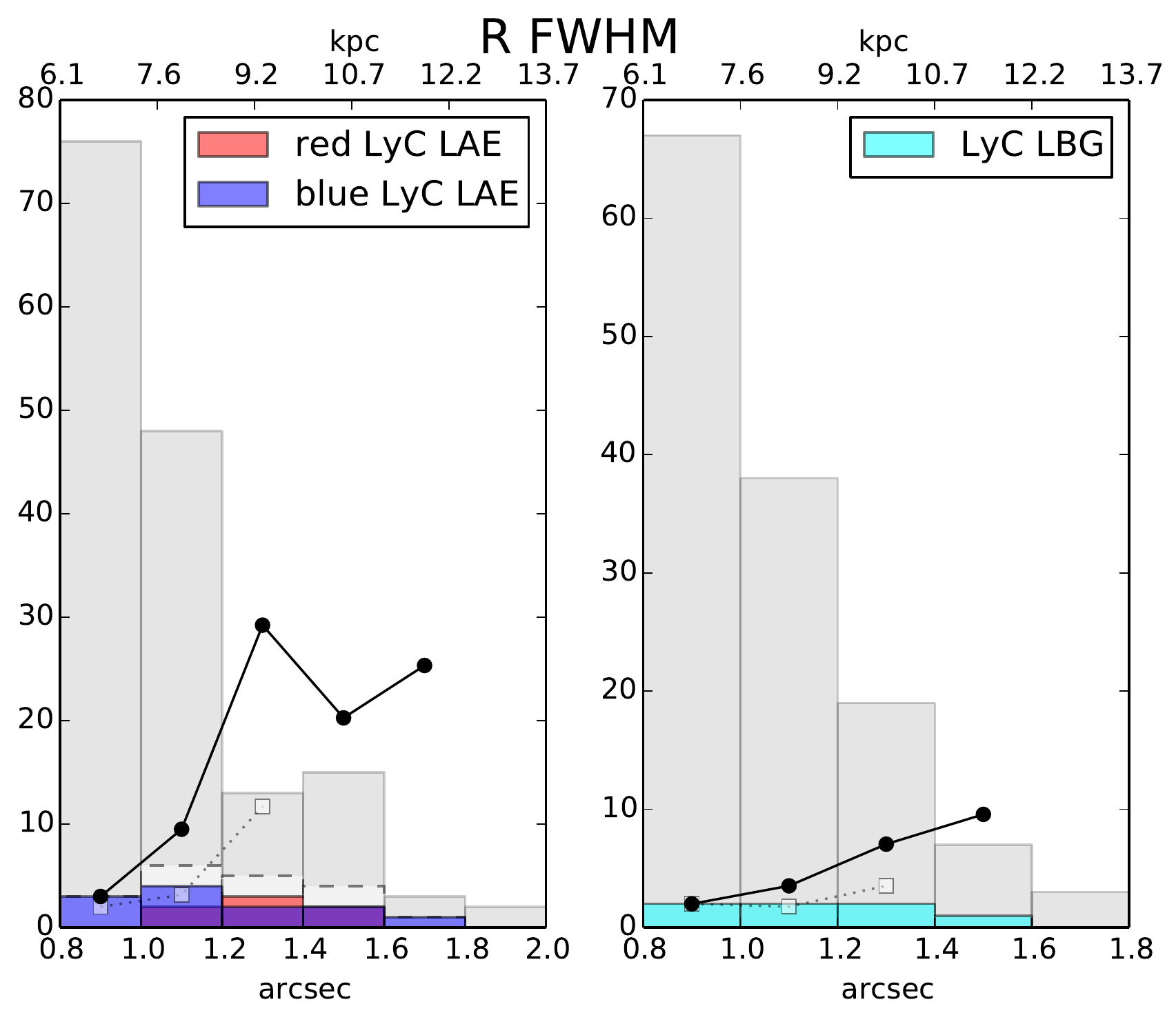}
\caption{Distribution for FWHMs for all LAEs (left panel, gray bars) and all LBGs (right panel, gray bars). The dashed histogram in the left panel is all \lyc~LAE. The black connected circles are the ratio of \lyc~to non-\lyc~galaxies plotted on arbitrary horizontal scales in the corresponding bin (bin size 0.2\arcsec). In both panels the square markers with dotted connecting line are the same ratio but excluding the possible (but unconfirmed) contaminants.}
\label{fig:fwhm}
\end{figure}

\subsection{Indications of mergers}
We have measured the FWHMs of the \lyc~candidates as described in Sec.~\ref{sec:photometry}. In all filters there is a trend for the ``large'' galaxies to be more often also \lyc~leakers compared to the smaller galaxies. We show this for the \filterr~band in Figure~\ref{fig:fwhm}. This is especially clear for LBGs but the trend is also there for LAEs. The Point Spread Function (PSF) of the images is $\sim1\arcsec$ and consequently any FWHM which measures $\lesssim1\arcsec$ simply indicates that we cannot resolve that object. Equally important is the fact that large FWHM does not indicate greater mass of the galaxy. Instead the large FWHMs come from the presence of offset substructures. Such substructures could indicate the infall/accretion of a minor object. For LBGs in the SSA22 proto-cluster a high merger fraction compared to field LBGs was recently found by~\citet{2016MNRAS.455.2363H}. We interpret the increased number of \lyc~candidates among the objects with large FWHMs as an indication that infall/accretion provides the necessary disturbance of the otherwise regular distribution of dust and gas in the galaxy, perhaps creating tunnels with low dust and gas density through which \lyc~escapes. \\

\begin{figure}
\includegraphics[width=84mm]{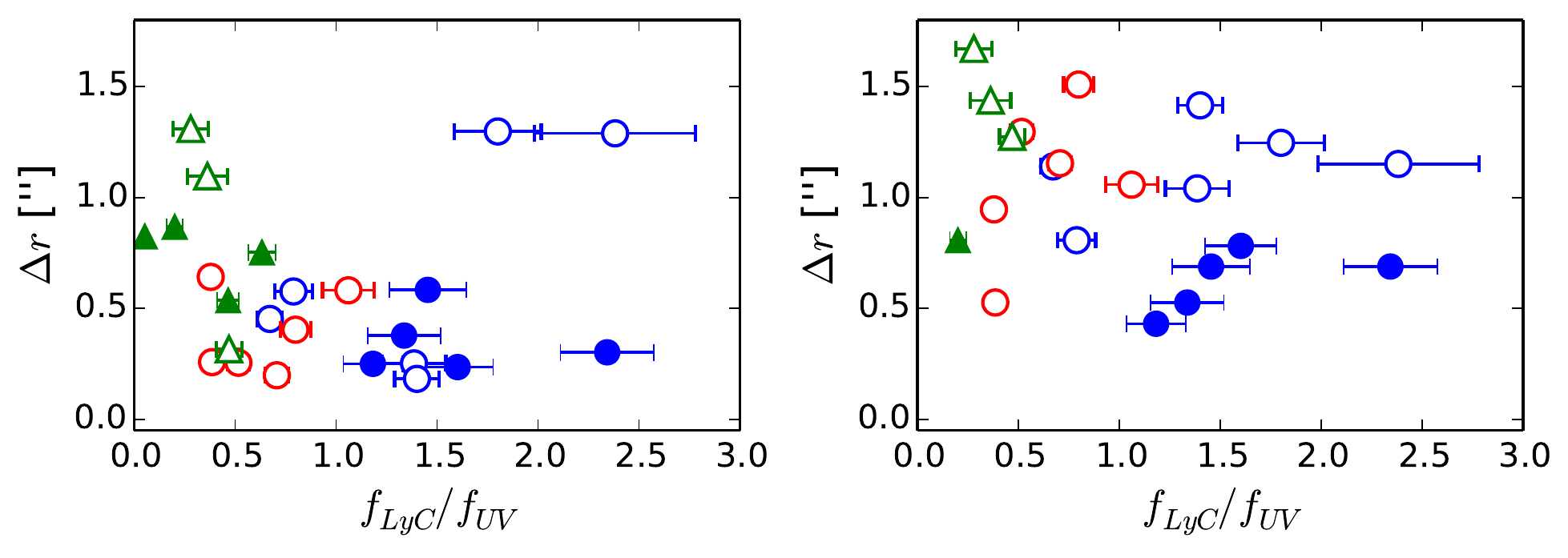}
\caption{Flux density ratio versus the offset $\Delta r$, measured between \filterr~and \filterlyc~(left panel), and between $($\filterlya$-$\filterbv$)$ and \filterlyc~(right panel). Marker legend is the same as in Fig.~\ref{fig:Vi vs R}.}
\label{fig:offsets}
\end{figure}

\noindent The probability for foreground contamination increases with offset between the \lyc~and UV positions~\citep{2010MNRAS.404.1672V}. If large $f_{\lyc}/f_{UV}$ ratios were due to foreground contaminants one may then expect large offsets for those objects as well. In Figure~\ref{fig:offsets} we do not see such a trend.

\subsection{$\left<f_{\lyc}/f_{UV}\right>_{obs}$ from stacking}\protect\label{sec:stack}
We estimate the average flux density ratio from stacking. We exclude the confirmed contaminants from this analysis, however since our candidate sample may contain foreground contaminants we use a MC method with 100 realizations to obtain statistically corrected averages. All measurements are from $\diameter=1.2\arcsec$ apertures and corrected for Galactic extinction. First, for all galaxies in the sample we make $10\arcsec\times10\arcsec$ cutouts from the \filterlyc~image. For each object we normalize the cutout by the corresponding \filterr~band flux. In each MC realization the number of contaminants $n_{cont}$ is randomly drawn from the probability mass function we obtained in Section~\ref{sec:contamination_stat}, shown in Figure~\ref{fig:foreg}. When stacking all LAEs (LBGs) a number of $N+n$ images are stacked, where $N=138$ ($N=127$) for LAE (LBG) non-detections in the \filterlyc~filter, and $n$ is the remaining number of \lyc~candidates in that run, $n=18-n_{cont}$ ($n=7-n_{cont}$) for LAEs (LBGs). Occasionally the number of contaminants $n_{cont}$ is greater than the number of candidates, and we set $n=0$ so that only $N$ images are stacked in that MC run. The uncertainty in the resulting averages is estimated in each MC run for each object by randomly placing $100$ apertures on empty sky, normalizing each one by the object's \filterr~band flux, and measuring the standard deviation of the aperture photometries at the 100 sky positions. As a sanity check of the uncertainty estimation we directly measured $\sigma$ from the final stacked images by placing a few apertures away from the central region and obtaining the standard deviation of the photometries. Although the available area is small, and the number of apertures therefore limited, these measurements are consistent with our MC estimation of the uncertainties.\\

\begin{figure}
\includegraphics[width=84mm]{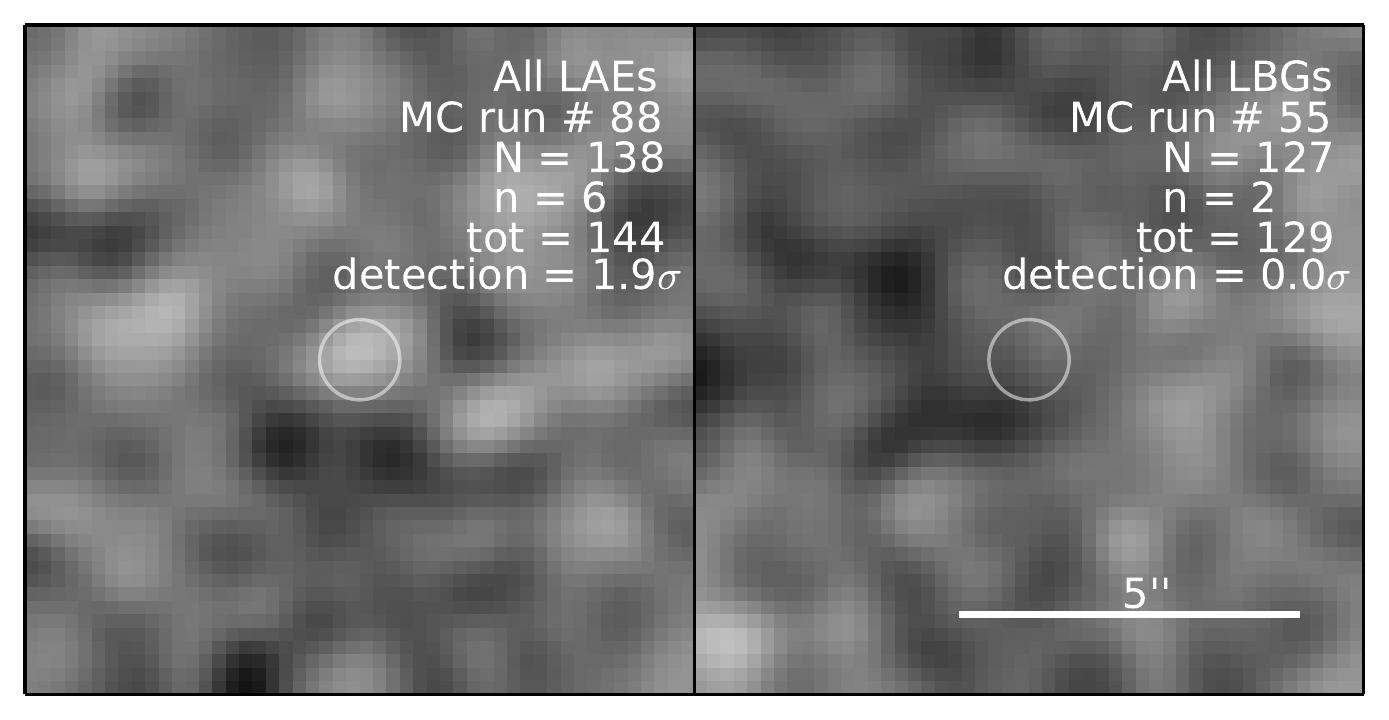}
\caption{Average stacks of $10\arcsec\times10\arcsec$ cutouts of \filterlyc$-$\filterr~from example MC runs for all (\lyc~and non-\lyc) LAEs and LBGs. $N$ is the number of non-detections in \filterlyc~objects, $n$ is the number of \lyc~candidates for that MC run, and the MC run number is the realization identifier. The aperture is indicated with a circle. The contrast levels are set to $\pm5\sigma$.}
\protect\label{fig:stacks}
\end{figure}

\noindent Figure~\ref{fig:stacks} shows random MC runs for the stacks of all LAEs and LBGs. These example runs show a marginal detection of $1.9\sigma$ for the stack of all LAEs (left panel), and a non-detection for the stack of all LBGs (right panel). Note that even if there is a detection for a stack from a single MC realization, we obtain the average flux density ratio of all MC realizations and the final result may therefore be a non-detection. \\

\begin{table*}
\begin{minipage}{184mm}
\caption{The average observed $3\sigma$ upper limits on \flycfuv~from stacking all LAEs and all LBGs. The observed \viobs~is used to estimate the dust attenuation by comparing to model $(\textit{V}-\textit{i}')_{i}$ predictions from SB99, with 100Myr. The relative and absolute escape fractions are shown for median IGM attenuation $\tau_{IGM}^{med}=0.74^{m}$. $(\textit{NB359}-\textit{R})_{i}$ is the intrinsic color. The average dust attenuation $\left<A_{UV}\right>$ is given for completion. The effect of foreground contamination is statistically corrected; see text for details. }\protect\label{tab:fesc}
\begin{tabular}{ l c c c c c c c }
\hline
stack & \flycfuv &\viobs & $(\textit{NB359}-\textit{R})_{i}$ & $(\textit{V}-\textit{i}')_{i}$ & $f_{esc}^{rel}$ & $\left<A_{UV}\right>$ & $f_{esc}$ \\
\hline
LAEs & $<0.077$&   $0.127$& $ 1.57$ & $-0.19$  &$< 0.685$ &$ 1.67$&  $< 0.147$\\
LBGs & $<0.020$&   $0.350$& $ 1.57$ & $ -0.19$ &$<0.181$  &$ 2.80$&  $<0.014$\\
\hline
\end{tabular}
\end{minipage}
\end{table*}

\noindent Table~\ref{tab:fesc} summarizes the results in terms of the average flux density ratios of $100$ such MC realizations. Since we have no $>3\sigma$ detection, we present $3\sigma$ upper limits instead.~\citet{2013ApJ...765...47N} obtain an observed emsemble average of $f_{\lyc}/f_{UV}$ of $0.086\pm0.036$ for LAEs and $0.017\pm0.011$ for LBGs in the SSA22 field with a smaller areal coverage of $42$ arcmin${}^2$. These are consistent with our $3\sigma$ limits of $<0.077$ for LAEs and $<0.020$ for LBGs. Our results are also consistent with~\citet{2013ApJ...779...65M}, who find an emsemble average of $0.060\pm0.029$ for LAEs and $0.0043\pm0.0024$ for LBGs in the proto-cluster {\sevensize HS1549+1933} at a comparable redshift of $z=2.85$. In fact, our upper limits are similarly consistent with all previous results in the literature, e.g.~\citet[][$z=3.6$ LBGs, $f_{\lyc}/f_{UV}<0.002$]{2010ApJ...725.1011V},~\citet[][$z=3.3$ LBGs, $f_{\lyc}/f_{UV}<0.013$]{2011ApJ...736...41B},~\citet[][$z=3.3$ LBGs, $f_{\lyc}/f_{UV}<0.006$]{2016A&A...585A..48G},~\citet[][$z=3.4$ LBGs, $f_{\lyc}/f_{UV}<0.026$]{2016A&A...587A.133G},~\citet[][$z=3.5$ LBGs, $f_{\lyc}/f_{UV}=4.66\cdot10^{-3}\pm1.20\cdot10^{-3}$]{2016arXiv160201555S}. We also tested stacking only non-detections in \lyc~but the result remained a non-detection, suggesting a possible bimodality of the \lyc~escape.\\

\subsection{Escape fraction estimation}\protect\label{sec:fesc}
The escape fraction of \lyc~is defined as:
\[
f_{esc}=\frac{L_{LyC,out}}{L_{LyC,int}}=\frac{\left(f_{LyC}/f_{UV}\right)_{obs}}{\left(f_{LyC}/f_{UV}\right)_{int}}10^{-0.4A_{UV}}e^{\tau_{IGM,LyC}}
\]
where $\left(f_{LyC}/f_{UV}\right)_{int}$ is the intrinsic flux density ratio at $<912$\AA~to $1500$\AA, $A_{UV}$ is the attenuation by dust at $1500$ \AA, and $\tau$ is the optical depth of the IGM for the \lyc~photons. The relative escape fraction is given by
\[
f_{esc, rel} = \frac{f_{esc}^{\lyc}}{f_{esc}^{UV}} = \frac{f_{obs}^{\lyc}}{f_{obs}^{UV}}e^{\tau_{IGM,LyC}}
\]
where $f_{obs}^{\lyc}$ and $f_{obs}^{UV}$ are the observed flux densities in \lyc~and UV continuum respectively. \\

\begin{figure*}
\includegraphics[width=184mm]{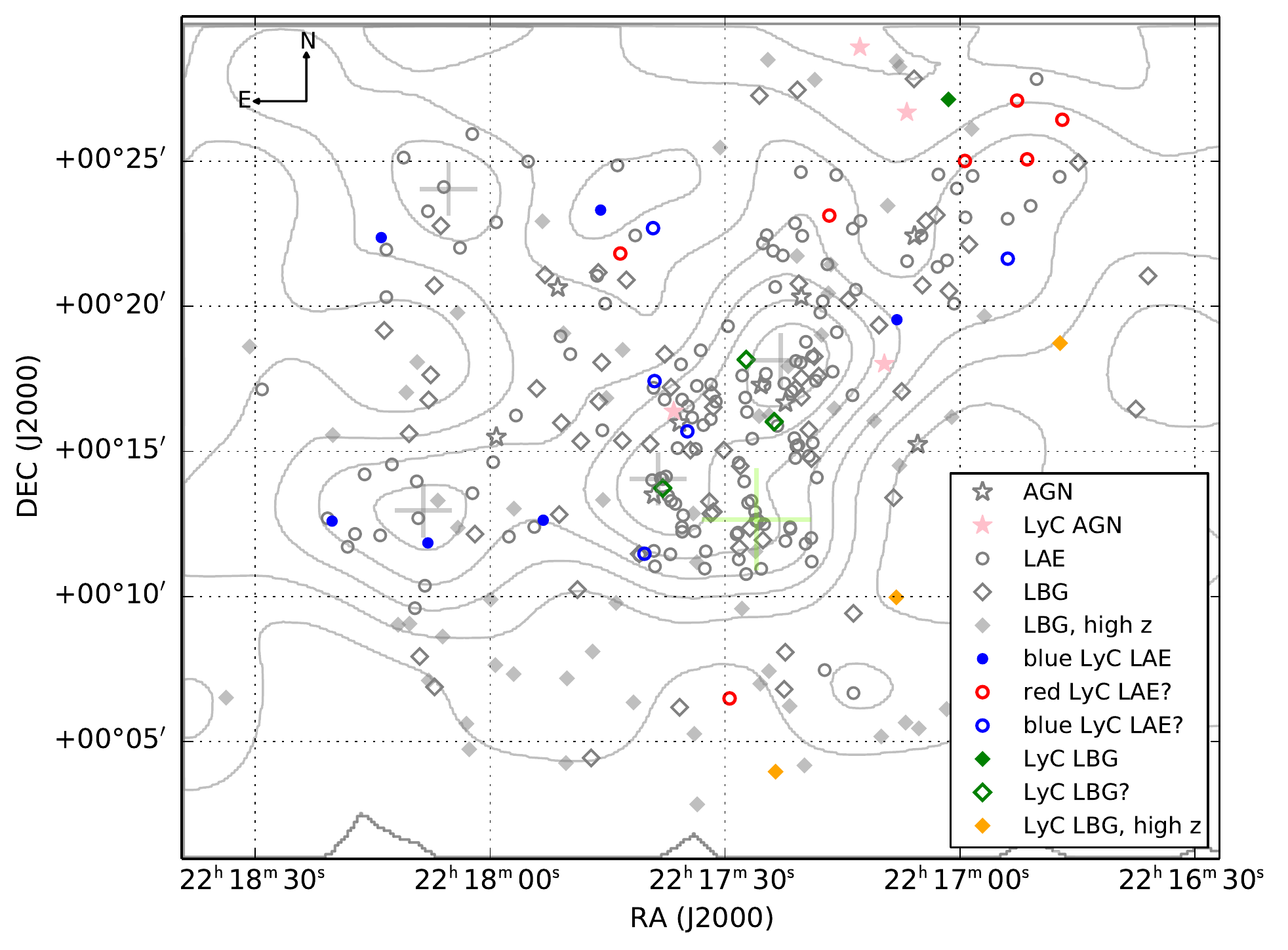}
\caption{Spatial distribution of the samples with~\citet{2012AJ....143...79Y} density contours. The position of LAB01 is marked with a green cross. Also marked are the 4 highest overdensities (gray crosses). Possible \lyc~contaminants are shown with open markers of the corresponding shape and color.}
\protect\label{fig:spacedist}
\end{figure*}

\noindent For the IGM attenuation we use the results of the new transmission model in~\citet{2008MNRAS.387.1681I}, but with an updated IGM absorbers' statistics described in~\citet{2014MNRAS.442.1805I}. This transmission model finds significantly smaller IGM attenuation of \lyc~than the~\citet{1995ApJ...441...18M} model. For example the cumulative probability to have an IGM attenuation of $<1$ mag in our \lyc~filter for galaxies at redshift $z\sim3.1$ is $\sim20\%$ for the Madau model but $\sim65\%$ for the~\citet{2014MNRAS.442.1805I} model. We use the median of all sightlines, $\tau^{med}_{IGM}=0.74^{m}$, which is appropriate for general LAEs and LBGs. \\
 
\noindent Direct observation of $\left(f_{LyC}/f_{UV}\right)_{int}$ is not possible so we must obtain it from population synthesis models. For $\left(f_{LyC}/f_{UV}\right)_{obs}$ we take the values from Section~\ref{sec:stack}, which have been statistically corrected for foreground contamination. We use a standard {\sevensize SB99} model with a Salpeter IMF ($\alpha=2.35$), with mass range $1$-$100$M${}_{\sun}$, low-metallicity ($Z=0.0004$), and $100$Myr age. The average relative escape fraction from LAEs and LBGs is shown in Table~\ref{tab:fesc}.\\

\noindent To obtain absolute $f_{esc}$, we estimate the amount of dust attenuation in the individual galaxies by comparing the average observed \filterv$-$\filteri~color of each stacking sample to dust-free model predictions. We assume the dust attenuation follows $A(1500)=10.33E(B-V)$~\citep{2000ApJ...533..682C}. With $E(B-V)=0.1$, the predicted reddening in $($\filterv$-$\filteri$)$ is then $0.199$. The average absolute $f_{esc}$ and the $A(1500)$ values are also listed in Table~\ref{tab:fesc}.\\

\subsection{Spatial distribution}
\begin{figure}
\includegraphics[width=85mm]{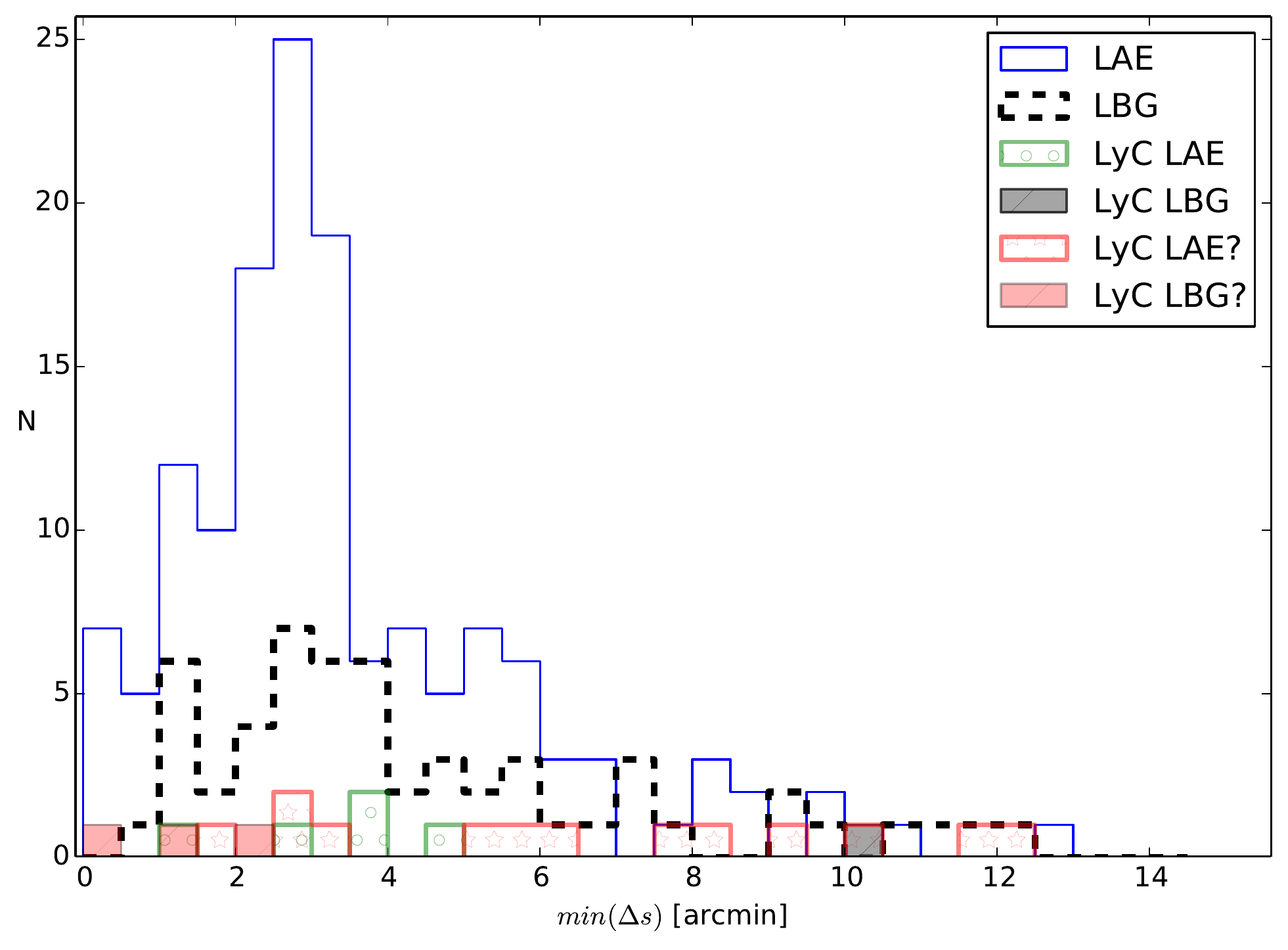}
\caption{Distribution of minimum distances $\Delta s$, measured between a galaxy and the position of the closest overdensity peak. Only galaxies associated with the proto-cluster are included.   }
\label{fig:mindist}
\end{figure}

In Figure~\ref{fig:spacedist} we examine the spatial distribution of all galaxies in the sample together with the \lyc~candidates, superimposed on density contours from~\citet{2012AJ....143...79Y}.  There are four major overdensity peaks, with maxima $6.3, 5.5, 4.3,$ and $3.9$ photometric LAEs per square arcminute. All viable \lyc~candidates seem to occupy regions either at the edges of these concentrations or far away from any overdensity peak. This can be shown more qualitatively by looking at the distribution of minimum distances to any one of the four overdensity peaks in Figure~\ref{fig:mindist}. The figure shows that viable \lyc~candidates seem to avoid overdense regions and are instead found at distances $\gtrsim 1\arcmin\approx450$ kpc in physical units away from any density maximum. The one \lyc~candidate found in the center of an overdensity peak, LBG07, is already flagged as a possible contaminant for morphological reasons in Section~\ref{sec:sampleselection}. The remaining \lyc~candidates are found in regions with lower spatial density compared to their parent sample. The average overdensity at the positions of the non-\lyc~LAEs is $\sim3.9$, while for the \lyc~LAEs it is $\sim2.7$. If we only consider viable \lyc~LAEs the average overdensity is $\sim3.0$. Note that the number of \lyc~LBG candidates unassociated with the proto-cluster is three which is comparable to four \lyc~LBGs proto-cluster members if we count possible contaminants, and is three times greater if we consider only viable candidates ($3:1$). Recently,~\citet{2016MNRAS.455.2363H} reported a higher merger fraction for LBGs in the SSA22 proto-cluster compared to field galaxies. A dynamical perturbation such as a merger could facilitate the building of low density tunnels and holes in the ISM which in turn possibly enables the escape of both \lya~and \lyc. It is then noteworthy that among LBGs \lyc~escape remains curiously unenhanced by the dynamic high density, high merger fraction environment that comprises the SSA22 proto-cluster, and we have in fact only one viable \lyc~LBG candidate (LBG02) at the redshift of the proto-cluster, located very far away from any overdensity peak ($\sim4.5$Mpc from the nearest peak). A similarly puzzling absence of \lyc~leaking proto-cluster LBGs can be noted in~\citet{2015ApJ...810..107M}, who revised their {\sevensize HS1549+1933} \lyc~detections with new multiband HST observations and found one secure \lyc~leaking LBG detection at a redshift of $z\sim3.14$, clearly unassociated with the proto-cluster at $z\sim2.85$. Higher {\sevensize HI} column densities in the IGM of proto-clusters~\citep[][Hayashino et al. in preparation]{2014A&A...570A..16C,2015MNRAS.453..311S} may be responsible for this apparent dearth of \lyc~LBGs in rich proto-cluster environments and could be an indication that one should instead be searching for them in the field. 

\section{Discussion}\protect\label{sec:discuss}

The mechanism that allows \lyc~to escape could be a major merger event in which extreme dynamical disturbance of the ISM clears escape paths for the ionizing photons. Another possibility is that the star formation is instead flocculent, with high mass stars ionizing the gas and blowing holes in the ISM via strong stellar winds and/or during their supernova (SN) stage. Alternatively, the ionizing high mass stars could be formed through cold accretion of pristine gas onto the main system, with the gas compressing and forming stars before it enters the main body of the galaxy where large quantities of neutral hydrogen prevent ionizing photons from escaping. This latter scenario could account for the observed offsets between main galaxy body (in UV) and the substructures associated with the \lyc~leakage~\citep{2011MNRAS.411.2336I}. The \lyc~emission in our candidate sample comes from relatively compact clumps for both LAEs and LBGs. Spatial offsets between these clumps and the position of the galaxy in the \filterr~band are predominantly detected for LBGs (for about half of the \lyc~LBG sample), while few LAE show similar offsets. If not all of them are just due to the chance coincidence of foreground objects, these offset substructures could be accretion or infall of a smaller galaxy with very low metallicity, forming stars for the first time. The fact that the offset substructures are present more often for LBGs than for LAEs possibly reflects the intrinsic differences in the properties of these two types of galaxies. LBGs are generally more luminous than LAEs in rest-frame UV continuum and more massive both kinematically and in stellar populations. Additionally, the data suggest a possible correlation between \lyc~and \lya~strength. While such a correlation seemingly indicates spatial coincidence of both emissions and, consequently, no offsets, \lya~is a resonant line and it is possible for \lya~to escape the galaxy but then be scattered in the circum-galactic medium, thus showing a spatial offset from the \lyc~and UV emissions. \\

\noindent The escape fraction of \lyc~appears to be bimodal in nature, with some galaxies showing very large \lyc~emission, while others have no discernible \lyc~signal. The stacking analysis indicates that there is a dearth of objects with marginal \lyc~detections. If all galaxies leak \lyc, these non-detections could be due to the viewing angle and the geometry of the distribution of the individual galaxy's dust and gas. Alternatively, the bimodality may instead reflect an intrinsic difference in the geometry of the ISM between \lyc~leaking and non-\lyc~galaxies. The geometry may be such that it does not allow for the formation of low density tunnels. We note that if the \lyc~emission is extended and diffuse it would be practically invisible in our images. It is also possible that the apparent bimodality may reflect the different nature of the stellar populations. Perhaps the \lyc~candidates we observe are only a few special LAEs/LBGs which have exotic stellar populations whose high \lyc~emissivity makes them detectable in \lyc.\\

\noindent The LBGs data indicate that the \lyc~candidates are dominated by a very young extremely metal-poor or \popiii~stellar population. The LAE data are difficult to interpret since they are inconsistent with models which assume a standard or even a top-heavy IMF. The debate on IMF variations in the upper mass end for high-$z$ galaxies ($z\ge3$) is ongoing in the literature~\citep[e.g.,][]{2004MNRAS.352L..21R,2008ApJ...680...32C,2011ASPC..440..361M}, and a top-heavy IMF at high-$z$ seems appropriate given the low metallicity environment. The situation is still unresolved, however, and constraints on the upper mass end have proven very hard to obtain. Our \lyc~LAEs data indicate that a dominant young and zero metallicity population seems to be a necessary feature, and further suggest that an exotic IMF comprising predominantly high-mass stars is necessary.  \\

\noindent Our \lyc~LBG detection rate ($6.6\%$) is a factor of $\sim2$ lower than that of the \lyc~LAEs ($13\%$), further strengthening a possible luminosity dependence for the \lyc~escape fraction. It may be harder for \lyc~to escape from LBGs, perhaps due to the lack of strong \lya. This is supported by the clearly lower average \ew~for LBGs in the protocluster (Figure~\ref{fig:lya vs R}) compared to LAEs. Some of the \lyc~candidates show offsets between the \lyc~detection and the \lya~emission, with no discernible counterpart of \lya~emission at the position of the \lyc~detection. These could be galaxies that have recently completely exhausted their HI supply and thus are no longer making new stars. If the star formation has only recently been quenched so that massive young stars are still present and producing ionizing radiation, the galaxies would still be detectable in UV continuum observations and the lack of circum-galactic gas would facilitate the escape of \lyc~without an accompanying emission in \lya.

\section{Conclusions}\protect\label{sec:conclusions}
We present the largest to date sample of \lyc~candidates at any redshift with $18$ \lyc~LAEs and $7$ \lyc~LBGs from a base sample of $156$ LAEs and $134$ LBGs, with very low statistical probability that all candidates are foreground contaminated. Many \lyc~candidates show a spatial offset between the rest-frame UV detection and the compact \lyc-emitting substructure. These vary in size (anywhere between $0$ to $\sim1.4\arcsec$), depend on how they are measured, and could indicate infall or accretion of a smaller object, which may facilitate the escape of \lyc. Although the sample number is small, especially if excluding also the possible but unconfirmed contaminants, we find evidence for a positive \lyc/\lya~correlation, which is consistent with the idea that both types of photons benefit from the same escape route through the galaxy's ISM. \\

\noindent The \lyc~emission seems to be bimodal - stacking non-detections reveals no significant \lyc~signal in either LAEs or LBGs - but are unable with the current data to decide on the nature of this seeming bimodality as it may be intrinsic or a product of the viewing angle. From stacking we obtain $3\sigma$ upper limits on the average flux density ratio, statistically corrected for foreground contamination. For LAEs we obtain $\left<f_{\lyc}/f_{UV}\right> < 0.08$, for LBGs $\left<f_{\lyc}/f_{UV}\right> < 0.02$. Assuming a Salpeter IMF ($\alpha=2.35$), a standard {\sevensize SB99} model with low metallicity $Z = 0.0004$, and age $100$ Myr the $3\sigma$ upper limits on the absolute escape fractions are $f_{esc}<0.15$ and $f_{esc}<0.01$ for LAEs and LBGs, respectively. \\

\noindent Many of the \lyc~LAE candidates in our sample are very strong \lyc~sources, with flux density ratios $f_{\lyc}/f_{UV}>1$, which are difficult to explain with standard models. They are also hard to explain as foreground contaminants, since their UV slope $\beta$ is unusually blue, $-2.7<\beta_{UV}<-3.2$.\\


\noindent There are indications that the dense rich environment of the proto-cluster is not hospitable to \lyc~LBGs, resulting in higher detection rate of viable \lyc~LBG candidates in the field. The spatial distribution of the \lyc~candidates revealed that they seemingly avoid the very dense regions around the overdensity peaks of the proto-cluster.

\noindent The possibility of foreground contamination prevents us from definitive discussion of the nature of strong \lyc~sources and further statistical analysis of \lyc~emissivity and escape fraction of star-forming galaxies at $z\gtrsim3.1$. Sensitive spectroscopy and high spatial resolution imaging for all candidates are necessary.

\section*{Acknowledgments}
GM and II are supported by JSPS KAKENHI Grant number: 24244018. GM acknowledges support by the Swedish Research Council (Vetenskapsr\aa det). AKI is supported by JSPS KAKENHI Grant number: 26287034.
\noindent We extend a warm thank you to Katsuki Kousai for providing us with VIMOS spectra.

\noindent Hubble Space Telescope data presented in this paper were obtained from the Mikulski Archive for Space Telescope (MAST) operated by the Space Telescope Science Institute / National Aeronautics and Space Administration (STScI/NASA) and from the Hubble Legacy Archive, which is a collaboration between the STScI, the Space Telescope European Coordinating Facility (ST-ECF/ESA) and the Canadian Astronomy Data Centre (CADC/NRC/CSA).

\noindent A part of this research has made use of the NASA/ IPAC Infrared Science Archive, which is operated by the Jet Propulsion Laboratory, California Institute of Technology, under contract with NASA.

\appendix

\section{Notes on individual \lyc~candidates}\protect\label{appendix:notes}
Here we briefly summarize our analysis of the individual objects and justify the decision to treat them as viable candidates or possible contaminants. In addition, in Figure~\ref{fig:hst} we show $5\arcsec \times 5\arcsec$ cut-out images of \lyc~candidates with available HST ACS, WFC3/UVIS or WFC3/IR imaging. The HST data are summarized in Table~\ref{tab:HST}. These data are only used for visual analysis of the visible substructure. Data from WFPC2 are too shallow to be useful. We have tried to match the coordinates of the HST image to the Subaru images with a full plate solution but in some images there seems to be a residual offset in the WCS. This is especially visible for LBG05 in the F336W image with overplotted NB359 contours. In Figure~\ref{fig:spectra} we also show complementary spectra for the \lyc~candidates which cannot be found in the literature. Most of our sample is based on the spectroscopic sample of~\citet{2003ApJ...592..728S}.\\

\begin{table}
\begin{minipage}{84mm}
\caption{Summary of the HST data${}^{\dagger}$ for the \lyc~candidates with object ID in our catalog (ID), filter name, instrument (Inst), exposure time (Exp) in seconds, HST proposal number (P. ID), and name of principle investigator (PI). Confirmed contaminants not included.}
\protect\label{tab:HST}
\tiny
\begin{tabular}{|l|l|l|l|l|l|}
\hline
ID&Filter&Inst&Exp&P. ID&PI\\
\hline
LAE08&\multirow{6}{*}{F814W}&\multirow{6}{*}{ACS}&\multirow{6}{*}{6144}&\multirow{6}{*}{10405}&\multirow{6}{*}{S. Chapman}\\
LAE17&&&&\\
LBG05&&&&\\
LBG06&&&&\\
\hline
\multirow{2}{*}{LAE12}&F160W&WFC3&2612&\multirow{2}{*}{11735}&\multirow{2}{*}{F. Mannucci}\\
&F625W&ACS&2208&&\\
\hline
LAE09&\multirow{4}{*}{F814W}&\multirow{4}{*}{ACS}&4670&\multirow{4}{*}{9760}&\multirow{4}{*}{R. Abraham}\\
LAE10&&&4670&&\\
LBG07&&&7320&&\\
\hline
LBG05&F110W&\multirow{4}{*}{WFC3}&2612&\multirow{4}{*}{11636}&\multirow{4}{*}{B. Siana}\\
LBG05&F160W&&2612&&\\
LBG05&F336W&&5300&&\\
LBG06&F336W&&5300&&\\
\hline
\end{tabular}
\end{minipage}
\medskip
$\dagger$ WFPC2 data not listed due to their very low exposure time.
\end{table}

\begin{figure}
\centering\tiny
\includegraphics[width=85mm]{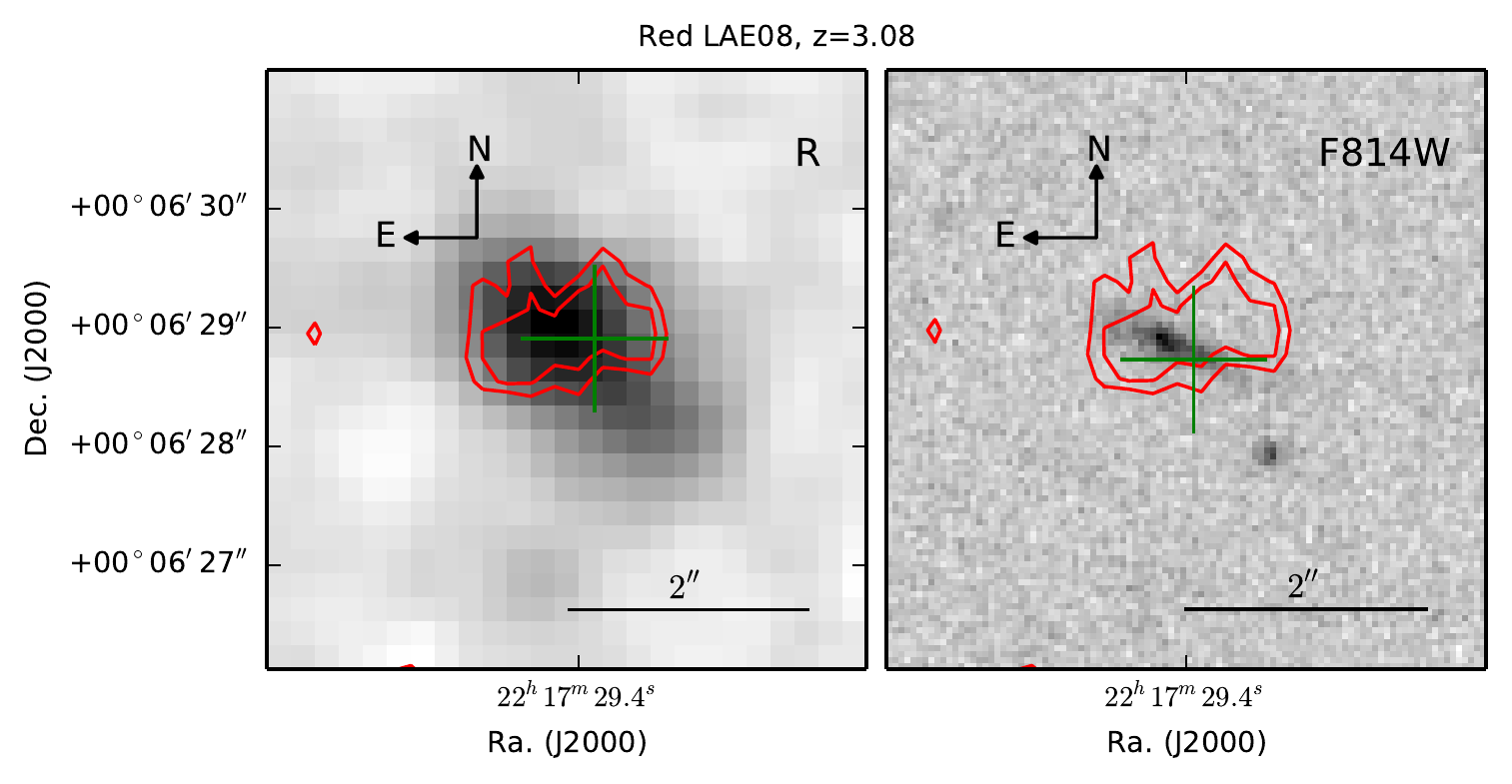}
\includegraphics[width=85mm]{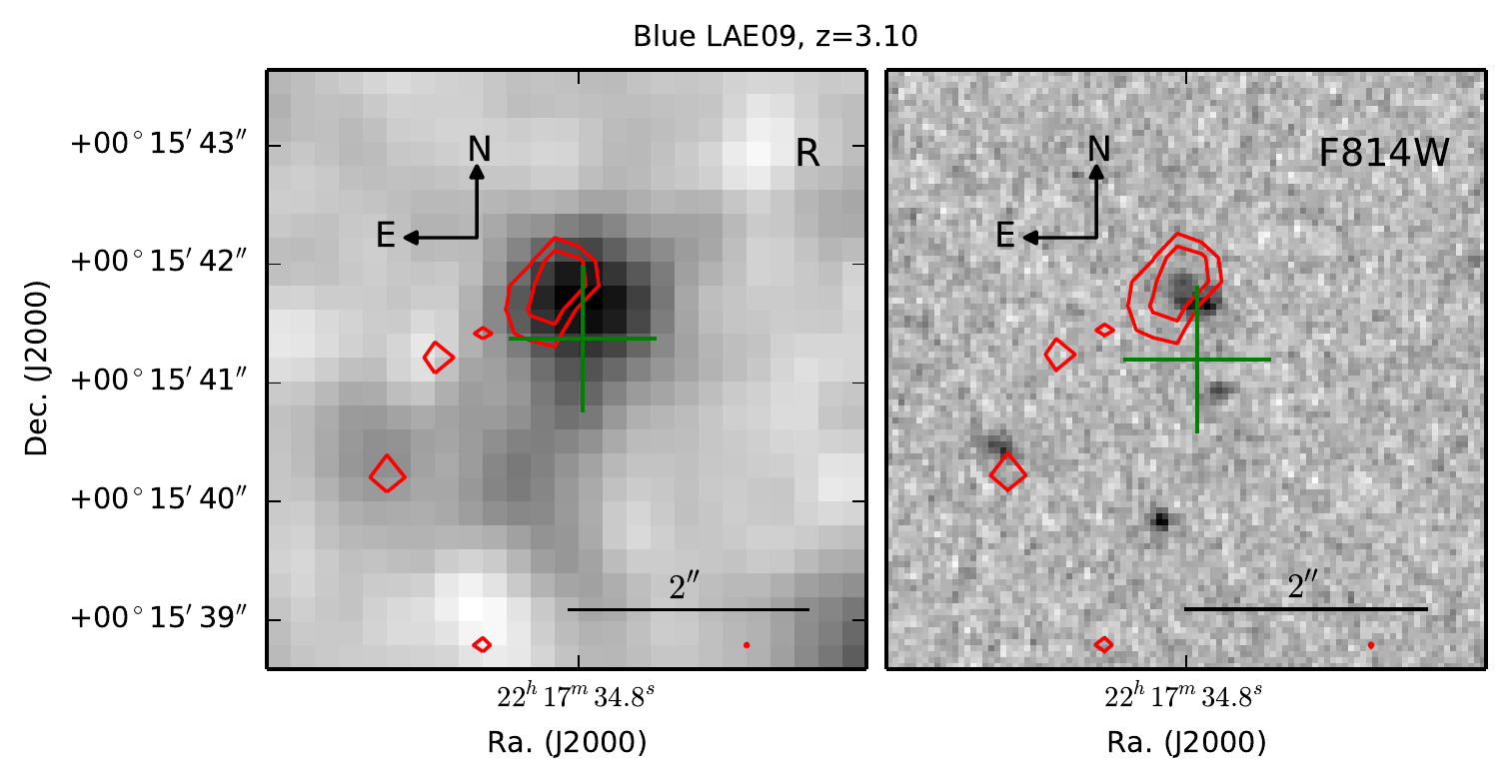}
\includegraphics[width=85mm]{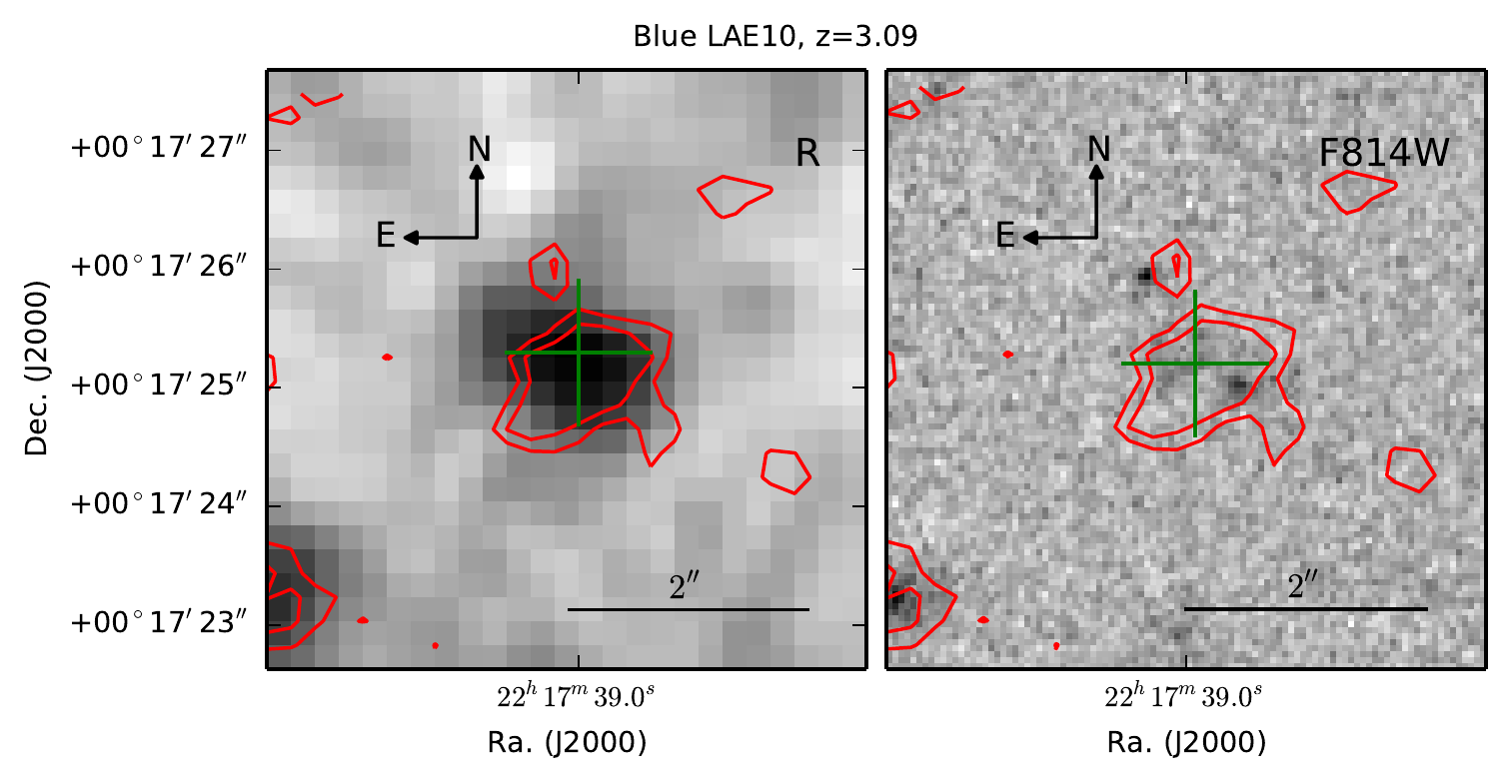}
\includegraphics[width=85mm]{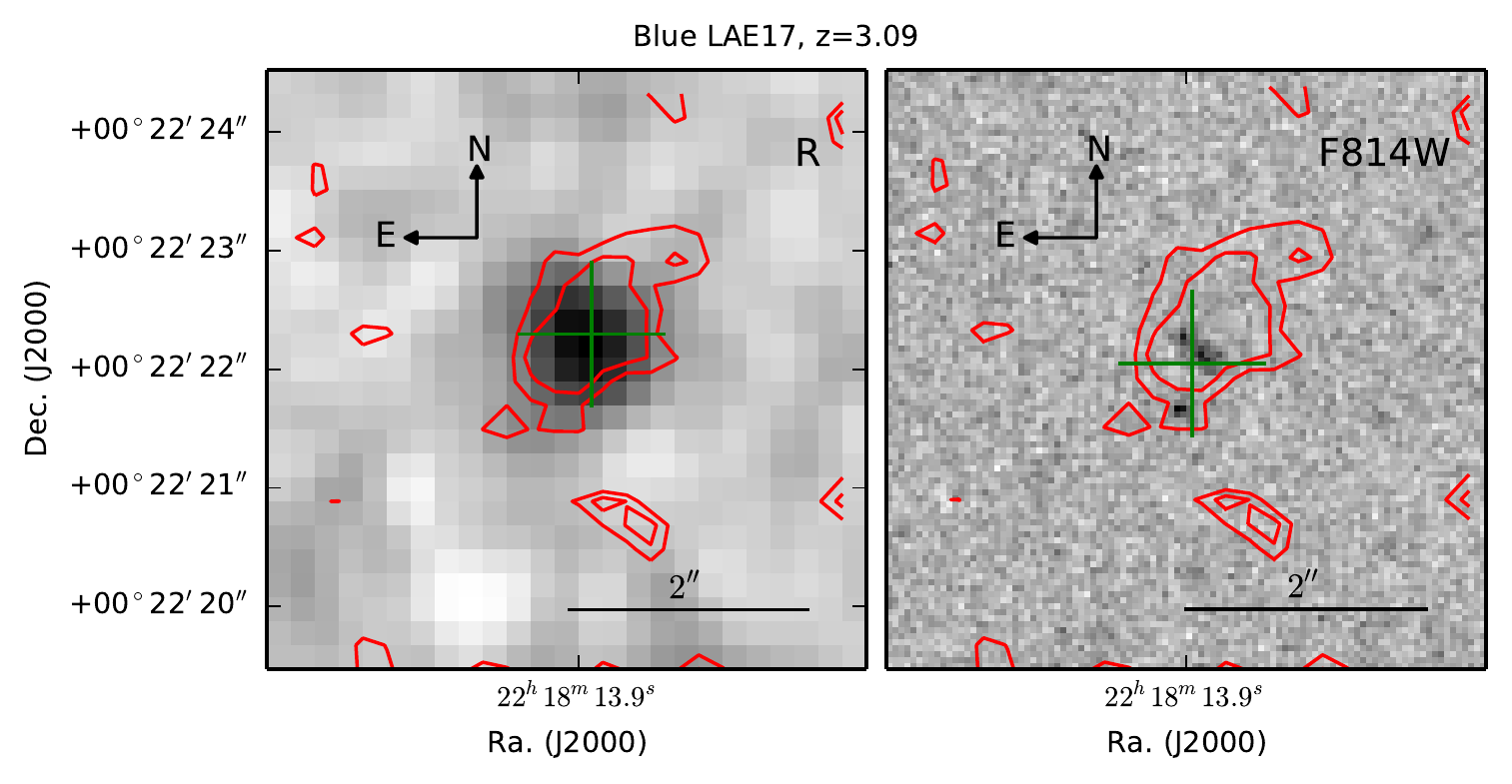}
\includegraphics[width=85mm]{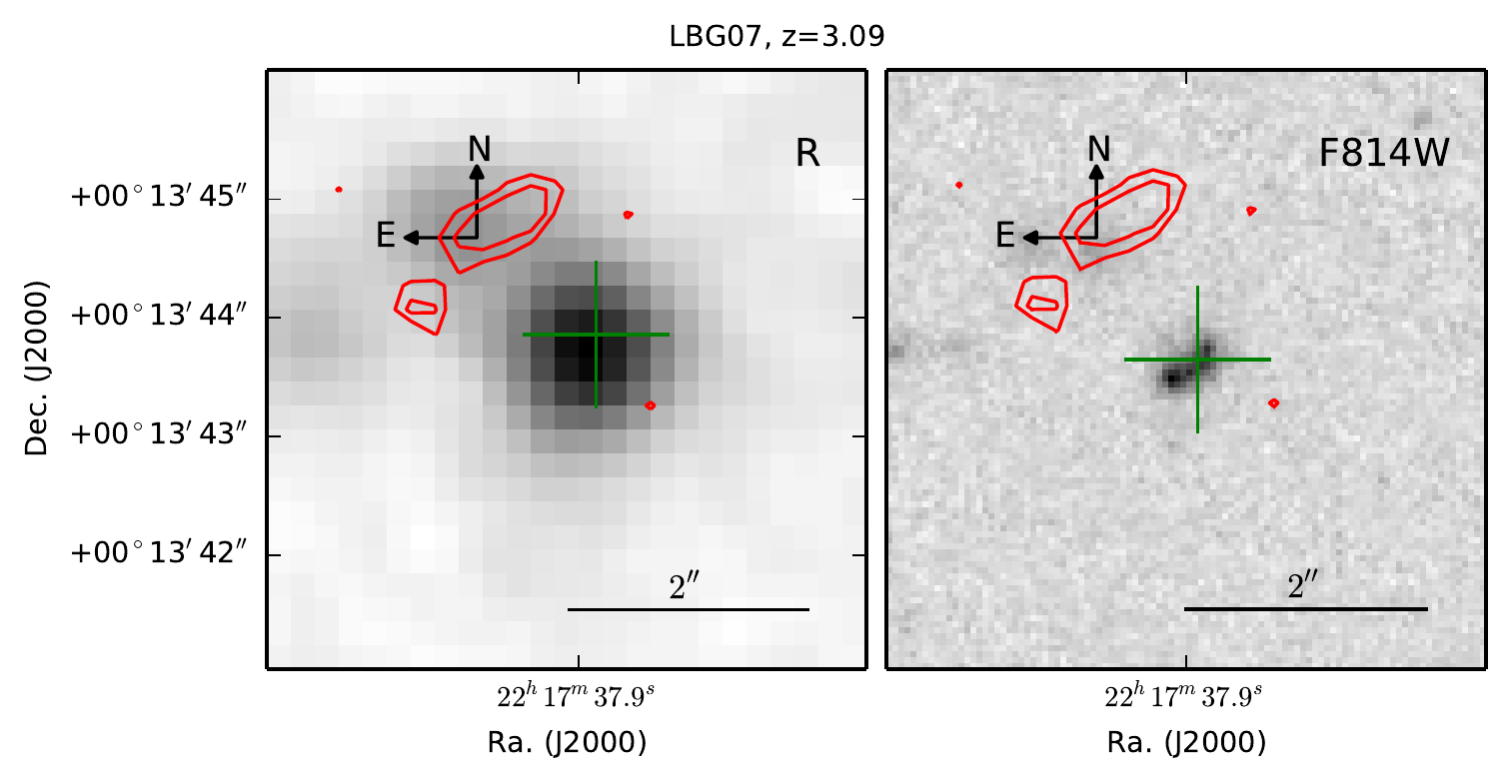}
\caption{HST data of the \lyc~candidates. The intensity scales of the display are individual for each image. The green cross marks the location of the \filterr~band detection in Subaru/Suprime-Cam. The red lines are 2 and 3$\sigma$ contour levels in \filterlyc. Passbands in HST and Subaru images are noted in the images. }\protect\label{fig:hst}
\end{figure}

\begin{figure}
\centering\tiny
\includegraphics[width=85mm]{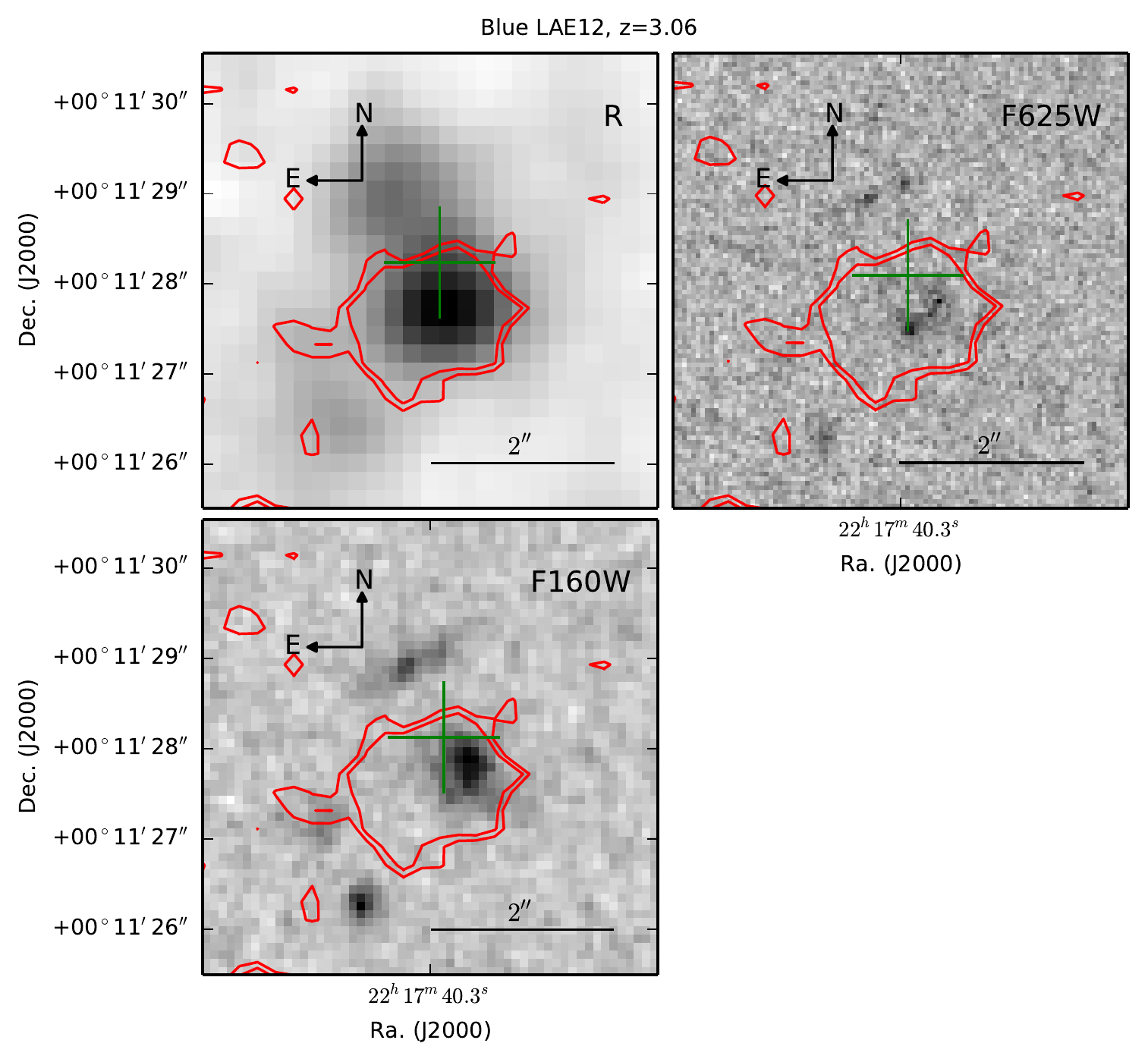}
\contcaption{}
\end{figure}
\begin{figure}
\centering\tiny
\includegraphics[width=85mm]{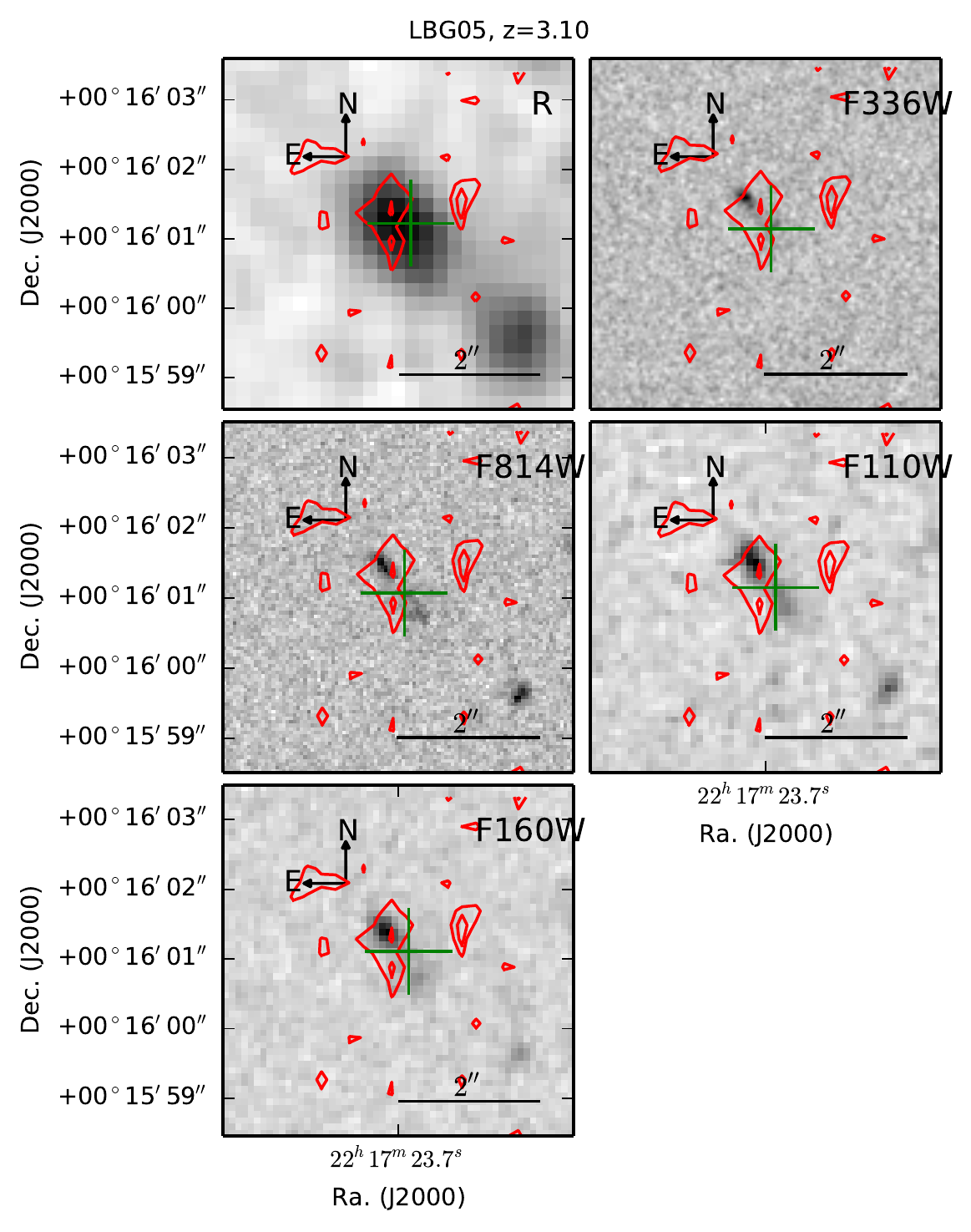}
\includegraphics[width=85mm]{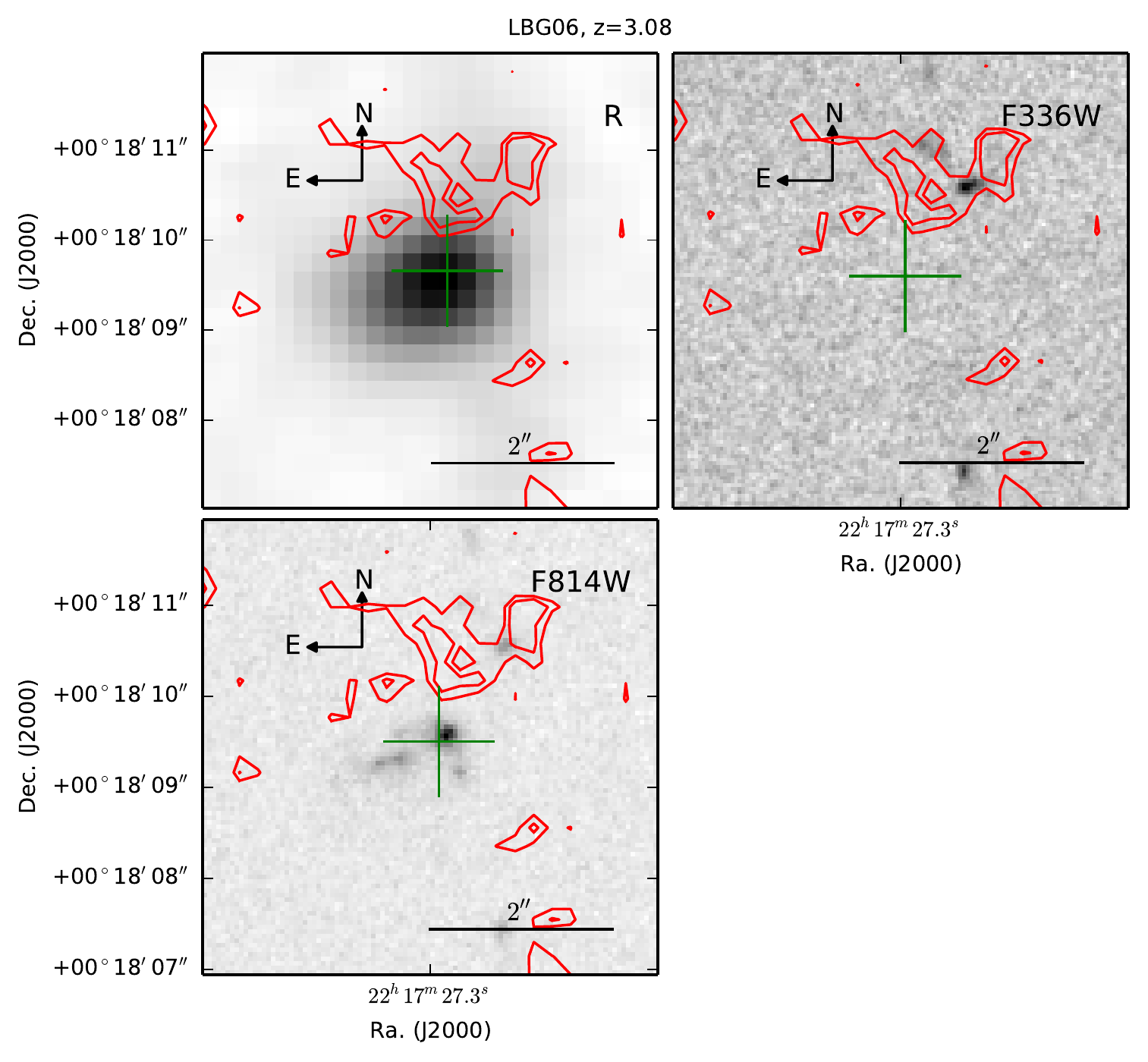}
\contcaption{}
\end{figure}

\noindent \textit{LAE01} This object has one of the largest offsets
between \lyc~(\filterlyc) and \lya~(\filterlya) ($1.3$\arcsec) and we therefore
flag it as a possibly but unconfirmed contaminated object.\\

\noindent \textit{LAE02} There are multiple components in the broadband
images, but no HST data is available for higher resolution. \lya~(\filterlya)
is extended in the South-East direction. In the FOCAS spectrum~(Figure~\ref{fig:spectra}) taken at
the position of \lyc~(\filterlyc) emission, the strongest line is detected at
$\lambda_{obs}=5021$\AA. If this is \lya~then $z=3.123$. There is
another detected line at $\lambda_{obs}=5734.8$\AA, which at this
redshift gives rest-frame $\lambda_{0}=1391$\AA, where no major line is
expected. If the strongest line is instead [OII]
$\lambda_{0}=3728.3$\AA~then the redshift would be $z=0.347$. The second
line would then be at $\lambda_{0}=4257$\AA, where again, no major line
is expected. Since we have an unidentified line in the spectrum we flag
this object as a possibly but unconfirmed contaminated object. \\

\noindent \textit{LAE03} The major line in the spectrum at the \lyc~(\filterlyc)
emission position at $\lambda_{obs}=4969.6$\AA~gives $z=3.088$ if it is
\lya~(Figure~\ref{fig:spectra}). There is a second line at $\lambda_{obs}=5583.9$\AA, which implies
$\lambda_{0}=1365.9$\AA. If the strongest line is instead [OII], then
$z=0.333$ and the second line would be at $\lambda_{0}=4186$\AA. No
major lines are expected at any of these positions. If the second line
is indicating a foreground contaminant, then this object would have to
be exactly aligned within the PSF with the position of the continuum in
the FOCAS spectrum. Since we have an unidentified line we flag this
object as a possibly but unconfirmed contaminated object. \\

\noindent \textit{LAE04} There is a visible counterpart in the
rest-frame non-ionizing UV image (\filterr~band) to the \lya~emission,
which is offset from the source with \lyc~emission by $1.3$\arcsec. Due
to the large offset we flag this as a possibly but unconfirmed
contaminated object.\\

\noindent \textit{LAE05} The primary emission line in the spectrum taken
at \lyc~(\filterlyc) position indicates $z=3.096$ if it is \lya~(Figure~\ref{fig:spectra}). Similarly to
LAE03, there is another line at $\lambda_{obs}=7480$\AA, giving
$\lambda_{0}=1826$\AA~at this redshift. If the primary line is instead
[OII], then $z=0.336$ and the secondary line is at
$\lambda_{0}=5598$\AA. No major lines are expected at any of these
positions. The secondary line is close to a sky line and may be simply
noise. Also, the contaminating object would have to be perfectly aligned
with the continuum (within the PSF). This object is flagged as a
possibly but unconfirmed contaminated one. \\

\noindent \textit{LAE06} This is object 'f'
in~\citet{2011MNRAS.411.2336I}. There are no significant offsets between
\lyc~and the reference position or the \lya~position, and the spectrum
is clean. This is a convincing object.  \\

\noindent \textit{LAE07} This is object 'e'
in~\citet{2011MNRAS.411.2336I}. There is a large offset between
broadband and \lya~($0.89$\arcsec), and also between \lya~and
\lyc~($1.06$\arcsec). We flag this object as possibly contaminated by a
foreground object.\\

\noindent \textit{LAE08} This is object 'h'
in~\citet{2011MNRAS.411.2336I}. There are two distinct objects in the
\filterr~image. The \lya~emission is associated with the object to the
South-West, as seen also from the HST F814W image. The \lyc~(\filterlyc)
emission seems more spatially coincident with the object to the
North-East. Since the \lya~and \lyc~sources are spatially different,
this may be contaminated by a foreground object and we flag it as
such. \\

\noindent \textit{LAE09} This object is a \lyc~candidate
from~\citet[][LAE038]{2013ApJ...765...47N}. We did not detect any
emission in our FOCAS 2008 spectrum.
The spectrum of the source spatially associated with the \lyc~emission
has no emission line and the peak of \lya~emission (\filterlya) is offset by
$0.66$\arcsec, where there is a corresponding substructure seen in the
HST image (Figure~\ref{fig:hst}). We therefore flag this object as a
possibly but unconfirmed contaminated object.  \\

\noindent \textit{LAE10} This is object 'b'
in~\citet{2011MNRAS.411.2336I}. \lya~is offset and appears to be
associated with a separate object seen in \filterb~and \filterv~bands,
and HST F814W. We mark it as possibly contaminated.\\

\noindent \textit{LAE11} The UV continuum (\filterr~band) and the
\lya~emission (\filterlya) are well aligned, with no offset. However, there
is a large offset ($1.1$\arcsec) between \lya~and \lyc~(\filterlyc), and
consequently we flag this object as a possibly but unconfirmed
contaminated object. \\ 

\noindent \textit{LAE12} The \lya~emission in our data is extended and
appears to spatially cover both the UV and \lyc~positions. HST data in
both F625W and F160W reveal a faint continuum ``edge-on'' source at the
position of \lya~(\filterlya). The offset between the centroid detections of
\lya~and \lyc~is large ($1.3$\arcsec) and we therefore flag this object
as a possibly but unconfirmed contaminated object. \\

\noindent \textit{LAE13} There is a large offset between \lyc~and
\lya~($1.0$\arcsec) and also between \lya~and rest-frame UV
($0.85$\arcsec). We therefore flag this object as possibly contaminated
by a foreground object.\\

\noindent \textit{LAE14} This is object 'c'
in~\citet{2011MNRAS.411.2336I}. Rest-frame UV (\filterr~band) and
\lyc~(\filterlyc) are spatially aligned, and although there is a small
($0.8$\arcsec) offset between \lya~and \lyc, the line is clearly visible
in the spectrum, which is otherwise clean. We retain this object as a
candidate due to lack of evidence to the contrary. \\

\noindent \textit{LAE15} This is object 'g'
in~\citet{2011MNRAS.411.2336I}. The \lyc~detection is marginal,
$3.3\sigma$ at the reference position, and $2.7\sigma$ at the position
of \lya. There is no high resolution data for this object and we keep it
as a candidate due to lack of evidence to the contrary.\\

\noindent \textit{LAE16} The offset between \lyc~and \filterr~is
relatively small, $0.59$\arcsec. \lya~appears to be extended and there
is no apparent offset from the broadband images. We keep this object as
a viable candidate.\\

\noindent \textit{LAE17} The HST F814W image shows a counterpart to the
\lya~emission. The offset between \lya~and the broadband images is,
however, small. We keep this object as a viable candidate.\\

\noindent \textit{LAE18} UV and \lya~are well aligned but there is a
large offset between \lya~and \lyc~($1.2$\arcsec). We therefore flag
this object as a possibly but unconfirmed contaminated one. \\

\noindent \textit{LBG01} We see no apparent substructure in the
broadband images, but the \lyc~emission comes from two clumps, one with
a larger offset. The spectrum is clean and this is a viable candidate.\\

\noindent \textit{LBG02} The \lyc~emission is associated with a
substructure (with a $0.8$\arcsec~separation) to the North. The
continuum subtracted \lya~is negative at the reference position but
there is positive signal surrounding the position, including the
location of the \lyc~emission. This is a viable candidate.\\

\noindent \textit{LBG03} The \lyc~emission comes from two clumps
surrounding the UV continuum. The spectrum shows strong \lya~at
$z=3.287$ and this is a viable candidate.\\

\noindent \textit{LBG04} The redshift for this object ($z=3.31$) was
established by~\citet{2003ApJ...592..728S}.
The \lyc~detection is offset from the broadband by $0.8$\arcsec~and it
is marginal even for measurements at the offset position
($3.1\sigma$). This object is retained as a viable candidate.  \\

\noindent \textit{LBG05} This object is
from~\citet{2013ApJ...765...47N}, whose optical spectrum shows a
\lya~line at $z=3.1$. They mention (but do not show) a near-IR spectrum
with a possible [OIII] emission at $z=2.88$, at $0.7$\arcsec
separation. This object was consequently removed from their
\lyc~candidate list. Our FOCAS 2008 spectrum, although noisy, confirms
the $z=3.102$ of the \lya~emission and shows no emission at \lya~for an
assumed redshift of $z=2.88$, which is why we retain this object as a
\lyc~candidate. The available F336W HST image shows an object well
aligned with the substructure we see in our narrowband \lyc~image, and
there is no offset between the UV peak and F336W or the narrowband
\lyc~image. However, this object was recently flagged as a foreground
contaminant by~\citet{2015ApJ...804...17S} based on a single emission
line in their near-IR spectrum of the \lyc~emitting substructure. We
flag this object as a possibly but unconfirmed contaminated one. \\

\noindent \textit{LBG06} This object was treated
in~\citet{2013ApJ...765...47N} as a confirmed \lyc~source, and a new
investigation by~\citet{2015ApJ...804...17S} could not definitively
demonstrate contamination. However, the offset between the UV continuum
and \lyc~emission is rather large ($1.1$\arcsec) and we flag it as a
possibly but unconfirmed contaminated object. \\

\noindent \textit{LBG07} Our LRIS 2010 spectrum for this object is of
poor quality and heavily affected by a nearby star, and no line was
identified. However, this object is placed at $z=3.09$ from spectroscopy
by both~\citet{2003ApJ...592..728S} and~\citet{2013ApJ...765...47N}, so
we consider its redshift confirmed. The HST F814W shows a diffuse
counterpart to the \lyc~emission, however we measure a large offset
($1.3$\arcsec) between UV and \lyc, and flag this object as a possibly
but unconfirmed contaminated one.  \\

\begin{figure}
\centering\tiny
\includegraphics[width=85mm]{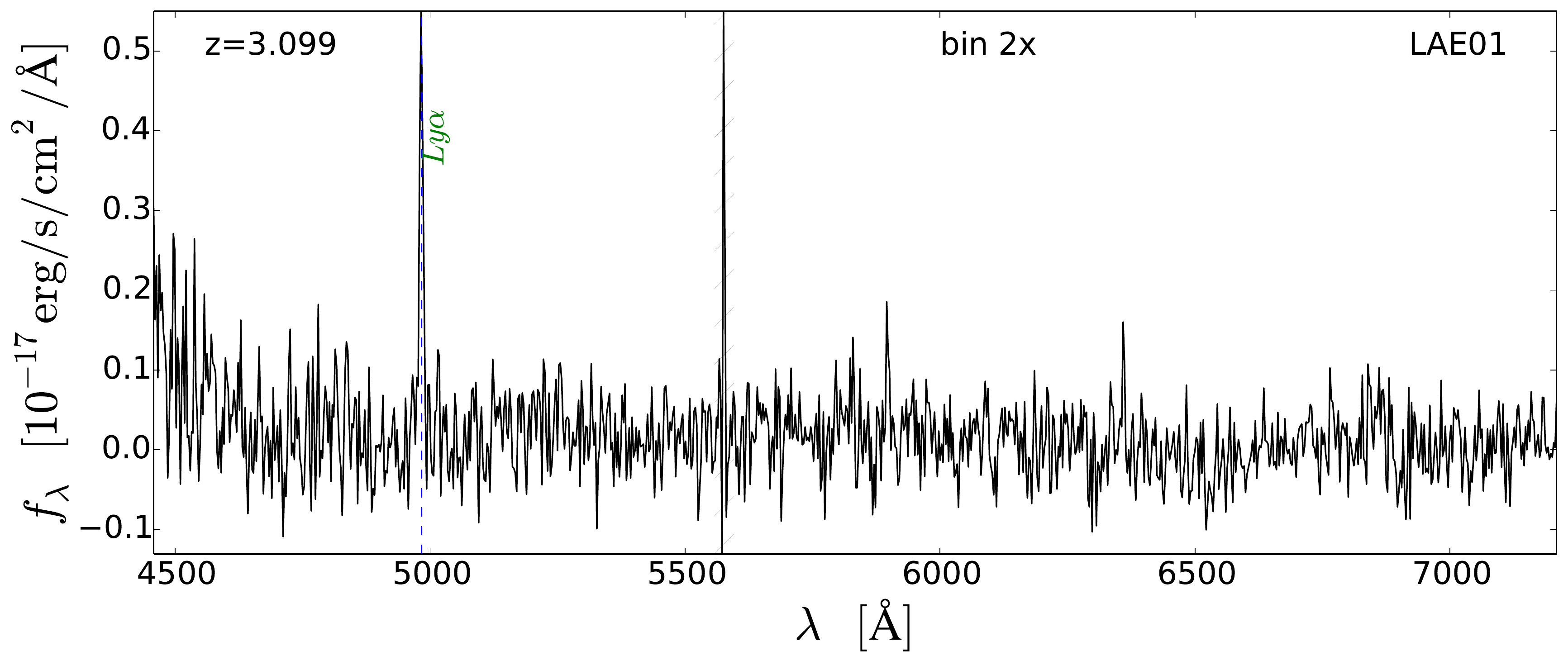}
\includegraphics[width=85mm]{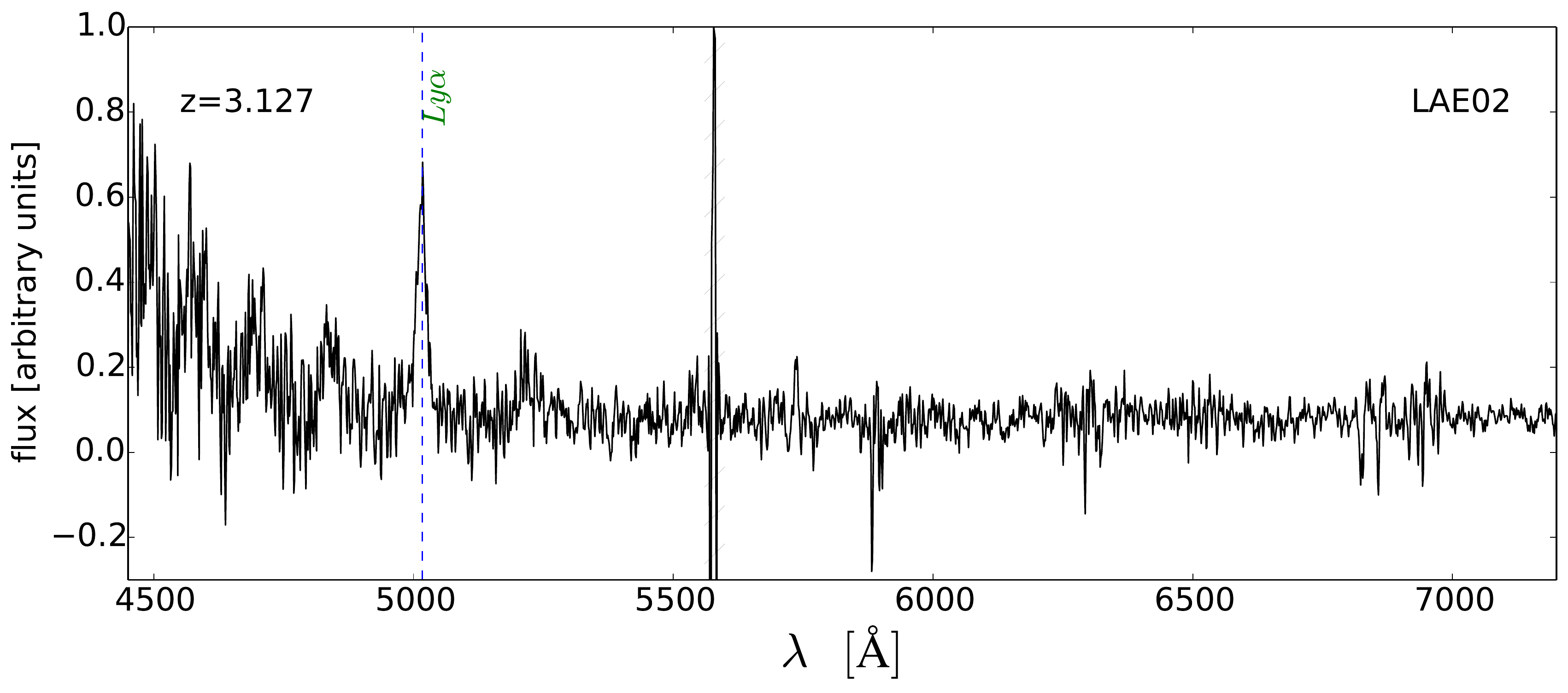}
\includegraphics[width=85mm]{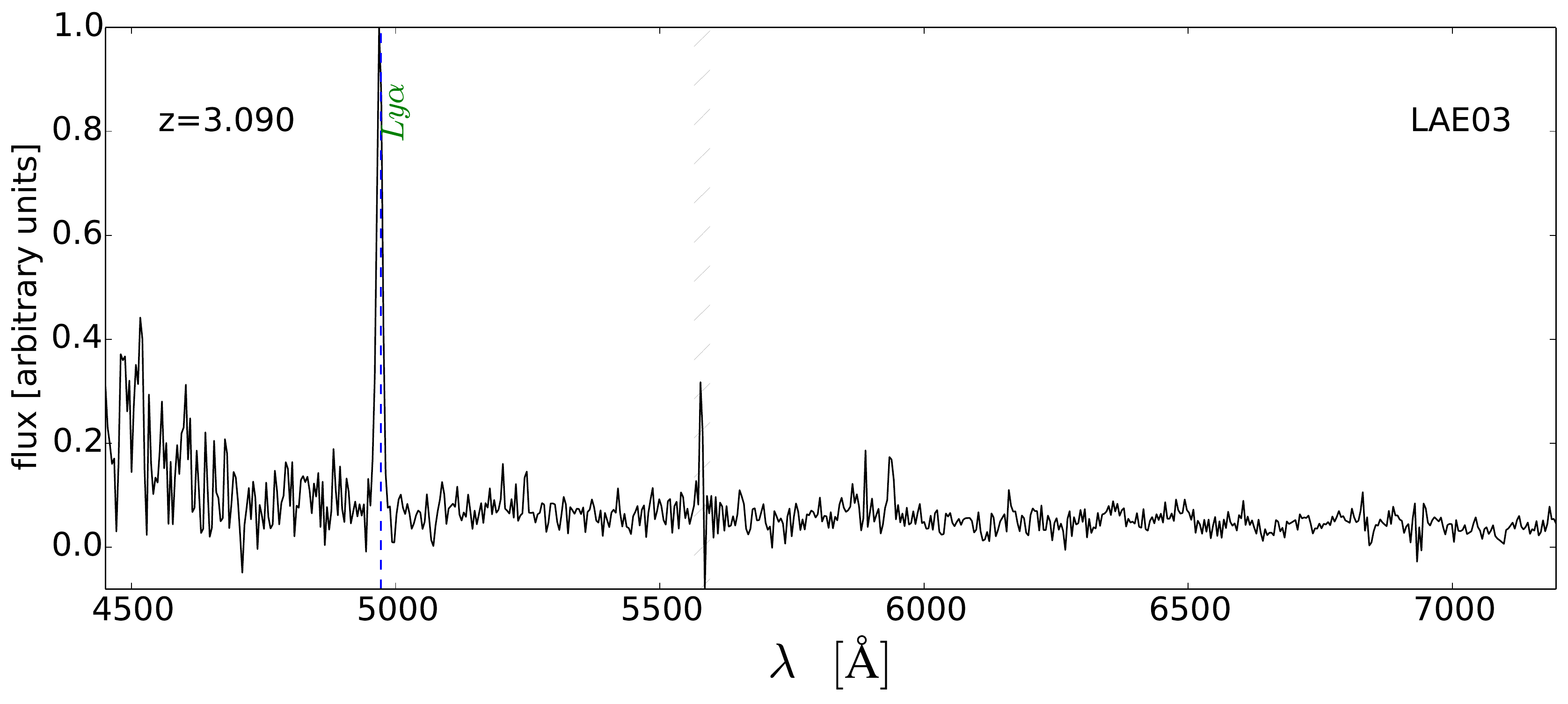}
\includegraphics[width=85mm]{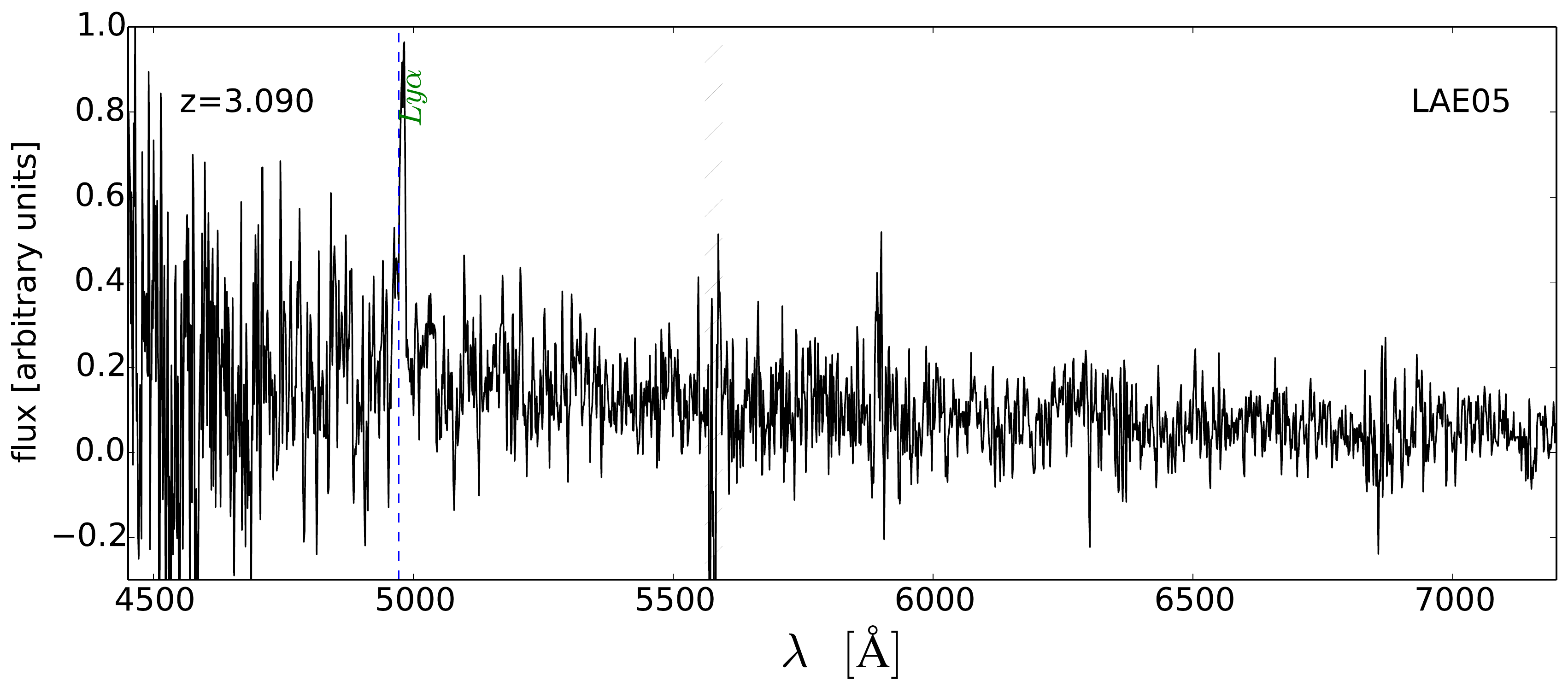}
\includegraphics[width=85mm]{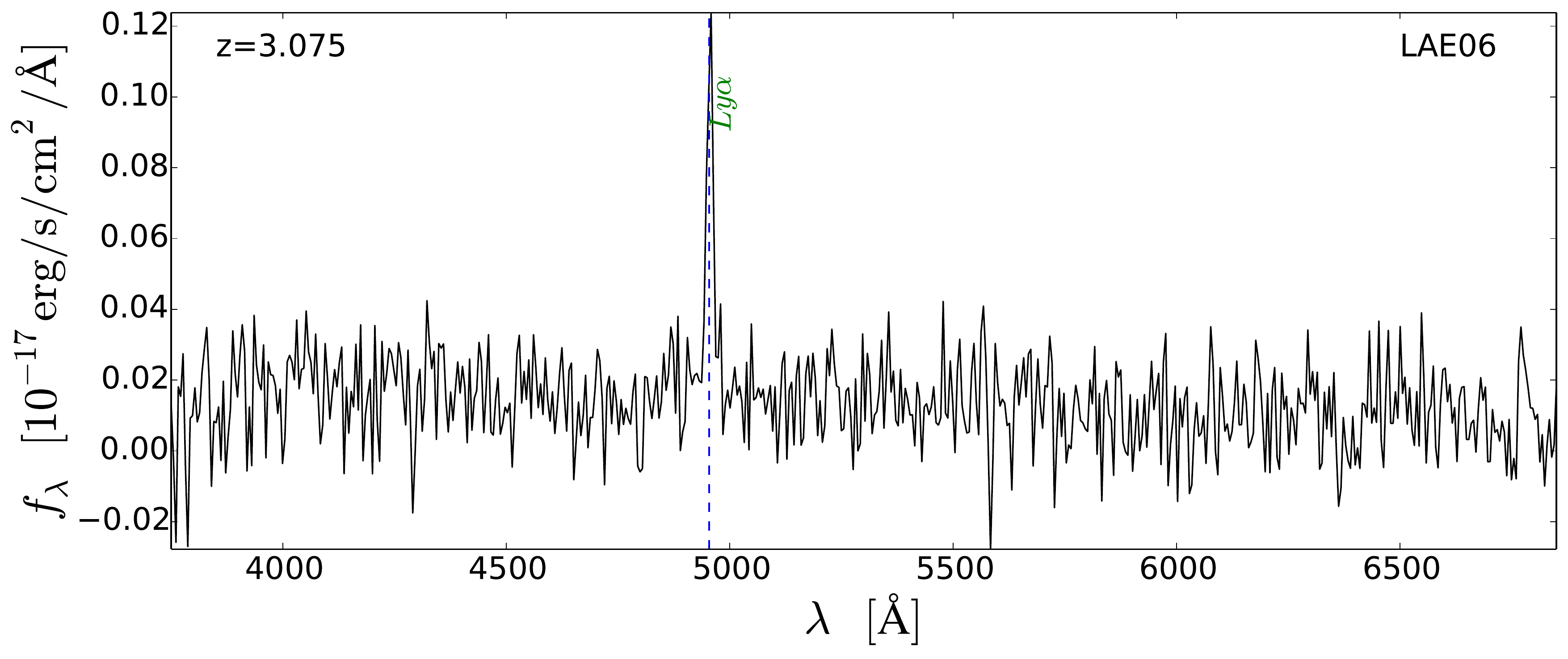}
\caption{Spectra of the \lyc~candidates from FOCAS $2003$ (LAE01), FOCAS $2010$ (LAE02, LAE03, LAE05), VIMOS $2008$ (LAE06, LAE07, LAE08, LAE10, LAE14, LAE15, LBG02), and VIMOS $2006$ (LBG01, LBG03).}\protect\label{fig:spectra}
\end{figure}

\begin{figure}
\centering\tiny
\includegraphics[width=85mm]{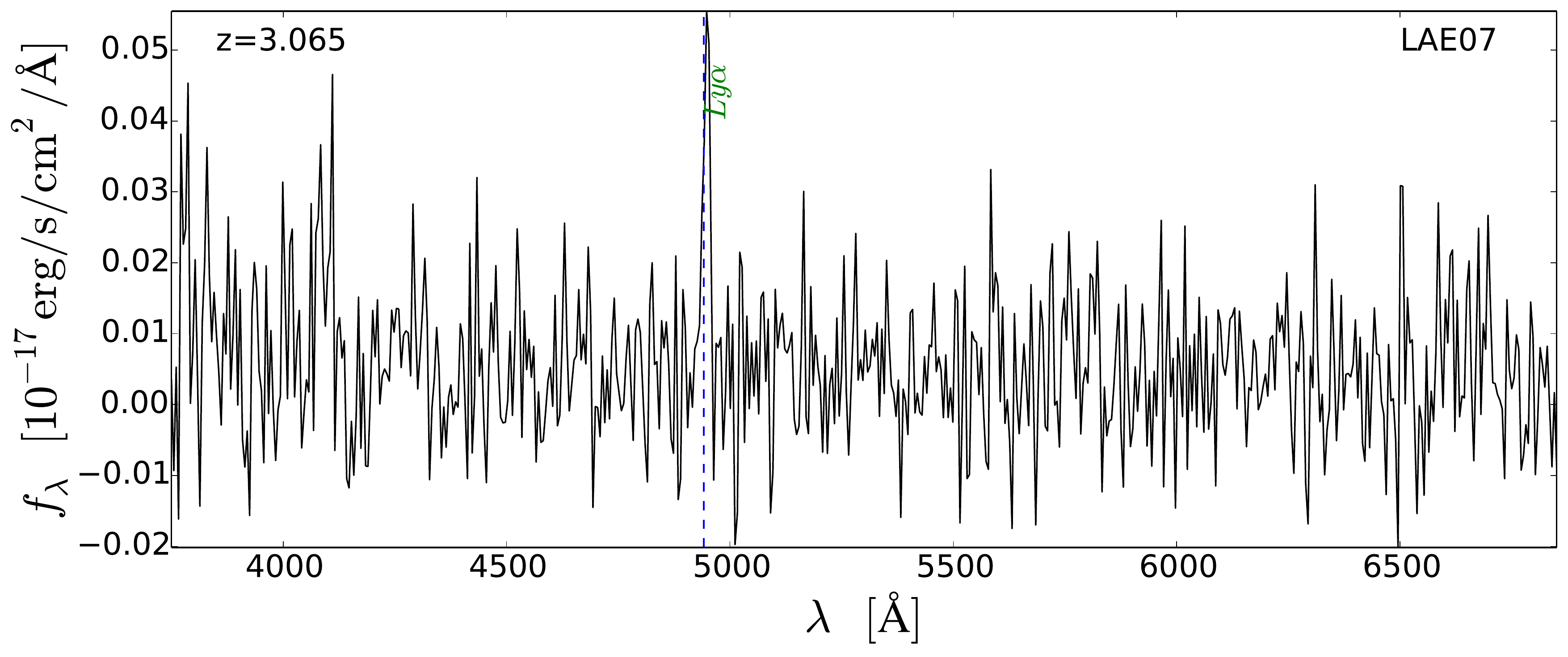}
\includegraphics[width=85mm]{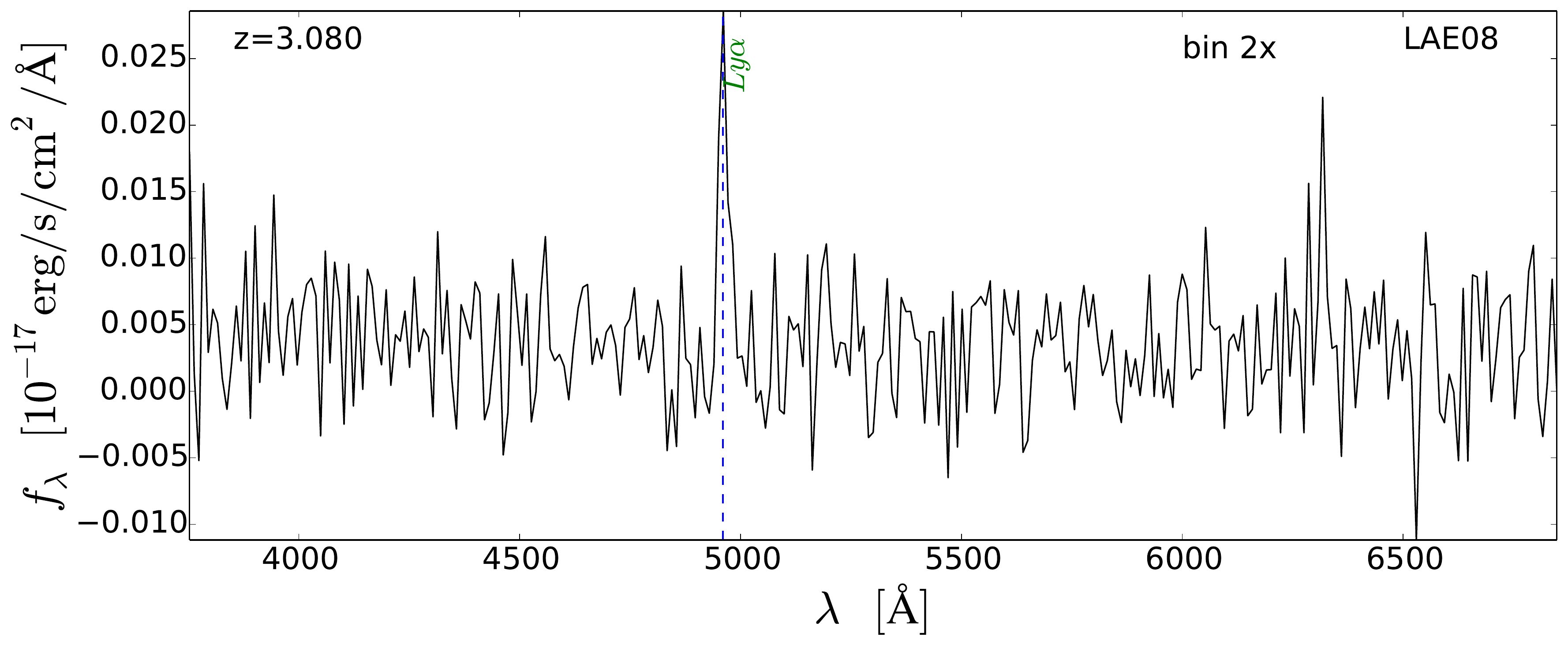}
\includegraphics[width=85mm]{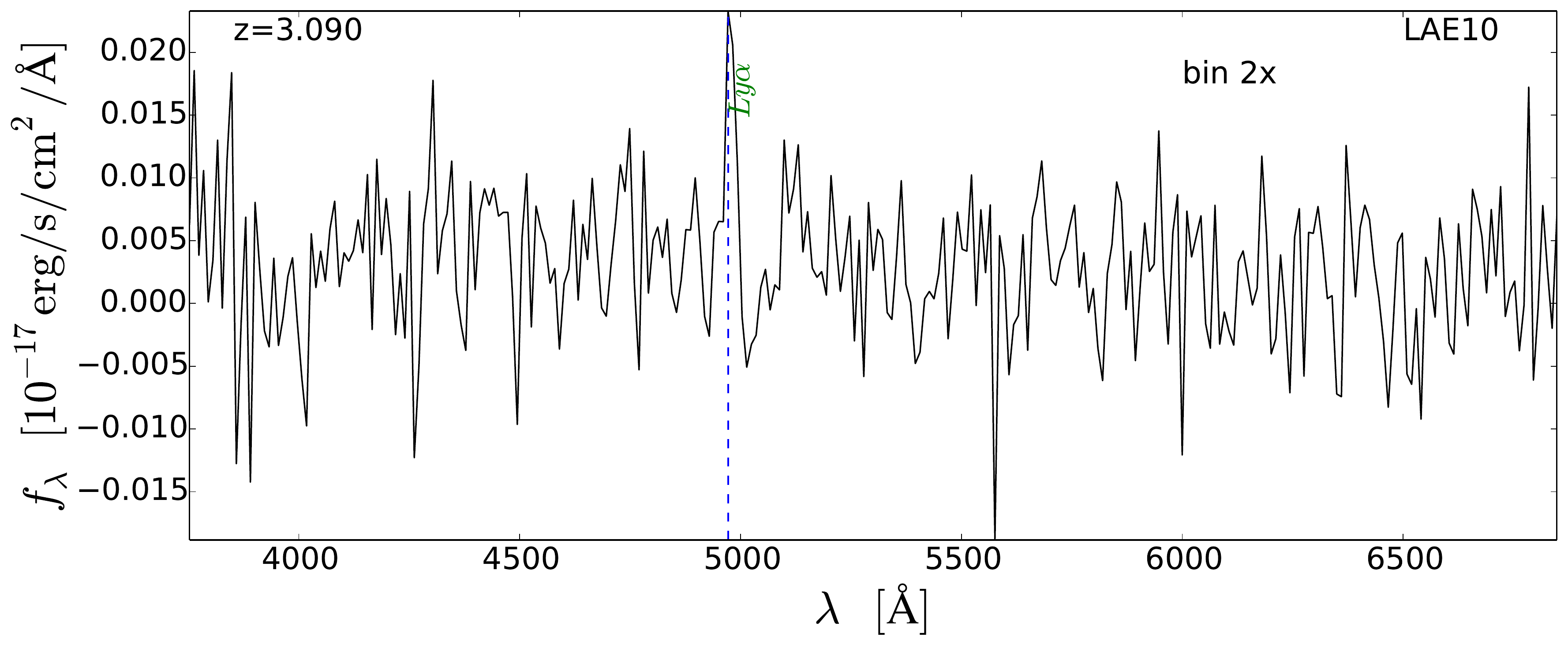}
\includegraphics[width=85mm]{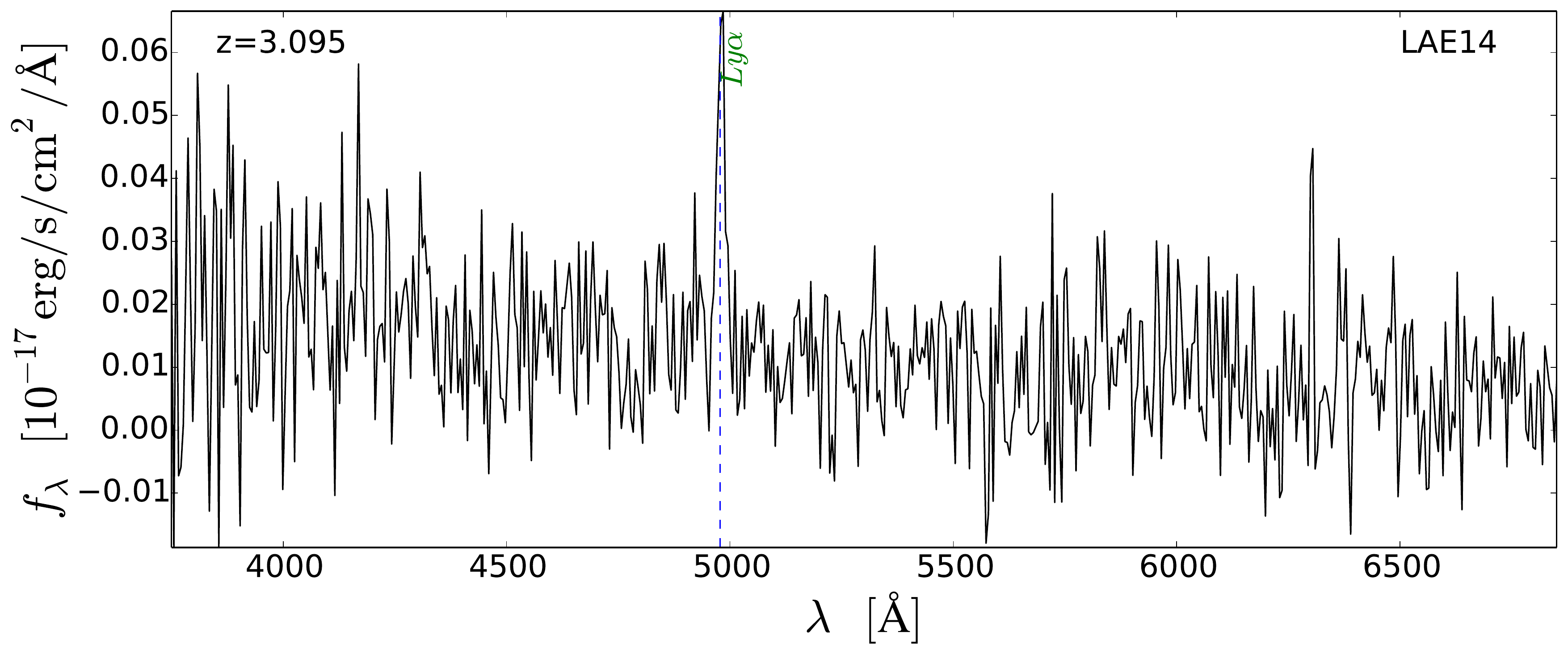}
\includegraphics[width=85mm]{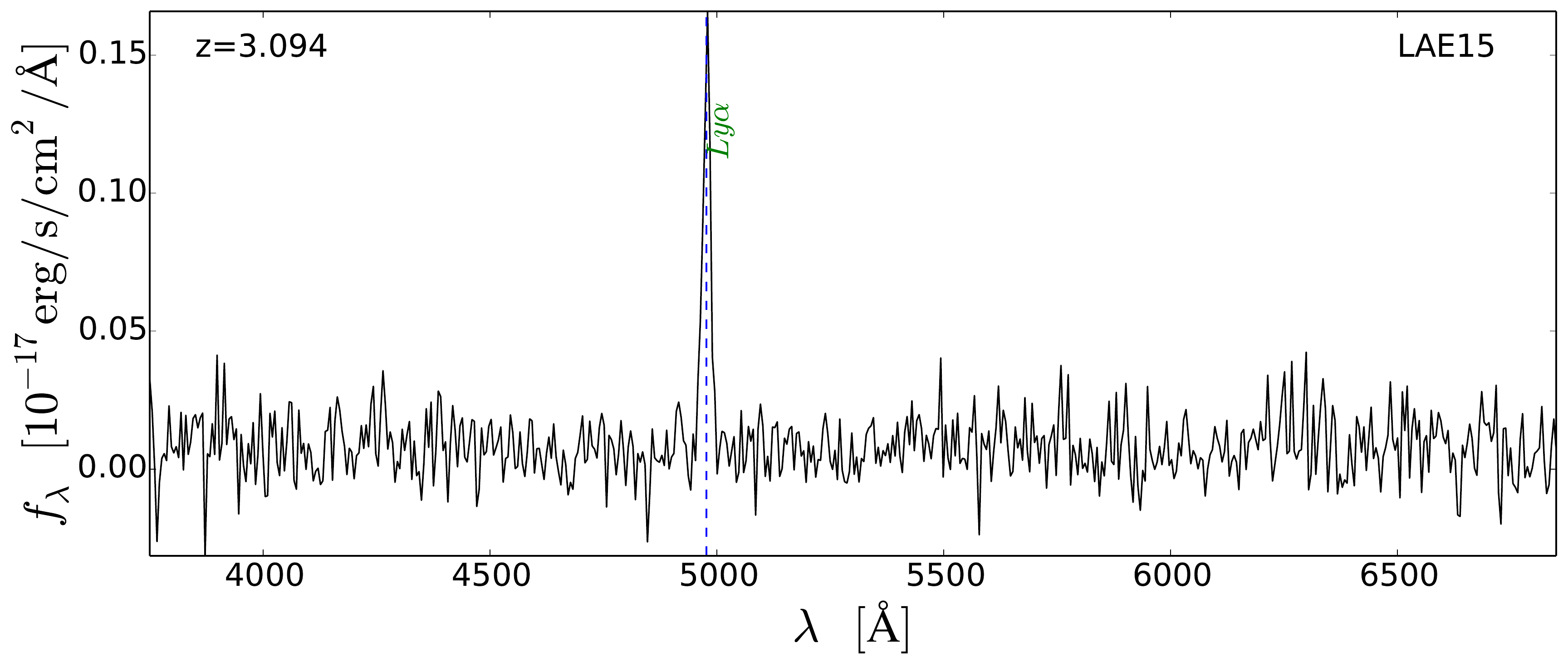}
\contcaption{}
\end{figure}

\begin{figure}
\centering\tiny
\includegraphics[width=85mm]{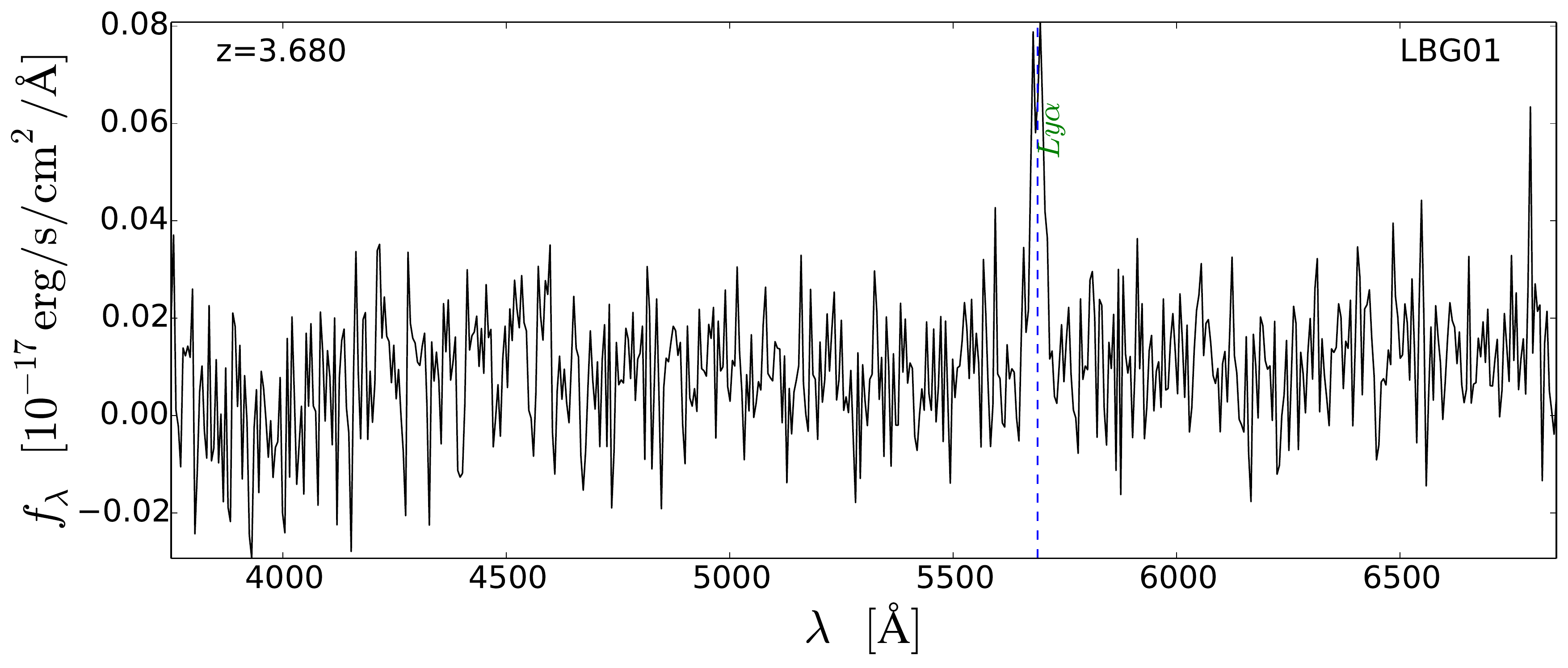}
\includegraphics[width=85mm]{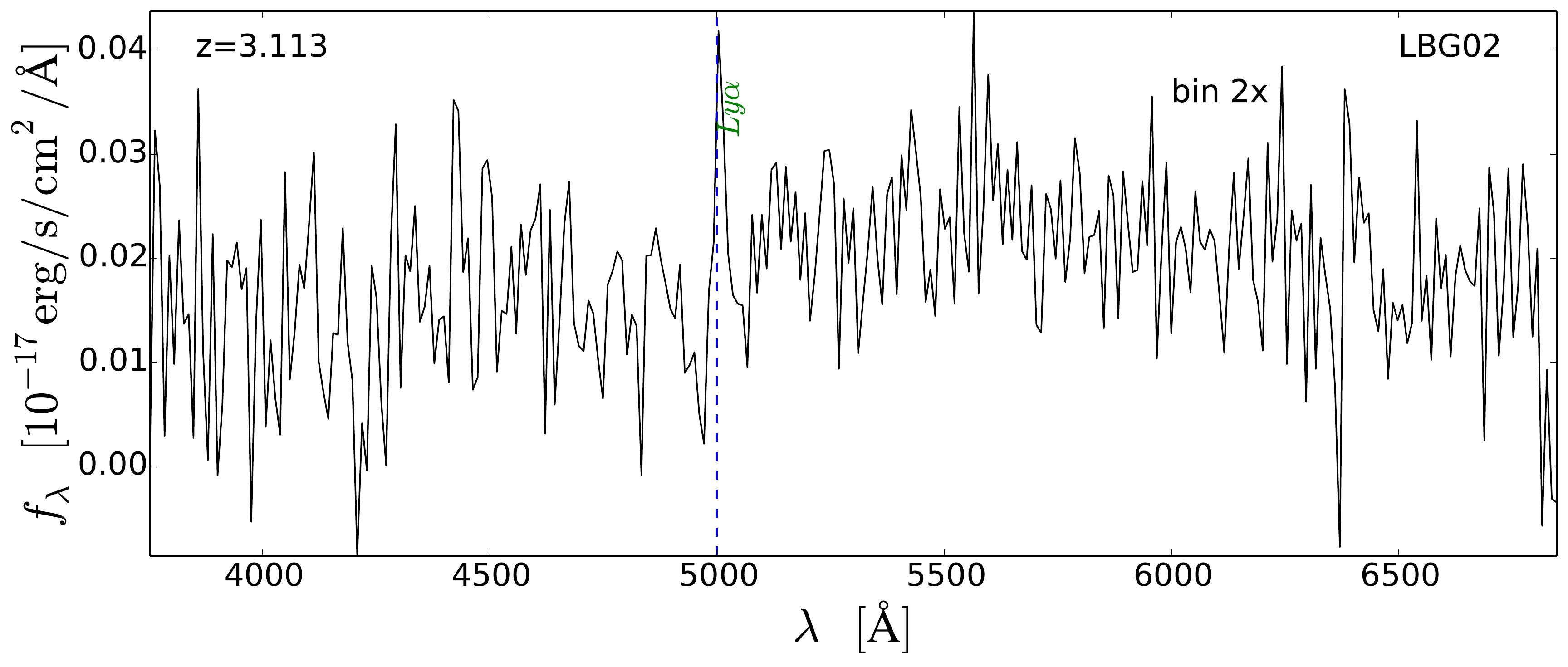}
\includegraphics[width=85mm]{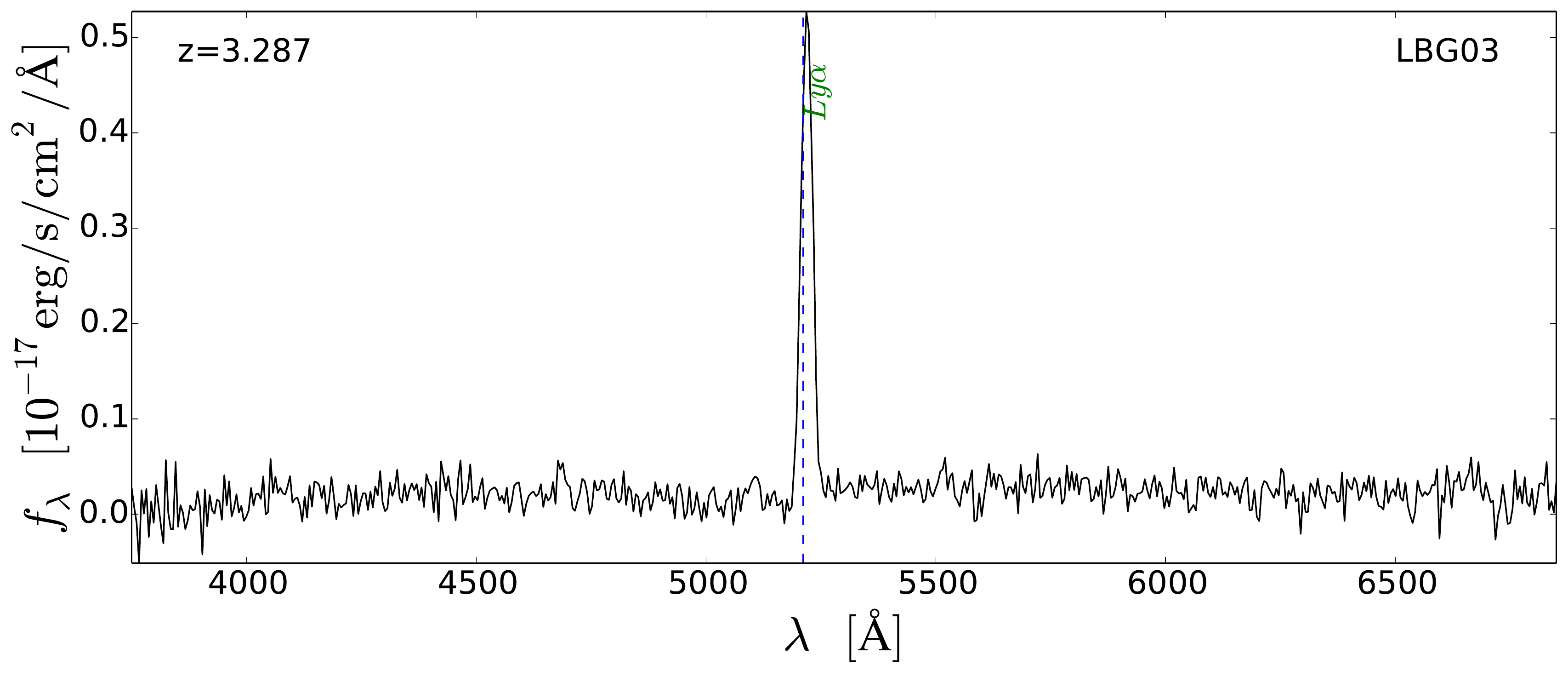}
\contcaption{}
\end{figure}

\section{Brightness variability}\protect\label{appendix:brightvar}
None of the non-AGN \lyc~candidates show any variability in brightness over the span of 6 years. The variability test we performed involved reducing \filterb~band Suprime-Cam data of the \ssa~field taken in the years $2002$, $2003$, $2007$, $2008$. After equalizing the zeropoint between the four years we compared the resulting four magnitudes per object, which revealed no variation of \lyc~candidates with the exception of two \lyc~AGN, which show variation with high significance. From the non-\lyc~objects only known X-ray AGN in our sample were found to vary. More details on the variability test and its results can be found in the forthcoming paper on AGN in the SSA22 field (Micheva et al. in preparation). We show the light curves of the LAE and LBG \lyc~candidates in Figure~\ref{fig:lightcurves}. LAE11 is too faint in the four individual frames and is not shown. LAE18 is too faint in the frame from $2003$, and is not shown.
\begin{figure}
\centering\tiny
\includegraphics[width=85mm]{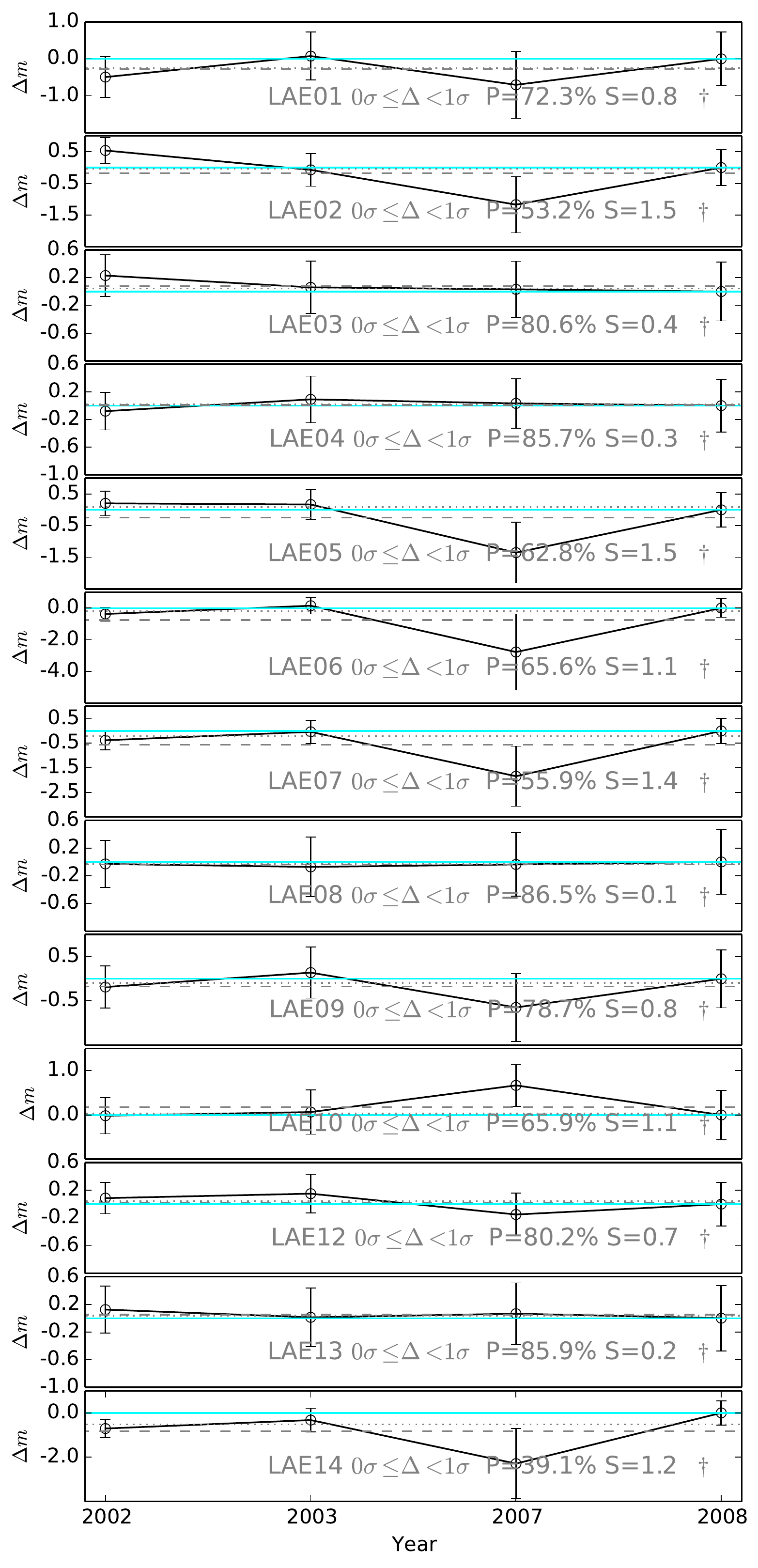}
\caption{Light curves for all \lyc~candidates. The source ID is indicated for each object, as well as the variability ``strength'' $\Delta$ in multiples of $\sigma$, the probability of a random match to the current profile (P in percent), and the formal variability significance (S) as defined in \citet{2007ApJ...665..225K}. Proto-cluster association is indicated by $\dagger$. Solid cyan line is at the reference year $2008$ of $\Delta m=0$. The dotted (dashed) line is the median (average) of the 4 points.  LAE11 and LAE18 are too faint in individual frames and could not be investigated.}\protect\label{fig:lightcurves}
\end{figure}
\begin{figure}
\centering\tiny
\includegraphics[width=85mm]{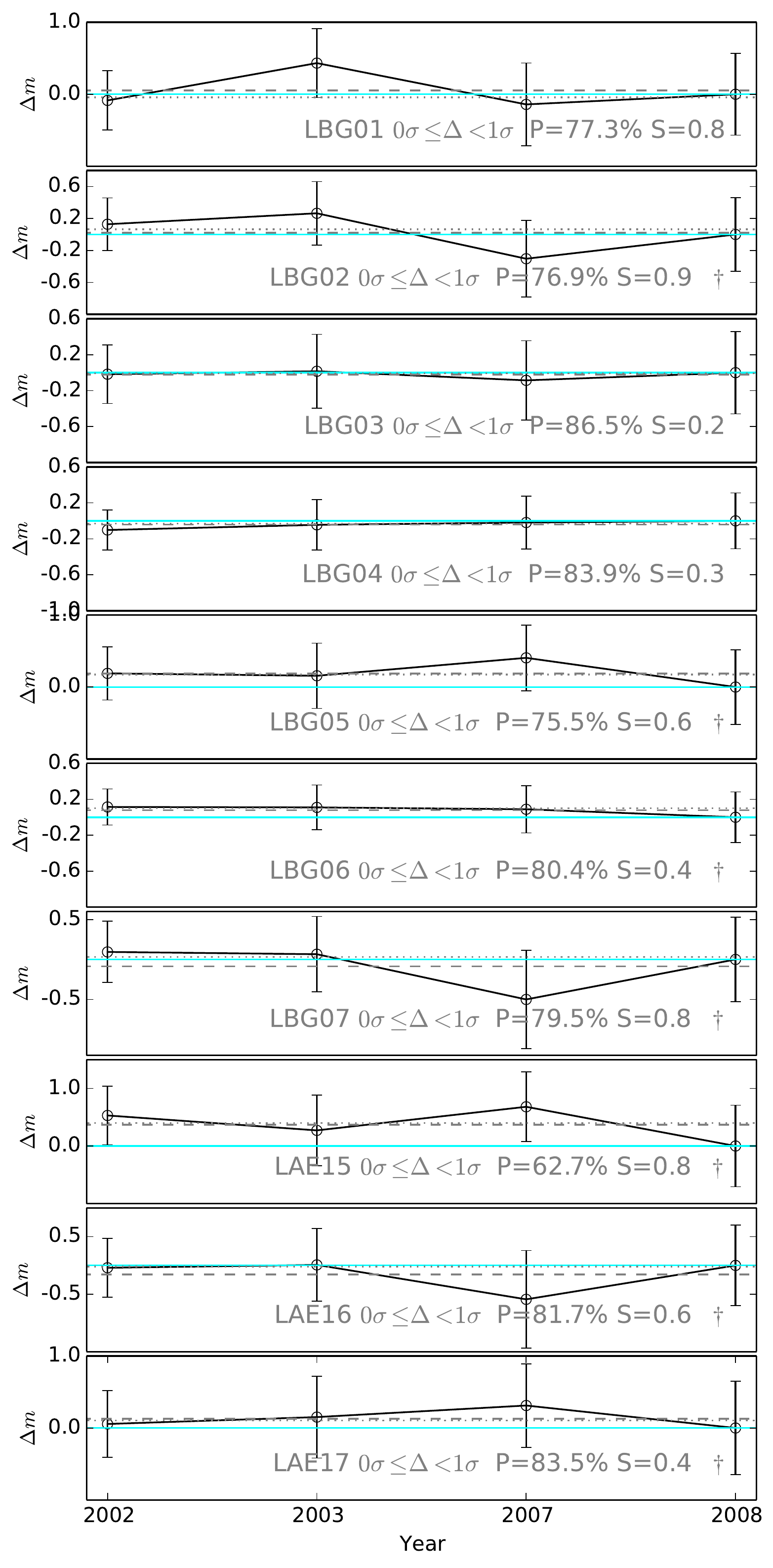}
\contcaption{}
\end{figure}

\label{lastpage}
\bibliographystyle{mn2e.bst}
\bibliography{lyc_ssa22}
\end{document}